\documentclass[useAMS,useusegraphicx]{mn2e}
\usepackage{graphics,graphicx}
\usepackage{aas_macros}
\usepackage{amssymb}
\usepackage{alltt}
\usepackage{epsfig}
\usepackage{subfigure}
\usepackage{textcomp}
\usepackage{epic,overpic,lpic}
\usepackage[usenames,dvipsnames]{color}
\newcommand{\be}{\begin{equation}}
\newcommand{\ee}{\end{equation}}
\newcommand{\ba}{\begin{eqnarray}}
\newcommand{\ea}{\end{eqnarray}}

\title[Mass and Light profiles of A1703, A370\& RXJ1347-11]{Detailed
  Cluster Mass and Light profiles of A1703, A370 \& RXJ1347-11 from
  Deep Subaru Imaging\thanks{Based on data collected at Subaru
    Telescope and obtained from the SMOKA, which is operated by the
    Astronomy Data Center, National Astronomical Observatory of Japan.
  }}

\author[E.~Medezinski et al.]{Elinor Medezinski,$^{1}\thanks{elinor@wise.tau.ac.il}$ Tom
  Broadhurst,$^{1}$ Keiichi Umetsu,$^{2}$ Masamune Oguri,$^{3}$
  \newauthor Yoel Rephaeli$^{1}$ and Narciso Ben\'\i tez$^{4}$\\
  $^{1}$School of Physics and Astronomy, Tel Aviv University, Tel Aviv 69978, Israel\\
  $^{2}$Institute of Astronomy and Astrophysics, Academia Sinica,
  P.~O. Box 23-141, Taipei 106, Taiwan, Repucblic of China\\
  $^{3}$Kavli Institute for Particle Astrophysics and Cosmology,
  Stanford University, 2575 Sand Hill Road, Menlo Park, CA 94025, USA\\
  $^{4}$Instituto de Astrof\'\i sica de Andaluc\'\i a (CSIC), Granada,
  Spain}

\begin{document}
\maketitle

\begin{abstract} 
  Weak lensing work can be badly compromised by unlensed foreground
  and cluster members which dilute the true lensing signal. We show
  how the lensing amplitude in multi-colour space can be harnessed to
  securely separate cluster members from the foreground and background
  populations for three massive clusters, A1703 ($z=0.258$), A370
  ($z=0.375$) and RXJ1347-11 ($z=0.451$) imaged with Subaru. The
  luminosity functions of these clusters when corrected for dilution,
  show similar faint-end slopes, $\alpha\simeq -1.0$, with no marked
  faint-end upturn to our limit of $M_R\simeq -15.0$, and only a mild
  radial gradient. In each case, the radial profile of the $M/L$ ratio
  peaks at intermediate radius, $\simeq 0.2r_{vir}$, at a level of
  $300-500(M/L_R)_\odot$, and then falls steadily towards $\sim
  100(M/L_R)_{\odot}$ at the virial radius, similar to the mean field
  level. This behaviour is likely due to the relative paucity of
  central late-type galaxies, whereas for the E/S0-sequence only a
  mild radial decline in $M/L$ is found for each cluster. We discuss
  this behaviour in the context of detailed simulations where
  predictions for tidal stripping may now be tested accurately with
  observations.
\end{abstract}                   


\section{Introduction}\label{Intro}

Galaxy Clusters are the largest virialised objects whose properties
reflect the initial spectrum of density perturbations and the
evolutionary history of dark matter (DM) and baryons.  Wide ranging
observations provide a wealth of information for examining the
relation between dark and visible matter and their radial profiles.
Unlike galaxies, where substantial cooling is predicted to have
concentrated baryons in the centres of DM halos, clusters are observed
to cool relatively inefficiently via thermal emission so that the gas
remains close to the virial temperature, making cluster mass profiles
easy to interpret, as corrections for gas cooling are not significant
\nocite{1986ApJ...301...27B, 2008MNRAS.390.1647B}({Blumenthal} {et~al.} 1986; {Broadhurst} \& {Barkana} 2008).

Lensing work is now able to achieve measurements of projected cluster
mass profiles with unprecedented detail, sufficient to usefully
address the distinctive prediction of a relatively shallow mass
profiles for CDM dominated halos (\nocite{1997ApJ...490..493N}{Navarro}, {Frenk} \&  {White}, 1997, hereafter
NFW; \nocite{2008MNRAS.390L..64D}{Duffy} {et~al.}, 2008). Combined
weak and strong lensing measurements have shown that the continuously
steepening form of the NFW profile is a reasonable description for the
mass profiles of three careful, independent studies
\nocite{2003ApJ...598..804K,2005A&A...443..793G,2005ApJ...619L.143B}({Kneib} {et~al.} 2003; {Gavazzi} 2005; {Broadhurst} {et~al.} 2005),
although with surprisingly high values derived for the profile
concentration parameter in each case \nocite{2005ApJ...619L.143B}({Broadhurst} {et~al.} 2005). More
recently, the concordance $\Lambda$CDM cosmology has been examined
critically with a somewhat larger sample of clusters, indicating that
the concentrations derived are significantly higher than predicted
over a wide range of cluster mass, after accounting for statistical
corrections for lensing-induced biases.  This is seen both in terms of
the size of the Einstein radius
\nocite{2008MNRAS.390.1647B,2009MNRAS.392..930O,2009MNRAS.396.1985Z}({Broadhurst} \& {Barkana} 2008; {Oguri} \& {Blandford} 2009; {Zitrin} {et~al.} 2009)
and the weak lensing (WL) mass profiles of several well known clusters
\nocite{2008ApJ...685L...9B,2008MNRAS.388.1759S,2009ApJ...695L.125L}({Broadhurst} {et~al.} 2008; {Sadeh} \& {Rephaeli} 2008; {Lapi} \& {Cavaliere} 2009).

Current detailed cluster mass and light profiles can now be used to
examine the ratio of $M/L$ all the way out to the virial radius.
Comparisons of the general DM distribution with the galaxy
distribution are potentially of great interest as the distribution of
galaxies is thought to be influenced by significant tidal forces which
may substantially modify massive galaxy halos reducing their masses
\nocite{1998MNRAS.300..146G,2000ApJ...544..616G,1999ApJ...523...32C,
  2001MNRAS.328..726S,2004MNRAS.348..333D,2004MNRAS.355..819G,
  2005ApJ...618..557N,2009ApJ...696.1771L}({Ghigna} {et~al.} 1998, 2000; {Col{\'{\i}}n} {et~al.} 1999; {Springel} {et~al.} 2001; {De Lucia} {et~al.} 2004; {Gao} {et~al.} 2004; {Nagai} \& {Kravtsov} 2005; {Limousin} {et~al.} 2009).  Mass loss calculated over
Hubble time in $\Lambda$CDM predict losses of $\sim 30\%$ for orbits
extending to the virial radius, with typically a loss of $\sim 70\%$
near cluster centres, so that the ratio of integrated galaxy mass to
the general DM distribution falls steadily towards the cluster centre
\nocite{2005ApJ...618..557N}({Nagai} \& {Kravtsov} 2005).

Stars are not expected to be stripped significantly in these
simulations because of their location deep in the central potential of
a galaxy. Hence, for clusters the galaxy light should simply trace
mass, assuming both are collisionless, so that a fairly constant $M/L$
is expected \nocite{2004MNRAS.355..819G,2005ApJ...618..557N}({Gao} {et~al.} 2004; {Nagai} \& {Kravtsov} 2005).  The
presence of high-mass galaxies in the inner core, resulting from both
higher merger rates and the effect of dynamical friction, explains the
reduced value of the integrated $M/L$ ratio within a radius of $\leq
200h^{-1}$~kpc for the highest mass clusters
\nocite{2005ApJ...618..557N,2001ApJ...560..636E}({Nagai} \& {Kravtsov} 2005; {El-Zant}, {Shlosman} \&  {Hoffman} 2001).

Reliable $M/L$ profiles of galaxy clusters have begun to emerge
recently based on dynamical
\nocite{2000AJ....120.2338R,2004AJ....128.1078R}({Rines} {et~al.} 2000, 2004) and lensing based
methods \nocite{2007ApJ...663..717M}({Medezinski} {et~al.} 2007).  For the local cluster A576, a
careful dynamically-based $M/L$ profile has been obtained with caustic
work by \nocite{2000AJ....120.2338R}({Rines} {et~al.} 2000), who note a puzzling declining
behavior, with a peak around $M/L\sim 600M/L_R$ at
$r\sim100h^{-1}$~kpc, declining to a limiting $100M/L_B$, at $r\sim
1h^{-1}$~Mpc and beyond. Other dynamical estimates from stacking
together samples of clusters also show this overall behavior, where
the $M/L$ ratio may peak around 10\% of the virial radius
\nocite{2004ApJ...600..657K}({Katgert}, {Biviano} \&  {Mazure} 2004) and declines thereafter
\nocite{2000AJ....120.2338R,2004AJ....128.1078R}({Rines} {et~al.} 2000, 2004).

Our earlier lensing based determination of the $M/L$ profile of
A1689 \nocite{2007ApJ...663..717M}({Medezinski} {et~al.} 2007) shows behaviour similar to
the above dynamical work, with a peak of $M/L_B\sim 400(M/L)_{\odot}$ at
$r\sim 100h^{-1}$~kpc, and a steady decline to $\sim 100M/L_B$ at the
virial radius, $r\sim 2h^{-1}$~Mpc \nocite{2007ApJ...663..717M}({Medezinski} {et~al.} 2007). In
the case of A576 \nocite{2000AJ....120.2338R}({Rines} {et~al.} 2000) and in our own work on
A1689 \nocite{2007ApJ...663..717M}({Medezinski} {et~al.} 2007), careful corrections are made for
the trend of $M/L$ of the stellar population with cluster radius,
using the measured colours of cluster member galaxies which favour
predominately redder, earlier-type galaxies in the central region.  In
both these differing analyses the colour trend was found to be weak
and not nearly sufficient to account for the measured variation of the
$M/L$ profile, leaving a puzzling result which is generally at odds
with the known result that $M/L$ is an increasing function of scale.
Clearly, it is important to extend the investigation of the $M/L$
profile to more clusters to examine the generality of our earlier work
and to examine possible trends with other observables in a more
statistically robust way.

Our previous work demonstrated the importance of carefully selecting a
background population to avoid contamination by the lensing cluster.
It is not sufficient to simply exclude a narrow band containing the
obvious E/S0 sequence, following common practice, because the lensing
signal of the remainder bluer objects is found to fall rapidly towards
the cluster centre relative to the background of red lensed galaxies
indicating that unlensed cluster members are present blueward of the
sequence diluting the true lensing signal \nocite{2005ApJ...619L.143B}({Broadhurst} {et~al.} 2005).
Due to this effect, many WL studies have underestimated the masses and
concentrations of cluster mass profiles \nocite{2009arXiv0912.2356L}({Lu} {et~al.} 2009), a
point now widely acknowledged
\nocite{2007ApJ...668..643L,2009ApJ...699.1038O,2009arXiv0903.1103O}({Limousin} {et~al.} 2007; {Oguri} {et~al.} 2009; {Okabe} {et~al.} 2009).
This effect resolved the long standing discrepancy between the strong
and WL effects, with the WL signal underpredicting the observed
Einstein radius by a factor of $\sim 2.5$
\nocite{2001A&A...379..384C,2005A&A...434..433B}({Clowe} \& {Schneider} 2001; {Bardeau} {et~al.} 2005), based only on a
minimal rejection of obvious cluster members using one or two-band
photometry. Red background galaxies behind clusters have since been
explored carefully with deep wide-field spectroscopy by
\nocite{2008AJ....135.1837R}{Rines} \& {Geller} (2008), who emphasize that the red population in
their cluster data comprises exclusively background objects.

This lensing ``dilution'' effect may be turned to our advantage as a
means of correcting the cluster population for background
contamination, extending the work of \nocite{2007ApJ...663..717M}{Medezinski} {et~al.} (2007). Here
we analyze very high quality Subaru images of A1703, A370, \&
RXJ1347-11 for which we have accurate WL measurements. For these
clusters, a wider range of colours is available than for A1689, which
we make full use here to improve the separation of cluster members
from foreground and background populations using the colour-colour
(CC) diagram and application of our dilution method.  Using the
dilution method has the advantage that it does not require the actual
subtraction of background galaxies which introduces the inherent
uncertainty caused by the clustering of the background signal and the
attendant requirement for the accurate photometry of control fields.

In \S~\ref{sec:data} we present the cluster observations, the data
reduction, and in \S~\ref{sec:samples} we explain how we select
cluster galaxies, background samples, and foreground samples. In
\S~\ref{sec:tracks} we examine evolutionary tracks to reinforce the
sample selection criteria, and compare with blank-field surveys to
determine redshift distributions in \S~\ref{sec:photo-z}. In
\S~\ref{dist_Sub} we describe the WL analysis, and in
\S~\ref{dilution} we describe how the WL measurements are used to
determine the cluster fraction using our WL dilution method. In
\S~\ref{lum_sec} we derive the light profile of each cluster, and
compare with the mass profile to obtain the $M/L$ profile in
\S~\ref{ML_sec}. In \S~\ref{lum_func} we investigate the shape and
variation of the cluster luminosity functions (LF) and its radial
dependence.  Finally, we summarize our findings in
\S~\ref{discussion}.

The $\Lambda$CDM cosmological model is adopted, with $\Omega_M=0.3$,
and $\Omega_{\Lambda}=0.7$, but with $h$ left in units of $H_0/100$ km
s$^{-1}$ Mpc$^{-1}$, for easier comparison with earlier work.

\section{Subaru Data reduction}
\label{sec:data}

%mnras table
\begin{table*}
  \caption{The Cluster Sample: Redshift and Filter Information} 
  \begin{tabular}{@{\hspace{0.01cm}}cc@{\hspace{0.1cm}}c@{\hspace{0.2cm}}c@{\hspace{0.2cm}}c@{\hspace{0.1cm}}c@{\hspace{0.1cm}}c@{\hspace{0.2cm}}c} 
    \hline
    {Cluster} & {$z$} & {$r_{vir}^1$} & {Filters} & {Total exposure}
      & {$m_{lim}$}$^2$
 &  {Seeing} &  {Detection}\\
    & &{($h^{-1}$~Mpc})&  &  time (sec) & (AB mag) & (arcsec)& {Band}\\    
    \hline
    A1689 & 0.183 &1.9& $V_{\rm J}$ & 1920& 26.5 & 0.82 & $i'$\\
          &       &   & $i'$        & 2640& 25.9 & 0.88 & \\
    A1703 & 0.258 &1.7& $g'$        & 1200& 27.4 & 0.89 & $r'$\\
          &       &   & $r'$        & 2100& 26.9 & 0.78 & \\
          &       &   & $i'$        & 1200& 26.1 & 0.8  & \\  
     A370 & 0.375 &2.1& $B_{\rm J}$ & 7200& 27.4 & 0.72 & $R_{\rm C}$\\
          &       &   & $R_{\rm C}$ & 8340& 26.9 & 0.6  & \\  
          &       &   & $z'$        &14221& 26.1 & 0.7  & \\  
 RXJ1347-11 &0.451&1.6& $V_{\rm J}$ & 1800& 26.5 & 0.7  & $R_{\rm C}$\\
          &       &   & $R_{\rm C}$ & 2880& 26.7 & 0.76 & \\  
          &       &   & $z'$        & 4860& 25.5 & 0.6  & \\  

    \hline\label{tab:cluster}
%\footnotesize{$^1$Limiting magnitude for a $3\sigma$ detection within a $2\arcsec$ aperture}
  \end{tabular}
  \flushleft \footnotesize{$^1$based on \nocite{2008ApJ...685L...9B}{Broadhurst} {et~al.} (2008)}
  \flushleft \footnotesize{$^2$limiting magnitude for a $3\sigma$ detection within a $2\arcsec$ aperture}
\end{table*}

We analyze deep images of three intermediate-redshift clusters, A1703
($z=0.258$), A370 ($z=0.375$) and RXJ1347-11 ($z=0.451$), observed
with the wide-field camera Suprime-Cam \nocite{2002PASJ...54..833M}({Miyazaki} {et~al.} 2002) in
several optical bands, at the prime focus of the 8.3m Subaru
telescope. The cluster A1703 was observed on 2007 June 15 and the WL
signal analyzed by \nocite{2008ApJ...685L...9B}{Broadhurst} {et~al.} (2008) and
\nocite{2009ApJ...699.1038O}{Oguri} {et~al.} (2009).  The observations of the other two
clusters are available in the Subaru archive,
SMOKA\footnote{http://smoka.nao.ac.jp.}. Subaru reduction software
\nocite{2002AJ....123...66Y,2004ApJ...611..660O}(SDFRED; {Yagi} {et~al.} 2002; {Ouchi} {et~al.} 2004) is used for
flat-fielding, instrumental distortion correction, differential
refraction, sky subtraction, and stacking.  Photometric catalogs are
created using {\sc SExtractor} \nocite{1996A&AS..117..393B}({Bertin} \& {Arnouts} 1996) in
dual-image mode, for which we use our deepest band as the detection
image. Since our work relies heavily on the colours of galaxies, we
prefer using isophotal magnitudes.  Astrometric correction is done
with the {\sc scamp} tool \nocite{2006ASPC..351..112B}({Bertin} 2006) using reference
objects in the NOMAD catalog \nocite{2004AAS...205.4815Z}({Zacharias} {et~al.} 2004) and the SDSS
DR6 \nocite{2008ApJS..175..297A}({Adelman-McCarthy} {et~al.} 2008) where available.  The clusters
observation details are listed in Table~\ref{tab:cluster}.

%Photometric calibration
Photometric zeropoints were calculated from associated standard star
observations and also from independent well calibrated photometry,
where available. In the case of A1703 and A370, no standard star
observations were available, so zeropoints were derived independently
in similar passbands by comparing to magnitudes of stars observed in
SDSS fields \nocite{2009ApJ...699.1038O}({Oguri} {et~al.} 2009). Conversions from SDSS $g,r$
bands to Subaru $B$, $R_{\rm C}$ bands were done using the relations
given in Lupton (2005) for the SDSS. For RXJ1347-11, standard stars
observed during the observation nights were used to derive photometric
zeropoints in the $B,R_{\rm C}$ and $z'$ bands. For the z'-band only
three Standard stars were unsaturated in the Subaru image. To better
constrain the z'-band, and as a consistency check for the $B_{\rm J}$ \&
$R_{\rm C}$-bands, unsaturated stars were compared with the $HST$/ACS
images of RXJ1347-11 in the $F475W, F814W$ \& $F850LP$ filters,
finding good consistency to within $\pm0.05$~mag.

\section{Sample selection from the colour-colour diagram}
\label{sec:samples}

For each cluster we use Subaru observations in three broad-bands,
which differ in their choice of passbands between the clusters, though
all observations are deep and taken in conditions of good seeing,
representing some of the highest quality imaging of any target by
Subaru in terms of depth, resolution, and colour coverage.  Using
three bands in terms of two-colour space for object selection will
help separate different populations and improve subsequent WL and
light measurements of the cluster, as we now demonstrate.

\subsection{Cluster Members}

\begin{figure}
  \begin{overpic}[scale=.50]{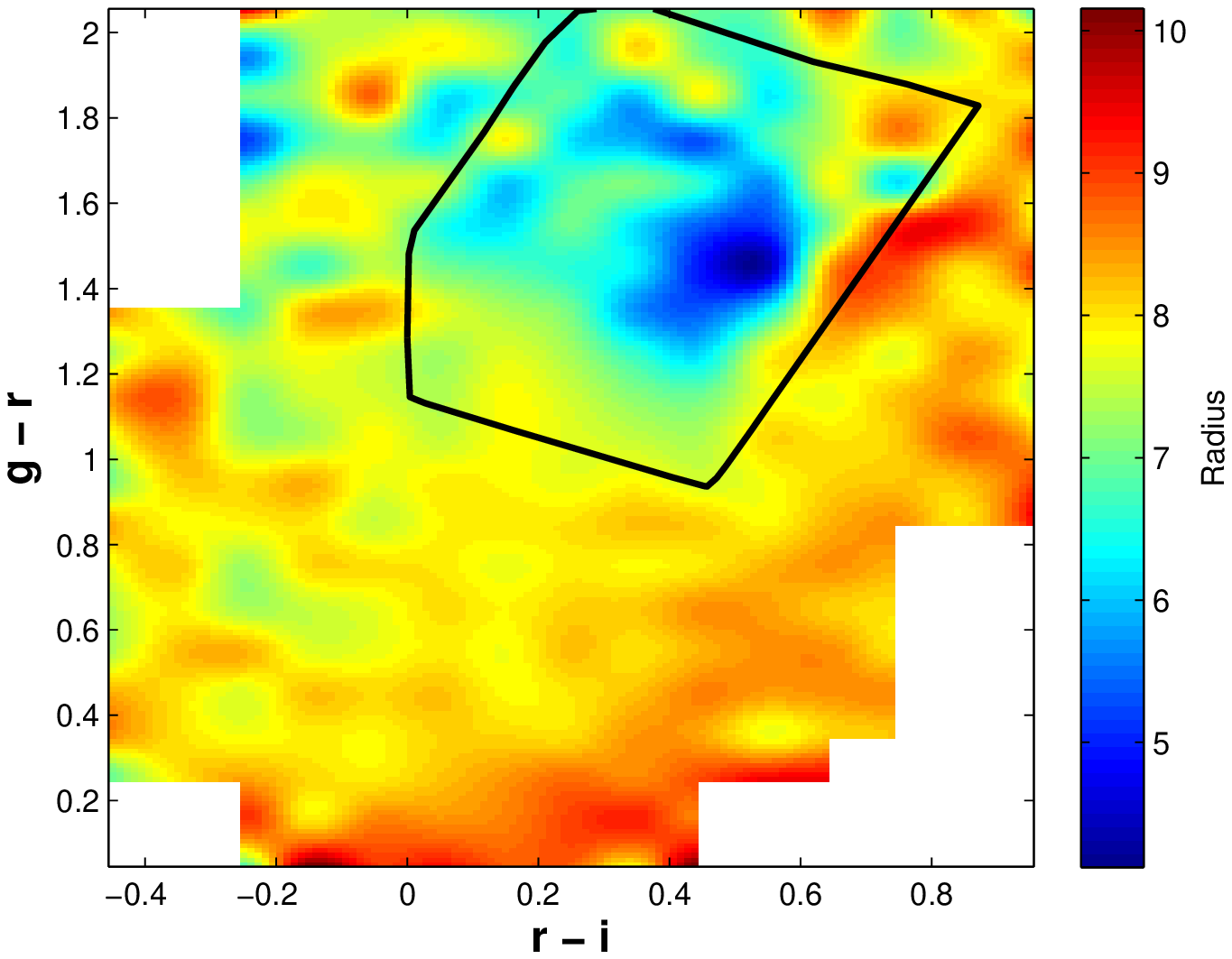}
    \put(15,64){\framebox(12,5){\bf A1703}}\end{overpic}
%  \begin{lpic}{A1703/A1703_colorcolor_rcells_090317(5cm)}
%    \lbl[tlW]{15,64\bf A1703}\end{lpic}
  \begin{overpic}[scale=.50]{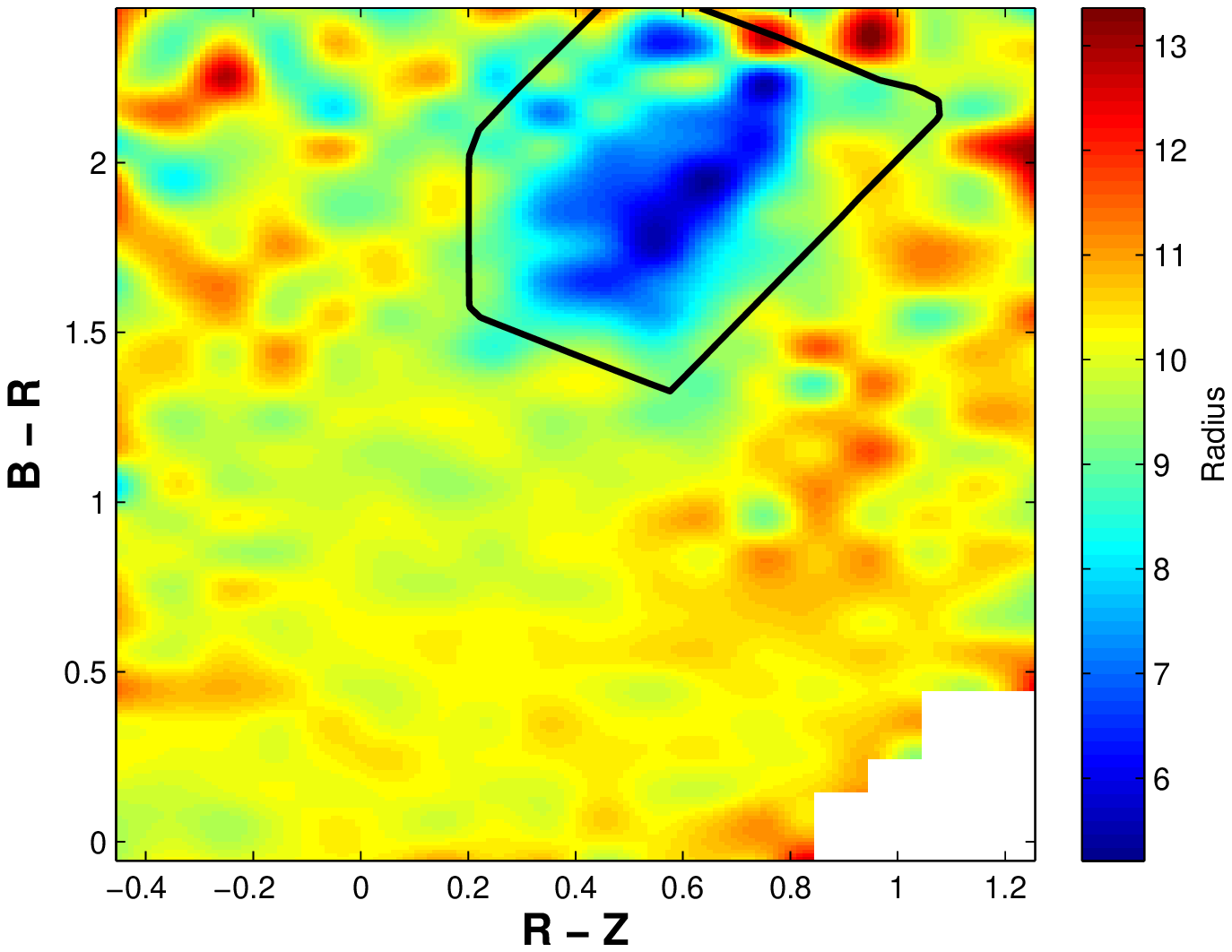}
    \put(15,64){\framebox(10,5){\bf A370}}\end{overpic}
  \begin{overpic}[scale=.50]{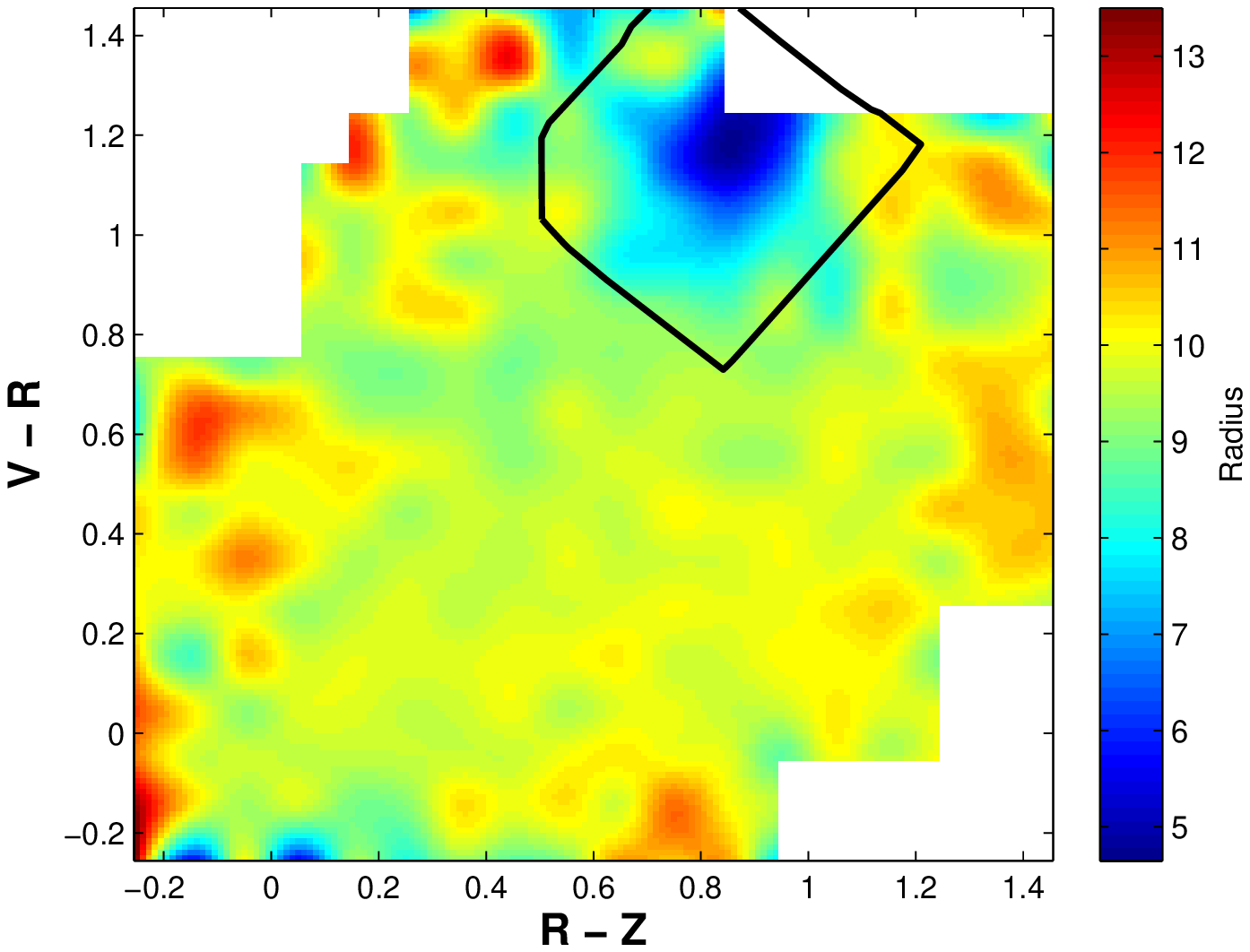}
    \put(15,64){\framebox(24,5){\bf RXJ1347-11}}\end{overpic}
  \caption{Averaged radius from cluster centre displayed in CC space
    for A1703, A370 \& RXJ1347-11 (Top to bottom).  The bluer (darker)
    colours imply average lower radius, hence correspond to the
    location of the cluster in CC space. The black box marks the
    boundaries of the green sample we select, which conservatively
    includes all cluster members.}
  \label{colorcolor_rcell}
\end{figure}

\begin{figure}
  \begin{overpic}[scale=.50]{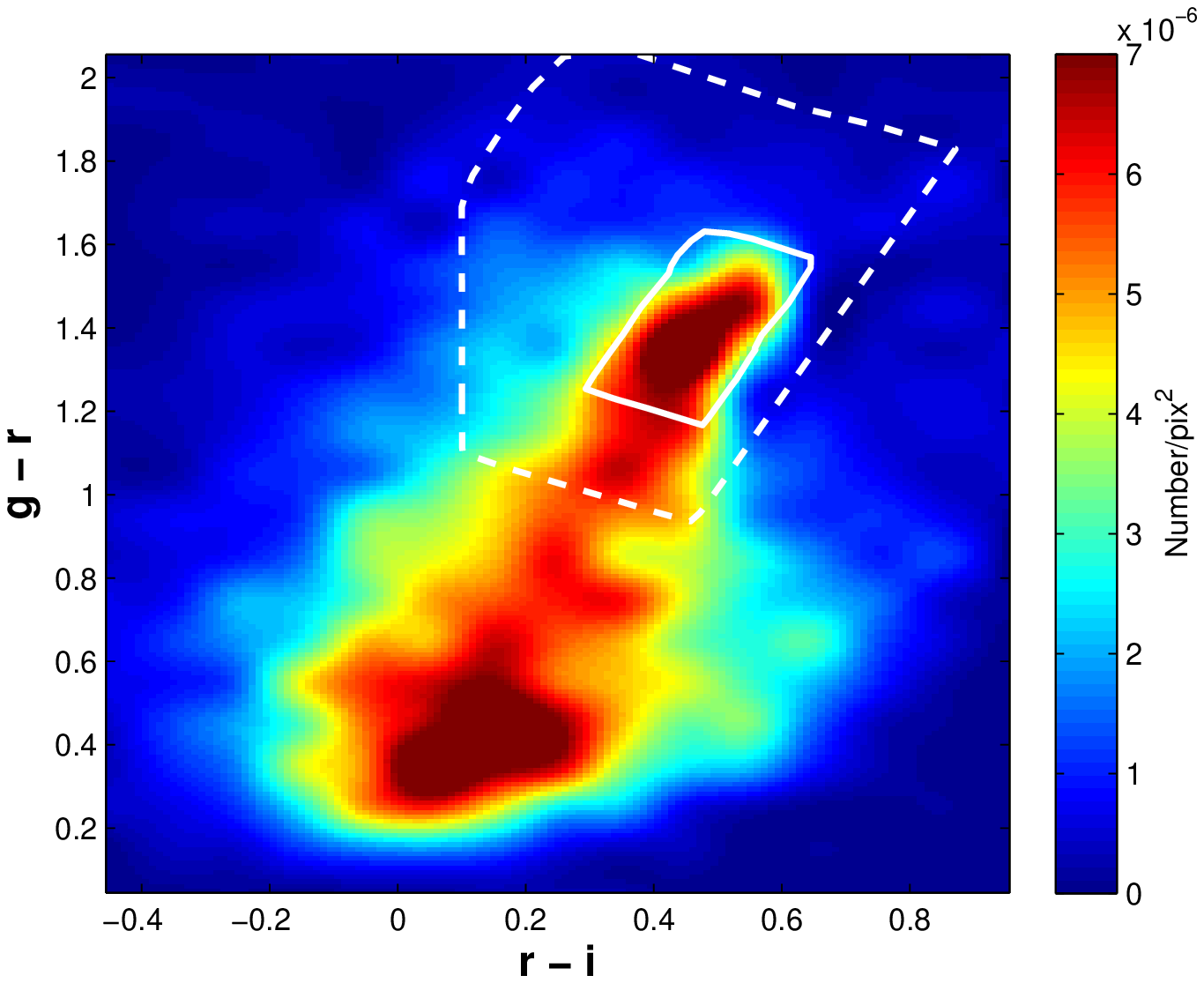}
    \put(15,64){\framebox(12,5){\bf A1703}}\end{overpic}
  \begin{overpic}[scale=.50]{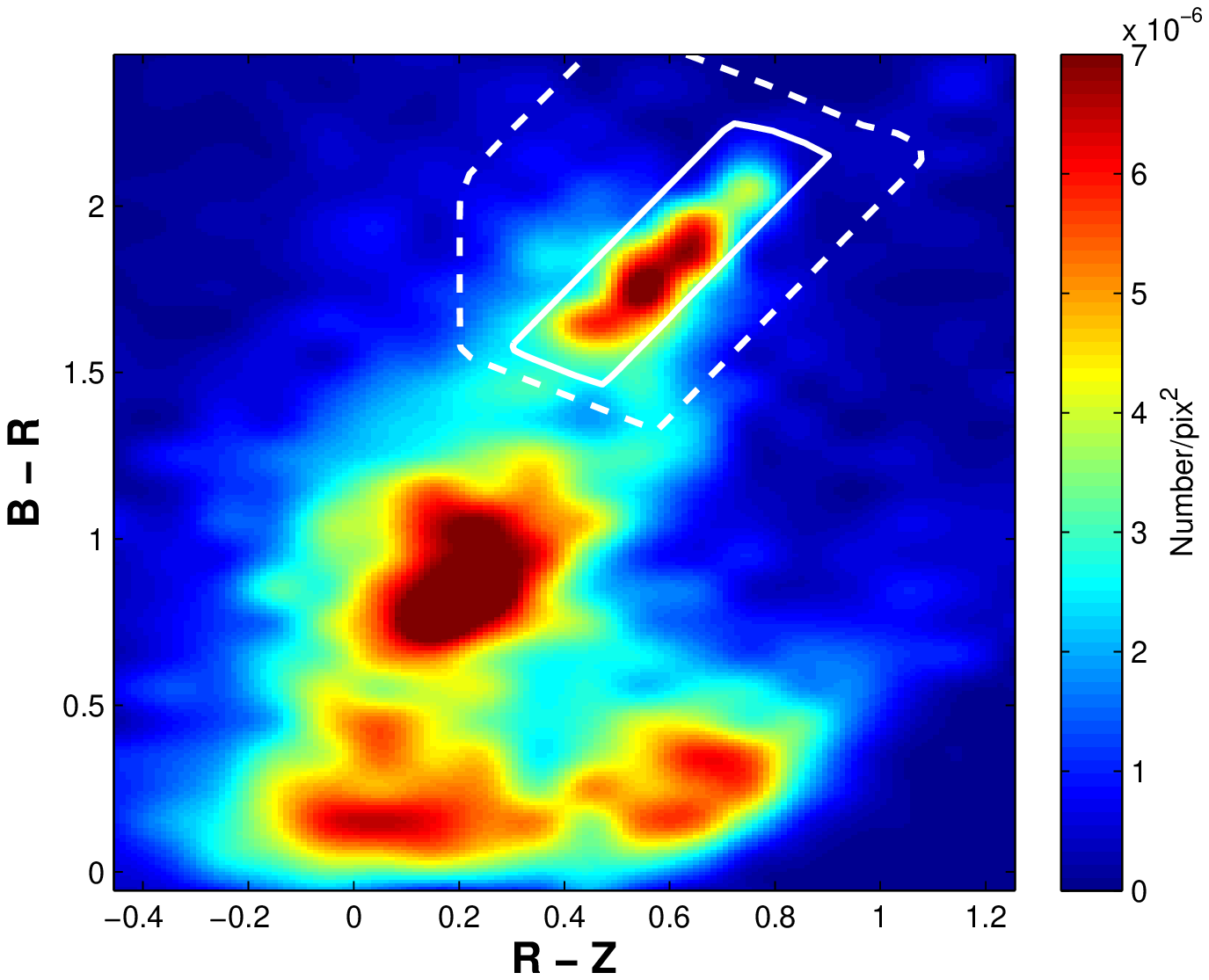}
    \put(15,64){\framebox(10,5){\bf A370}}\end{overpic}
  \begin{overpic}[scale=.50]{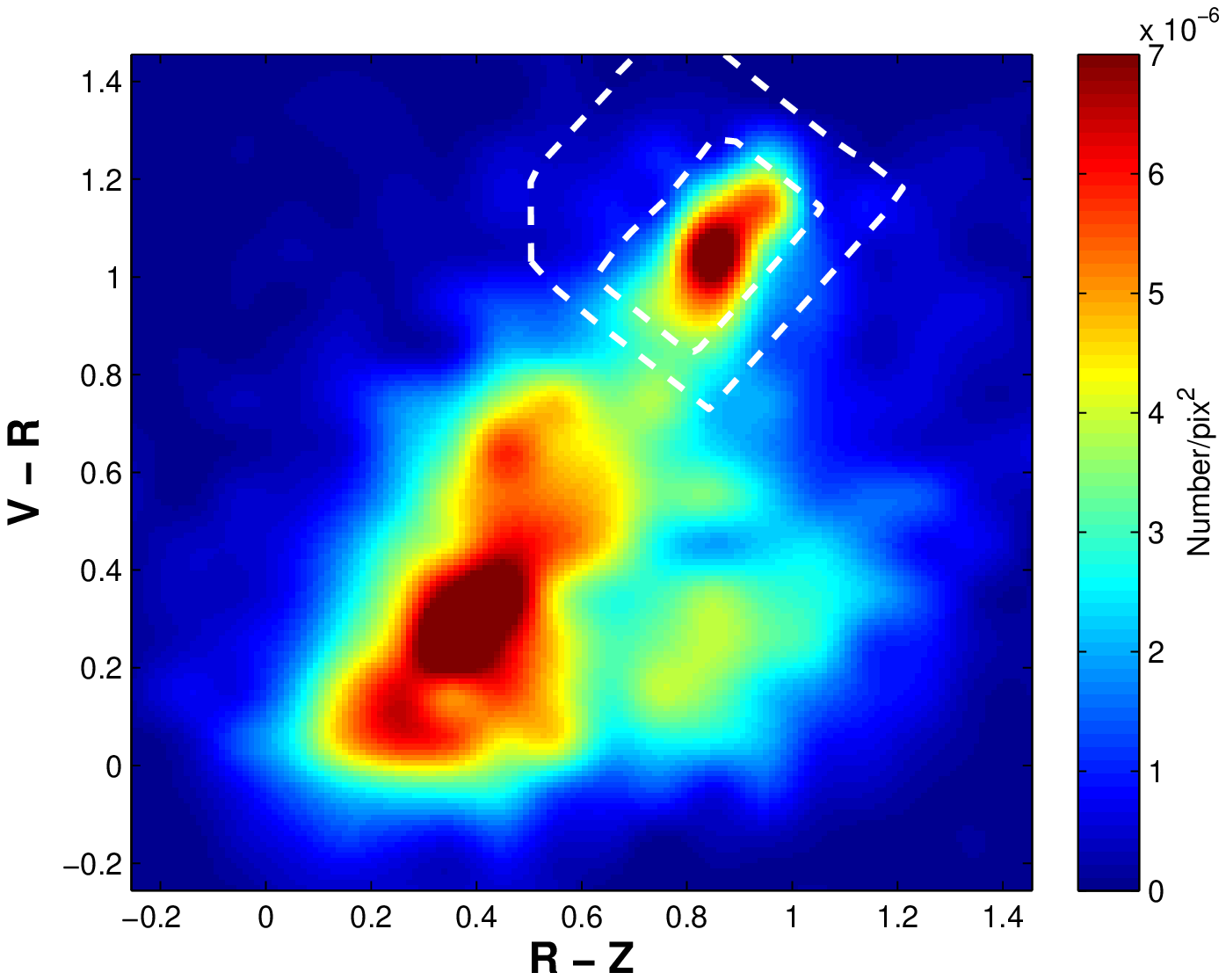}
    \put(15,64){\framebox(24,5){\bf RXJ1347-11}}\end{overpic}
  \caption{Number density in CC space for A1703, A370 \& RXJ1347-11
    (Top to bottom). The four distinct density peaks (seen especially
    in the A370 plot) are shown to be different galaxy populations -
    the reddest peak in the upper right corner of the plots (dashed
    white line) depicts the overdensity of cluster galaxies, whose
    colours are lying on the red sequence; the middle peak with
    colours bluer than the cluster shows the overdensity of foreground
    galaxies; the two peaks in the bottom part (bluest in $B-R$) can
    be demonstrated to comprise of blue and red (left and right,
    respectively) background galaxies.}
  \label{colorcolor_ncell}
\end{figure}

For all three clusters we construct a CC diagram and first identify in
this space where the cluster lies.  We do this by calculating the mean
distance of all objects from the cluster centre in a given CC cell, as
shown in Fig.~\ref{colorcolor_rcell}. This turns out to be a very
clear way of finding the cluster as the mean radius is markedly lower
in a well defined region of CC space for each cluster (region of
bluer (darker) colours, the approximate boundary of which is marked in black).
This small region clearly corresponds to an overdensity of galaxies in
CC space comprising the red sequence of each cluster and a blue trail
of later type cluster members (Fig.~\ref{colorcolor_ncell}, dashed
white curve). Of course, some background galaxies must also be
expected in this region of CC space, and the proportion of these will
be established in \S~\ref{dilution} by looking at their WL signal --
which should be consistent with zero in the case of no background
contamination. In general, a positive signal will be expected at some
level, in proportion to the fraction of lensed background galaxies
occupying that same region of CC space as the cluster.  Indeed, it is
clear that the level of the tangential distortion seen in
Fig.~\ref{gt_sub} is consistent with zero almost to the outskirts of
each cluster radial span, and only rises to meet the background level
at large radius where the proportion of cluster members is small
compared with the background counts. In Fig.~\ref{colorcolor_rcell}
(black curve), in Fig.~\ref{colorcolor_ncell} (dashed white curve) and
in Fig.~\ref{sample_analysis} (left panels, green points) we mark the
boundaries of this region, selected relative to the red sequence in
this CC space, and conservatively to safely encompass the region
occupied by the cluster.

The above sample, which includes all cluster member galaxies, we term
the ``green'' sample, as distinct from the well separated populations
of red and bluer galaxies discussed below (Fig.~\ref{sample_analysis},
left panels). In defining this green sample we apply a magnitude limit
of $m\simeq 26.5$ in the reddest band available for each cluster
($i'$-band for A1703 and $z'$-band for A370 and RXJ1347-11), to which
the data are complete for all three clusters. Beyond this limit
incompleteness creeps into the bluer bands, complicating colour
measurements, in particular of red galaxies. We also set a bright
magnitude limit, according to the magnitude of the brightest cluster
member in the same equivalent band.


\subsection{Red Background Galaxies}\label{subsec:red}

The mean radius used above is useful for locating the cluster in CC
space but does not help in the definition of background or foreground
galaxy populations, which are of course relatively uniform over the
field of view, apart from the relatively small magnification bias
which tends to deplete the number of red background galaxies towards
the centre of mass, as described in \nocite{2008ApJ...685L...9B}{Broadhurst} {et~al.} (2008) and is
visible here in Fig.~\ref{sample_analysis} (central panels, black
squares) as a decrease in number counts towards the centre.

To define the foreground and background populations we look instead at
the combination of the WL signal and the distribution of galaxies in
the CC plane. For each cluster, we see a dense cloud of red objects
(Fig.~\ref{colorcolor_ncell}, lower right overdensity).  This is
particularly obvious for A370 where we have the widest wavelength
coverage, $B_{\rm J}$, $R_{\rm C}$ \& $z'$.  The cloud is relatively
red in $R_{\rm C}-z'$ but blue in $B_{\rm J}-R_{\rm C}$, so very well
separated from the cluster member region defined above.  This red
cloud is known from wide-field survey work \nocite{2007ApJS..172...99C}({Capak} {et~al.} 2007)
and established to comprise early- to mid-type galaxies spanning a
broad redshift range, $0.5<z<2$, as discussed below
(\S~\ref{sec:photo-z}).

For this red sample we define a conservative diagonal boundary
relative to the red sequence as for the green sample (see
Fig.~\ref{sample_analysis}, left panels, red points), to safely avoid
contamination by cluster members and also foreground galaxies as
discussed below (see \S~\ref{subsec:blue/for}). We can check if there
is any significant contamination of this red sample by unlensed
galaxies by measuring the WL amplitude of these red galaxies as a
function of distance from this boundary.  As shown in
Fig.~\ref{gt_sub} (red triangles), no evidence of dilution of the
lensing signal is visible. We also define a blue colour limit to
separate this red population from what appears to be a distinct
density maximum of very blue objects -- the so called ``blue cloud''
(see Fig.~\ref{colorcolor_ncell}, in the extreme blue corner of CC
space (see below \S~\ref{subsec:blue/for}). We further limit the red
sample to $m<26$ AB mag in the reddest band for each cluster to avoid
incompleteness and so that we can rely on good detections in all bands
when defining our colour samples.

The boundaries of the red sample as defined above are marked on
Fig.~\ref{sample_analysis} (red points), and can be seen to lie well
away from the ``green'' cluster sample previously defined (green
points). For this red sample a clearly rising WL signal is seen all
the way to the smallest radius accessible ($r\lesssim 1\arcmin$) for
each cluster (Fig.~\ref{gt_sub}, red triangles), with no sign of a
central turnover which would indicate the presence of unlensed cluster
members, as discussed fully in \S~\ref{dist_Sub}.

\subsection{Blue Background and Foreground Galaxies}\label{subsec:blue/for}

Above we have been able to define the region of CC space occupied by
cluster members by virtue of their clustered distribution. We have
also defined with confidence a sample of red background galaxies lying
well separated from cluster members and for which there is a clear WL
signal. For bluer galaxies the situation is more complex as these may
comprise blue cluster members, foreground objects and background blue
galaxies.  This is of particular concern where only one colour is
available, and which we have shown in our earlier work can lead to a
dilution of the WL signal of blue selected objects relative to the red
background galaxies by unlensed foreground galaxies and cluster
members \nocite{2005ApJ...619L.143B,2007ApJ...663..717M}({Broadhurst} {et~al.} 2005; {Medezinski} {et~al.} 2007).

\begin{figure*}\centering
  \subfigure[A1703]{ 
    \includegraphics[scale=.35]{A1703/A1703_colorcolor_090823.eps3}
    \includegraphics[scale=.35]{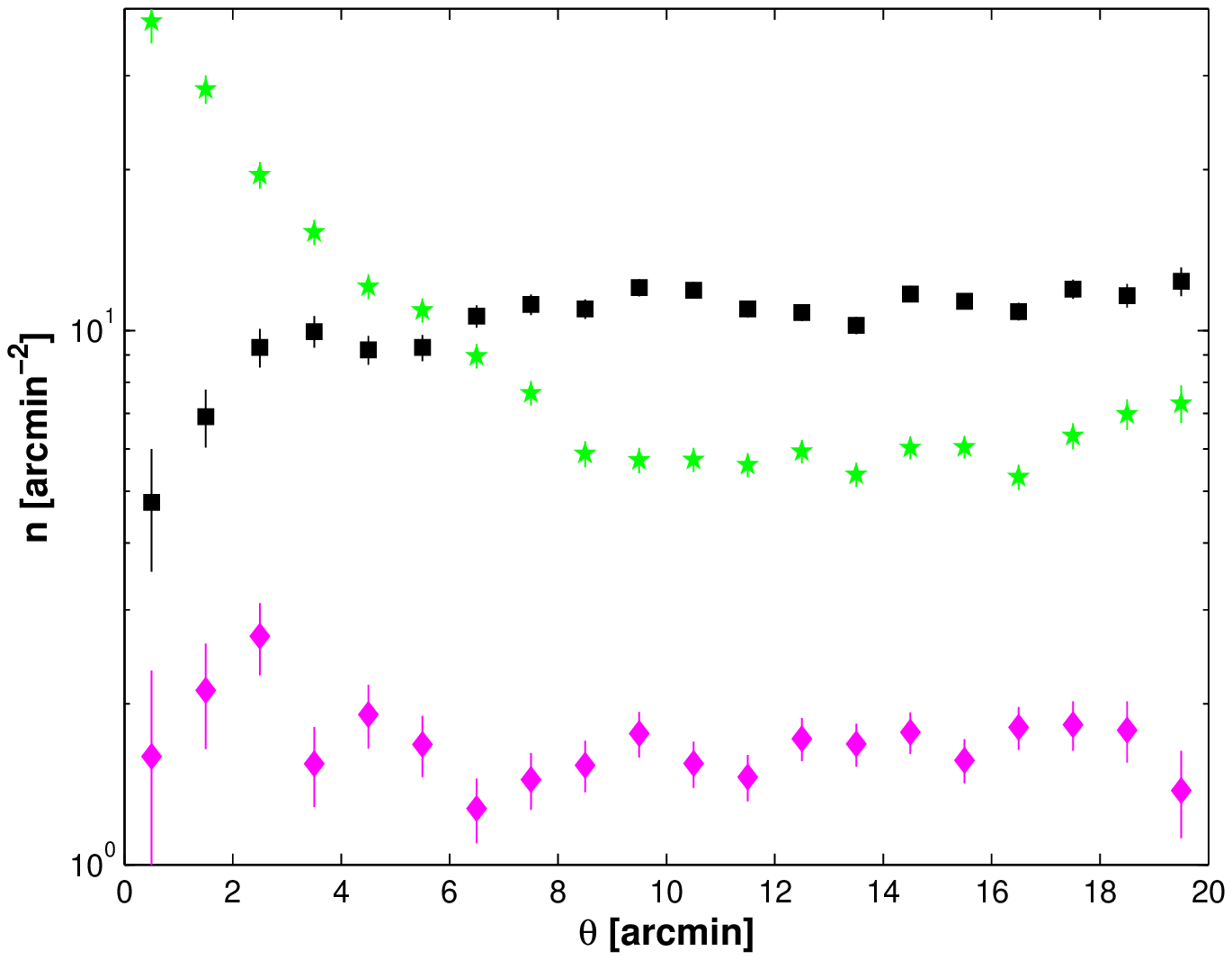}
    \includegraphics[scale=.35]{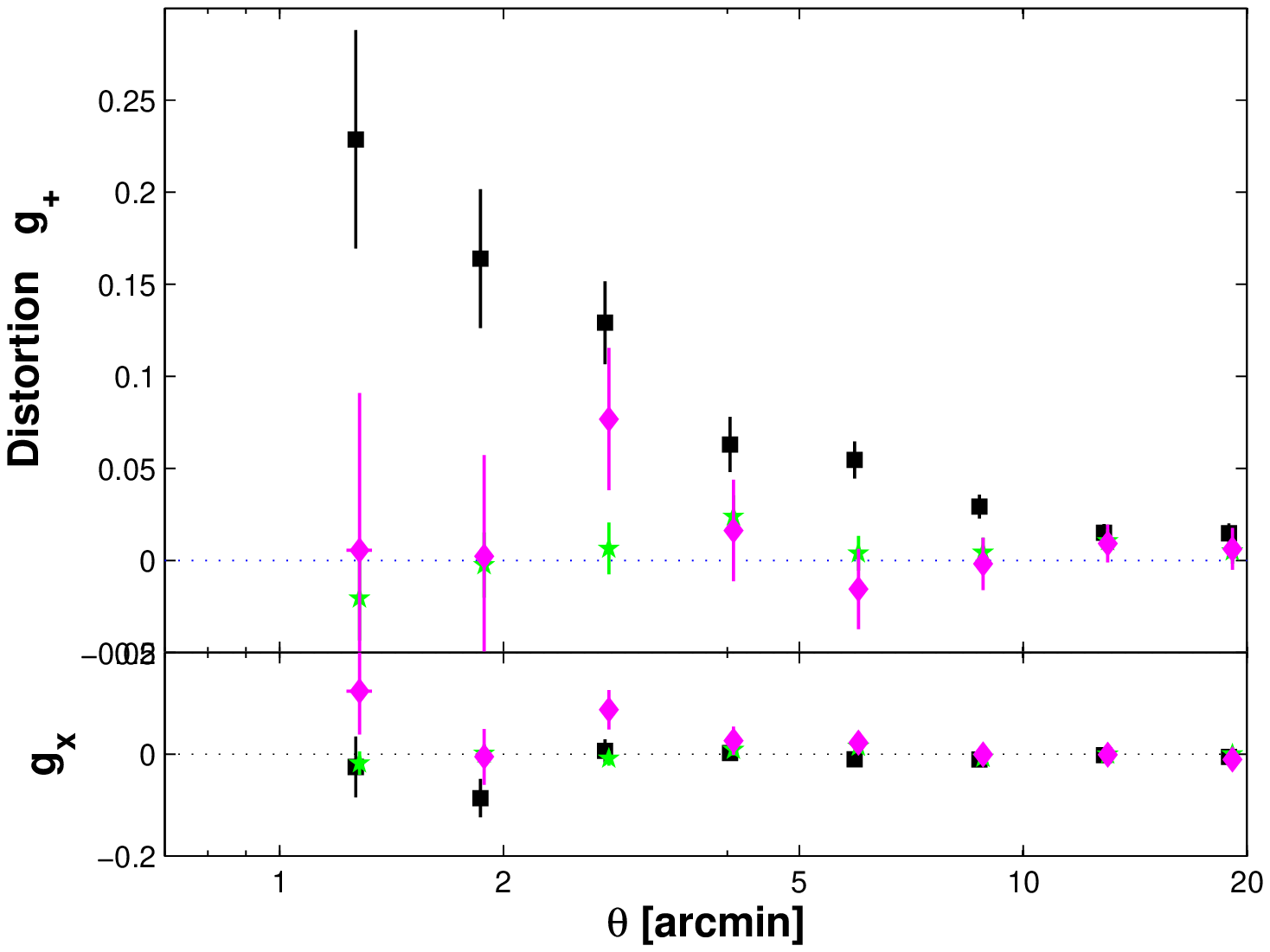} }
  \subfigure[A370]{
    \includegraphics[scale=.35]{A370/A370_colorcolor_090823.eps3}
    \includegraphics[scale=.35]{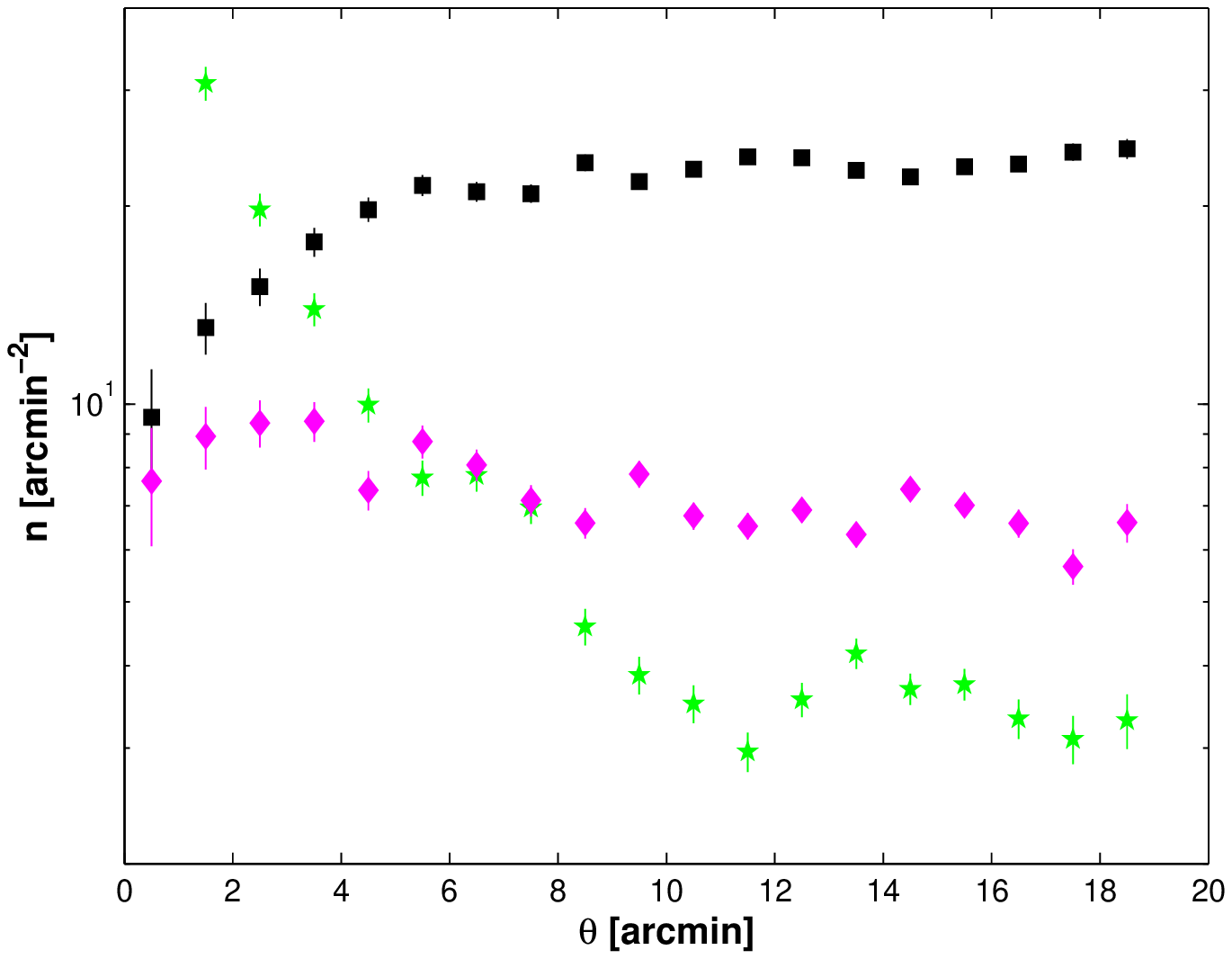}
    \includegraphics[scale=.35]{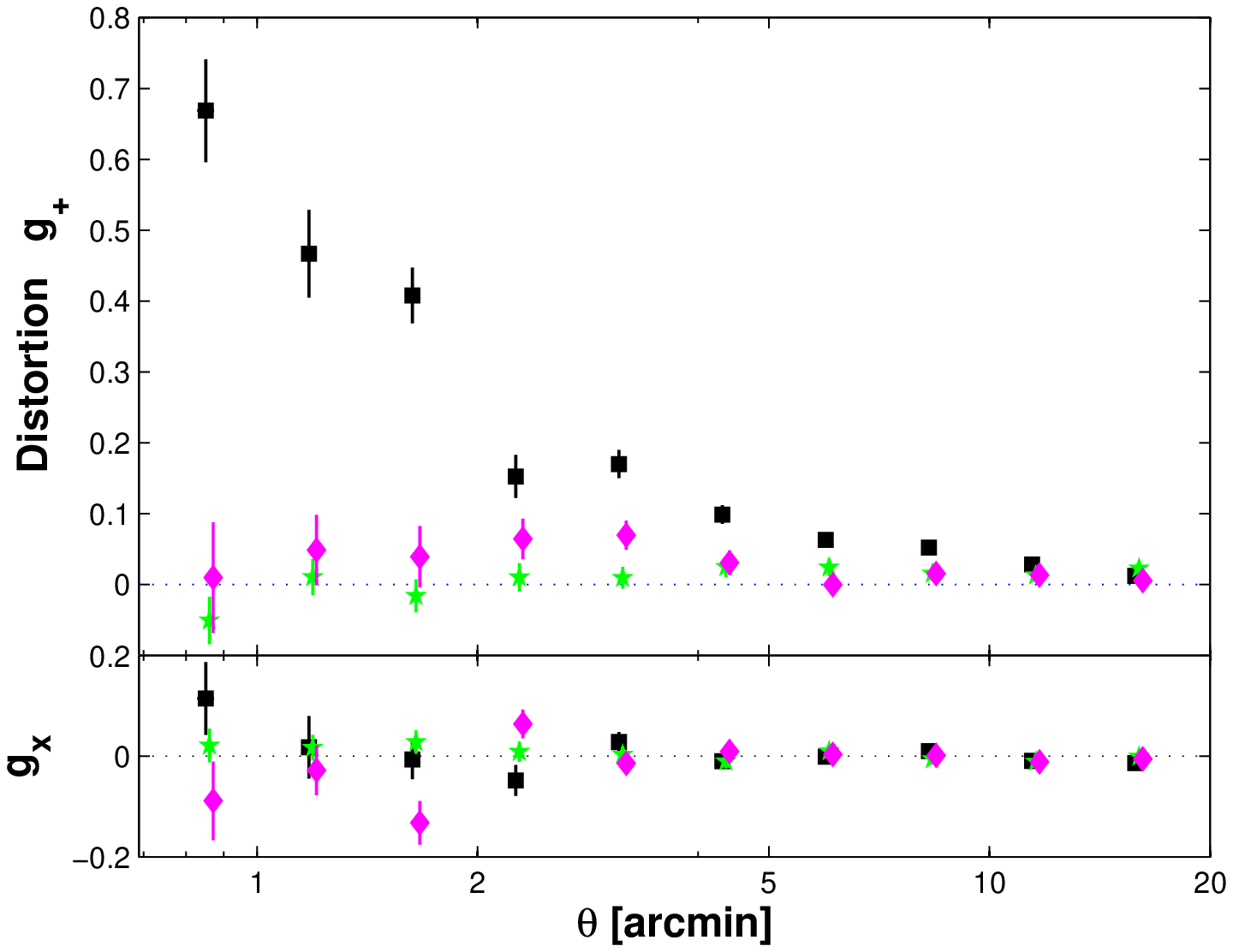} }
  \subfigure[RXJ1347]{
    \includegraphics[scale=.35]{RXJ1347/RXJ1347_colorcolor_091124.eps3}
    \includegraphics[scale=.35]{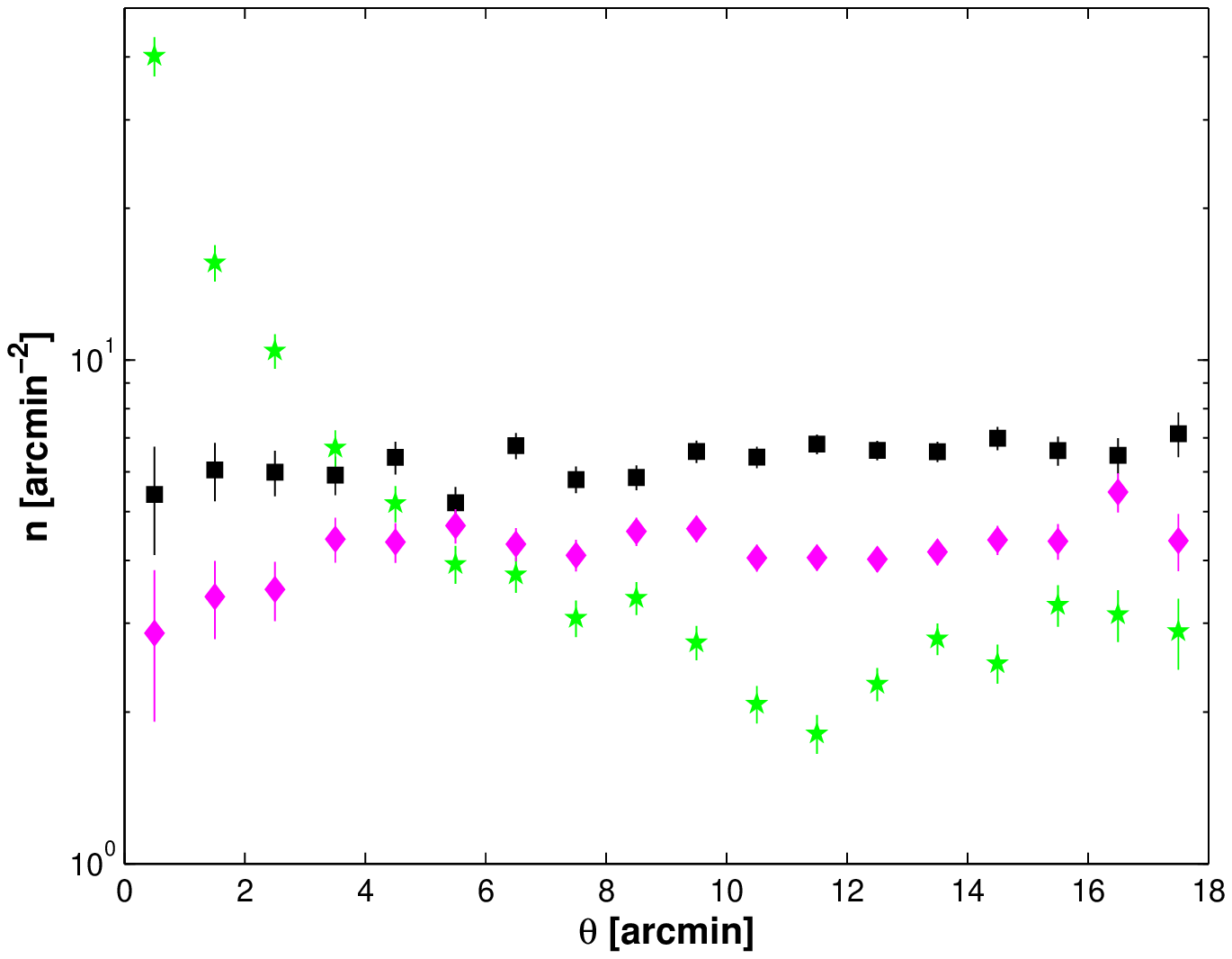}
    \includegraphics[scale=.35]{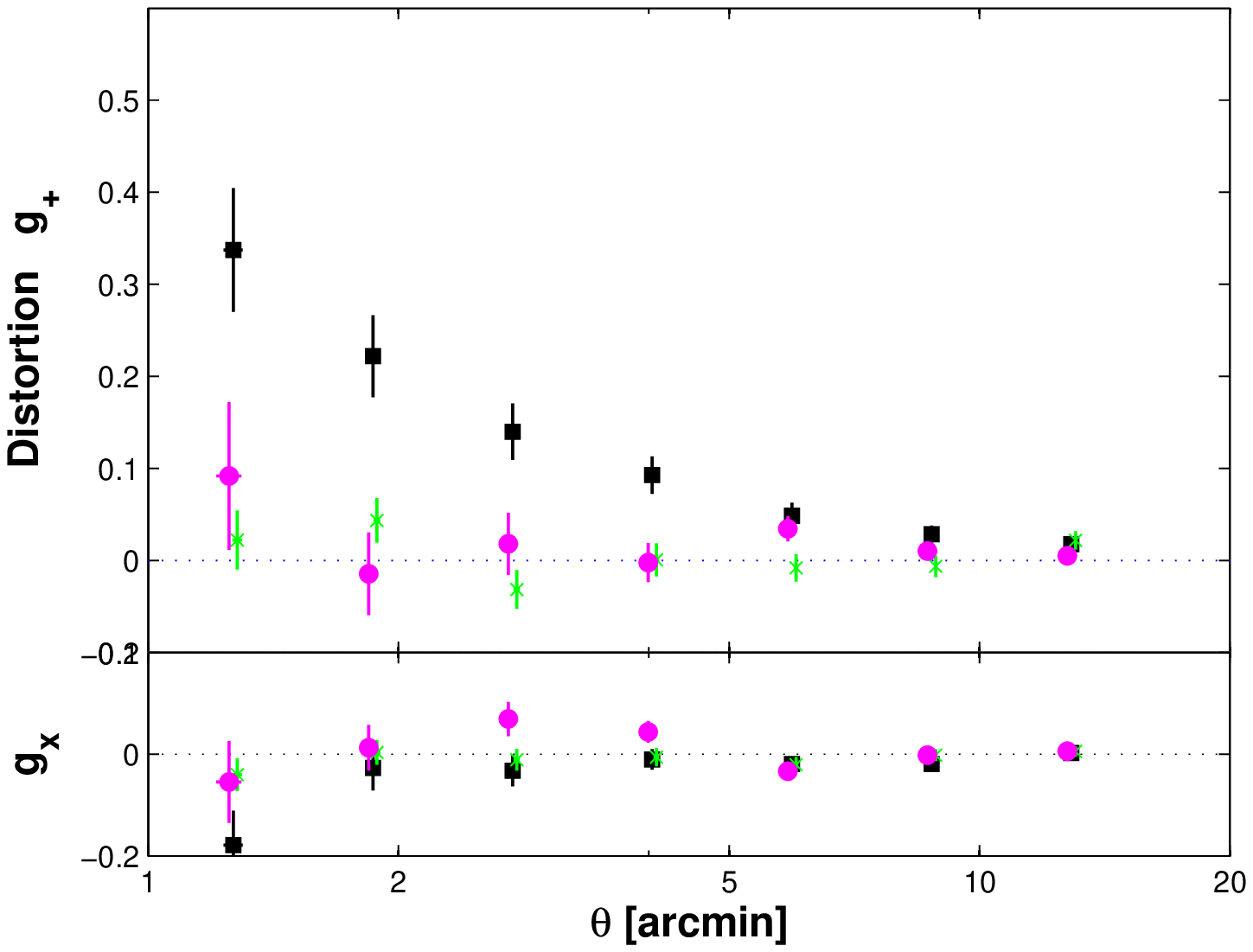} }
  \caption{ Sample selection and analysis. {\it Left}: CC diagrams,
    for A1703 (a), A370 (b) and RXJ1347-11 (c) (top to bottom),
    displaying the green sample, comprising mostly cluster member
    galaxies, and the red and blue samples, comprising of background
    galaxies. The galaxies that we identify as predominantly
    foreground lie in between the cluster and background galaxies are
    marked in magenta. {\it Centre}: Surface number density of
    galaxies vs. radius. Background (red+blue galaxies) density (black
    squares) is fairly constant with radius, except for a slight
    decrease in the central region (depletion due to lensing
    magnification effect). The green sample (green pentagrams) rises
    steeply toward the centre as expected for cluster members.  The
    magenta sample (magenta diamonds) has a flat density profile,
    meaning these are not part of the cluster. {\it Right}: Tangential
    distortion vs. radius. The background (black squares) rises
    continuously toward the centre, representing real WL signal. The
    green sample distortion (green pentagrams) is essentially in
    agreement with zero, as expected from unlensed cluster galaxies,
    and the magenta sample distortion (magenta diamonds) is
    significantly lower than the background WL signal, indicating
    object selected in this CC range are mostly unlensed foreground
    galaxies.  However, the magenta sample also contains to some
    extent background galaxies, therefore the WL signal is slightly
    higher than zero.}
  \label{sample_analysis}
\end{figure*}

We can see in the CC plane of each cluster that there are two regions
lying blueward of the cluster within which most blue objects lie. For
A370, where we have the deepest dataset, these density peaks are
located around $R_{\rm C}-z'\sim0$ and $B_{\rm J}-R_{\rm
  C}\sim0.8,1.5$.  The redder of these two blue peaks lies in the
centre of the CC plane, relatively close to the blue-end of the
cluster sequence (see Fig.~\ref{sample_analysis}, left panels, magenta
points). The spatial distribution of objects in this CC region shows
no significant clustering about the cluster centre (see
Fig.~\ref{sample_analysis}, middle panels, magenta circles). In
addition, the WL signal of these objects (see
Fig.~\ref{sample_analysis}, right panels, magenta circles) shows very
little signal, indicating this is a predominately unlensed population,
and together with the lack of clustering means this blue peak is
dominated by galaxies lying in the foreground of the cluster. The
lensing signal here is not strictly zero but marginally positive,
implying a relatively low level of $10-20\%$ lensed background
galaxies. This region of CC space is clearly of no use for our
purposes as it does not contain cluster members and is mostly
unlensed. It is convenient, however, that the foreground unlensed
population is so well defined in CC space as a clear overdensity and
therefore can be readily excluded from our analysis. The redshift
distribution of this population using deep field surveys (discussed in
\S~\ref{sec:photo-z}) supports our finding that the bulk of this
population lies at low redshift.

The bluer of the two peaks, in the extreme blue corner of CC space
seen in Fig.~\ref{colorcolor_ncell} (lower left peak) is also
unclustered on the sky, indicating no significant cluster members lie
in this region (Fig.~\ref{sample_analysis}, middle panels, black
squares). This blue overdensity is the well known ``blue cloud''
identified in deep field images. For each cluster, a clear WL signal
is found for objects lying in this blue cloud (Fig.~\ref{gt_sub}, blue
circles), rising toward the centre of each cluster with a radial trend
very similar to that of the red population ((Fig.~\ref{gt_sub}, red
triangles), indicating minimal foreground or cluster dilution of the
WL signal. Objects lying in this region are expected to fall in the
redshift range $1<z<2.5$ (discussed below in \S~\ref{sec:photo-z}).
Hence, we can safely conclude that these objects lie in the background
with negligible cluster or foreground contamination, which would
otherwise drag down the central WL signal.  The boundaries we have
chosen for this blue sample and plotted in Fig.~\ref{sample_analysis}
(left panels, blue points), are extended to include objects lying
outside the main blue cloud but well away from the foreground and
cluster populations defined above, in order to maximise the size of
the blue sample but steering well clear of contamination by unlensed
low redshift objects.  We further limit the blue sample magnitude in
the range $22<m_{AB}<26$ to avoid contamination by very bright
foreground galaxies, and to avoid incompleteness at the faint end.

\section{Evolutionary colour tracks}\label{sec:tracks}

\begin{figure}
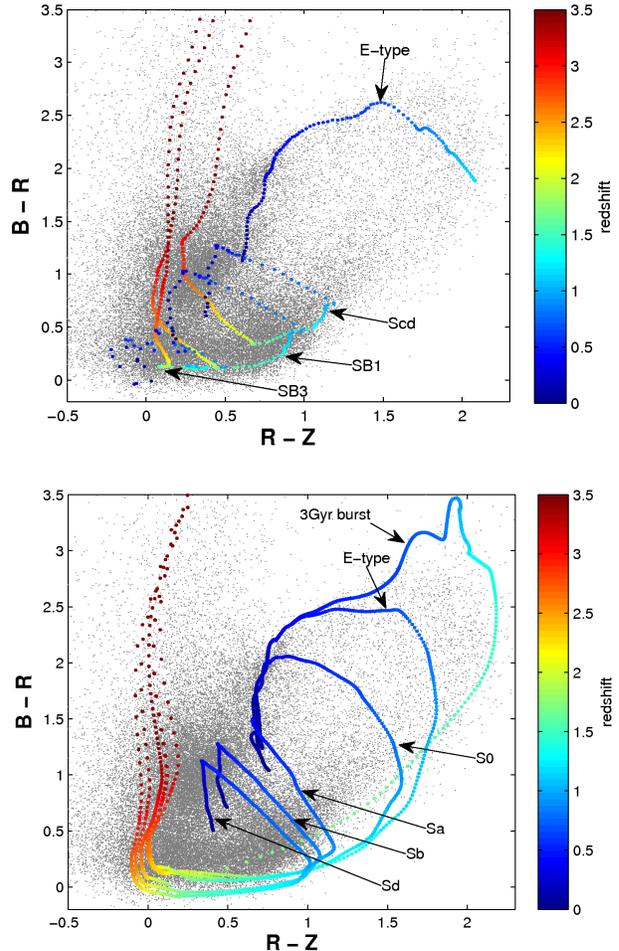

  \centering
  \includegraphics[width=9cm]{figs/A370_BRZ_BPZ_evtracks.eps3}
  \includegraphics[width=9cm]{figs/A370_BRZ_Galev_evtracks_100112.eps3}
  \caption{{\it Top}: Empirical colour tracks overlaid on the galaxy
    distribution of A370 in $B_{\rm J}-R_{\rm C}$ vs. $R_{\rm C}-z'$
    plane, for a range of galaxy templates: elliptical, Scd, SB1 and
    SB3 SED's. {\it Bottom}: Synthetic colour tracks including
    evolution, calculated with the Galev code for a single 3~Gyr
    burst, elliptical, S0, Sa, Sb and Sd type models.}
  \label{tracks}
\end{figure}

In order to examine in more detail the sample selection technique, we
overlay evolutionary and empirical tracks on the CC plane, for
different classes of galaxies. We first calculate non-evolving
magnitudes in Subaru filters $B_{\rm J},R_{\rm C},z'$ as a function of
redshift using different galaxy templates constructed empirically for
elliptical, Scd and starburst types from the Kinney-Calzetti library
\nocite{1994ApJ...429..582C,1996ApJ...467...38K}({Calzetti}, {Kinney} \&  {Storchi-Bergmann} 1994; {Kinney} {et~al.} 1996).  The resulting colour
tracks are overplotted in Fig.~\ref{tracks} (top) on the $B_{\rm
  J}-R_{\rm C}$ vs. $R_{\rm C}-z'$ plane of A370. These templates show
how the foreground population of galaxies in the centre of the CC
plane quickly jumps above a redshift of $\sim1$ to the lower blue
density peak. For the E/S0 it appears that even the no-evolution track
is a good fit to a distribution of red galaxies which fall on a
well-defined loop visible in the CC plane with a maximum red colour of
$B_{\rm J}-R_{\rm C}=2.5,R_{\rm C}-z'=1.5$ at a redshift of
$z\sim0.8$.  The remainder of the loop is not well constrained (and we
therefore leave it out), likely where evolution kicks in making the
SED of early-type objects bluer.

To generate evolving tracks representing the evolution of galaxy
SED's, the Galev\footnote{http://www.galev.org/}
\nocite{2009MNRAS.396..462K}({Kotulla} {et~al.} 2009) code was used to represent galaxy types
with different star-formation histories -- a short 3~Gyr starburst
($Z_\odot$), E-type (exponentially declining SFR with $Z_\odot$), S0
(gas-related SFR with $Z_\odot$), Sa (gas-related SFR with
$2.5Z_\odot$), Sb (gas-related SFR with $0.4Z_\odot$) and Sd (constant
SFR with $0.2Z_\odot$) (Fig.~\ref{tracks}, bottom). It is evident that
at low redshifts the modeled galaxies coincide with the high central
peak we recognized above with our WL technique to be foreground, and
as they are redshifted the colours of E/S0 galaxies agree at
$z\sim0.4$ with the cluster colours. At higher redshifts the early
types objects become bluer due to the effect of evolution, and also
the late-type galaxies jump quickly across the valley and fall close
to the observed trail of red and blue background galaxies at $1<z<2$.
At even higher redshifts ($z>2.5$), they fade out from the $B_{\rm
  J}$-band and become redder in $B_{\rm J}-R_{\rm C}$, agreeing with
observed colours of dropout galaxies. Accounting for evolution, the
tracks follow more carefully the trend around the lower-left blue
corner ($B_{\rm J}-R_{\rm C}\sim0.1, R_{\rm C}-z'\sim0$), seen clearly
from the galaxy colour distribution.

In general, similar behaviour is seen for both methods, giving
compelling evidence that in this CC plane the foreground and cluster
galaxies are rather well-separated from the higher-redshift background
galaxies, and support the picture we presented above for the safe
selection of background galaxies for a WL study. The slight
differences in colour (offset seen from the Galev models to bluer
$B_{\rm J}-R_{\rm C}$ colours for example) is interesting and may
arise from the inclusion of evolution, the presence of dust or minor
model uncertainties. The shape of the early-type galaxy loop is
certainly worth exploring in greater detail and may help to constrain
the general evolutionary history of early-type galaxies.

\section{Depth Estimation from SDF/COSMOS}\label{sec:photo-z}
\begin{figure}
  \centering \includegraphics[width=9cm]{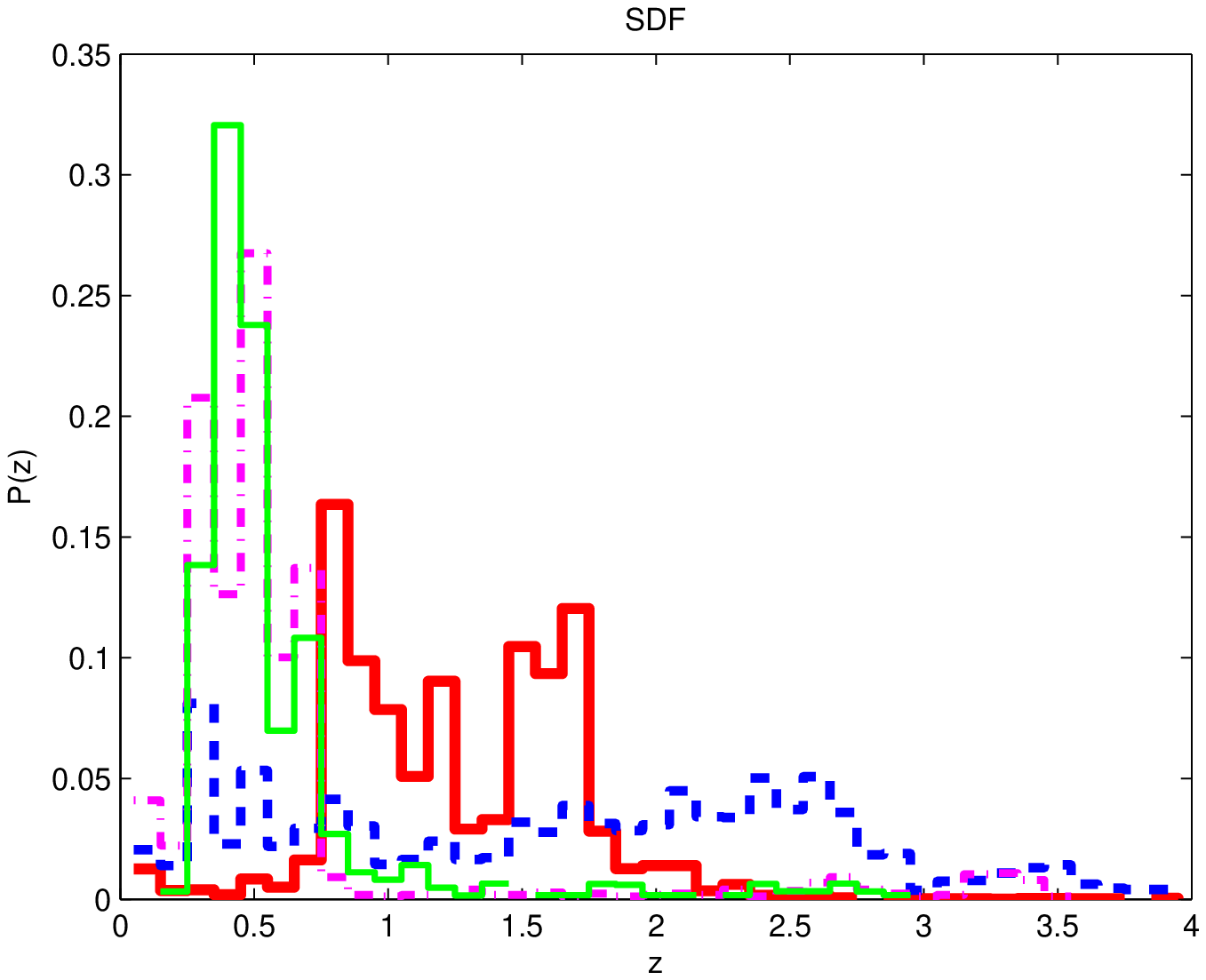}
  \includegraphics[width=9cm]{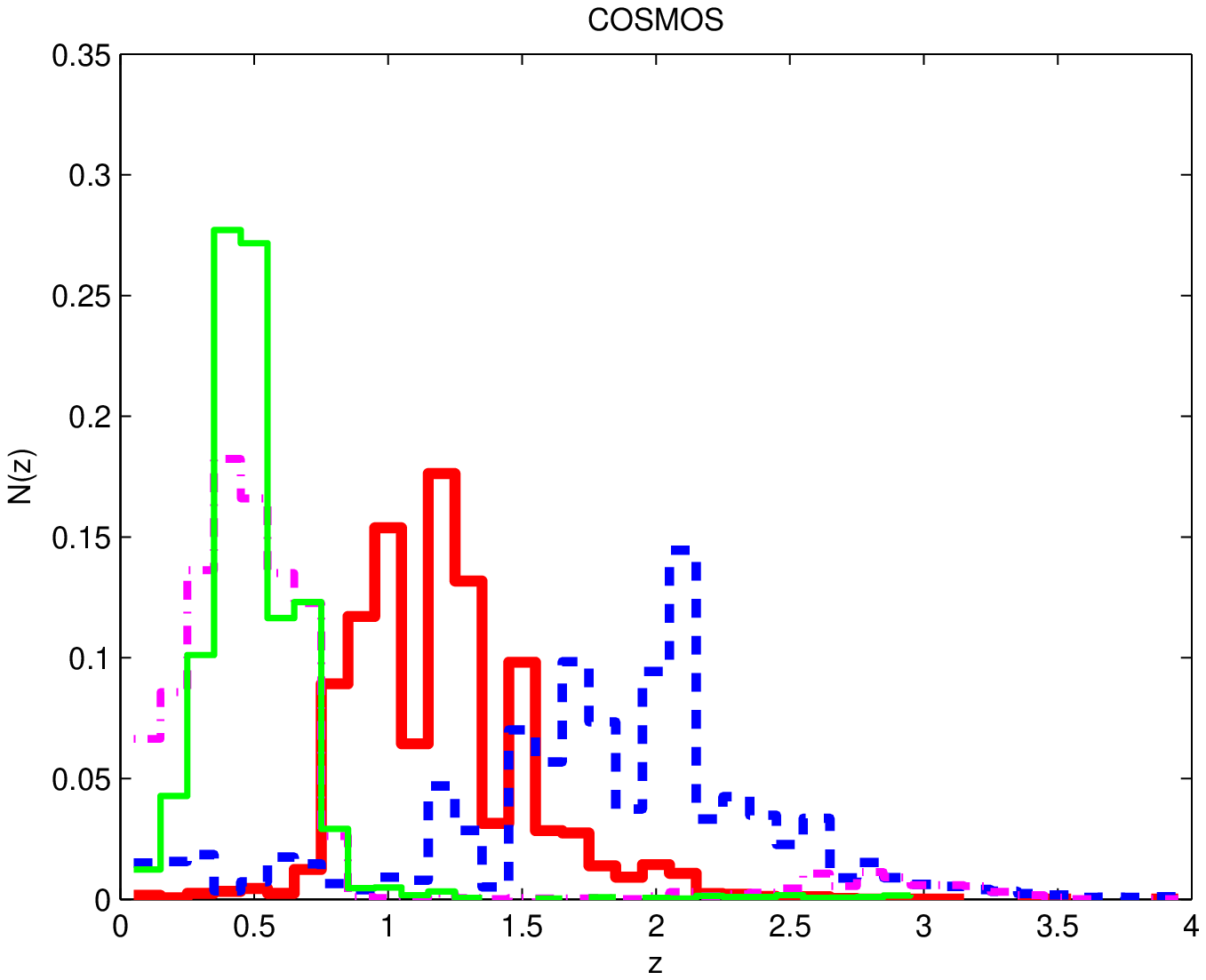}
%fix caption
  \caption{Redshift distributions of CC-selected samples. Two deep
    field surveys are used to estimate mean redshifts and depth: SDF
    (top), and COSMOS (bottom).  The green (green thin solid line), red
    (red thick solid line), blue (blue thick dashed line) and
    foreground (magenta dotted-dashed line) samples are selected
    according to A370 CC/magnitude limits (see
    Fig.~\ref{sample_analysis}).}
  \label{redshift}
\end{figure}

\begin{table*}
  \centering
  \caption{CC-selected Sample Properties} 
  \begin{tabular}{ccccccccc} 
    \hline
    {Cluster} & {Sample} & {magnitude limits} &  {N} &{$\bar{n}$} &  \multicolumn{2}{c}{$\bar{z_s}$} &
    \multicolumn{2}{c}{$<{D_{\rm ls}/D_{\rm s}}>$} \\ 
    \cline{6-7} \cline{8-9} 
    & & & & arcmin$^{-2}$ & \tiny{COSMOS} & \tiny{SDF} & \tiny{COSMOS} & \tiny{SDF} \\ 
    \hline
    A1703 & green & $16.3< i'<26.5$ & 5678 & 6.2 & 0.54 & -    & 0.45 & - \\
          & red   & $21< i'< 26$    & 4742 & 5.2 & 1.00 & -    & 0.67 & - \\
          & blue  & $22< i'< 26$    & 4467 & 4.9 & 1.59 & -    & 0.74 & - \\
          \hline
    A370  & green & $16.5< z'<26.5$ & 5369 & 5.3 & 0.46 & 0.69 & 0.25 & 0.29 \\
          & red   & $21< z'< 26$    & 13372& 13.7& 1.14 & 1.21 & 0.57 & 0.57\\
          & blue  & $22< z'< 26$    & 8481 & 8.7 & 1.74 & 1.65 & 0.64 & 0.56\\
          \hline
RXJ1347-11& green & $17.5< z'<26.5$ & 2512 & 3.5 & 0.67 & 0.79 & 0.28 & 0.31\\
          & red   & $21< z'< 26$    & 2943 & 4.1 & 1.09 & 1.26 & 0.49 & 0.52\\
          & blue  & $22< z'< 26$    & 1691 & 2.3 & 1.45 & 1.67 & 0.44 & 0.46\\
          \hline\label{tab:samples}
   \end{tabular}
\end{table*}

The lensing signal is dependent on the source distance, scaling
linearly with $D_{\rm ls}/D_{\rm s}$, the lensing distance ratio.  We
thus need to estimate and correct for the respective depths of the
different samples.  To this end we rely on accurate photometric
redshifts derived for two deep, multi-band field surveys, the Subaru
Deep Field \nocite{2004PASJ...56.1011K}(SDF; {Kashikawa} {et~al.} 2004), and COSMOS
\nocite{2007ApJS..172...99C}({Capak} {et~al.} 2007).  For the SDF, photometric redshifts are
calculated with BPZ \nocite{2000ApJ...536..571B}({Ben{\'{\i}}tez} 2000), using a new template
library, generated using a set of PEGASE templates
\nocite{1997A&A...326..950F}({Fioc} \& {Rocca-Volmerange} 1997) which approximates the library described
in \nocite{2004ApJS..150....1B}{Ben{\'{\i}}tez} {et~al.} (2004) and \nocite{2006AJ....132..926C}{Coe} {et~al.} (2006). This new
library has an accuracy of $\delta z/(1+z)=0.037$ with the FIREWORKS
catalog \nocite{2009ApJ...696..348W}({Wuyts} {et~al.} 2009) and a low catastrophic error rate.
Using those photo-$z$, and the spectral classification provided by
BPZ, we generate magnitudes in the required Subaru filters to match
the cluster observations. We select samples according to CC/magnitude
limits described above (Fig.~\ref{sample_analysis}) for A370 and
RXJ1347-11 (the SDF filters are not matched with the filters used for
A1703). The resulting redshift distributions of the green (thin solid
line), red (thick solid line), blue (dashed line) and foreground
(dash-dotted line) samples selected to match A370 samples using SDF
are displayed in Fig.~\ref{redshift} (top).

For COSMOS, photometric redshifts have been derived by
\nocite{2009ApJ...690.1236I}{Ilbert} {et~al.} (2009) using 30 bands in the UV to mid-IR.  Since
the COSMOS photometry does not cover the Subaru $R_{\rm C}$ band, we
need to estimate $R_{\rm C}$-band magnitudes for it.  We use the
HyperZ \nocite{2000A&A...363..476B}({Bolzonella}, {Miralles} \&  {Pell{\'o}} 2000) template fitting code to obtain the
best-fitting spectral template for each galaxy, from which the $R_{\rm
  C}$ magnitude is derived with the transmission curve of the Subaru
$R_{\rm C}$-band filter \nocite{2009arXiv0908.0069U}(see {Umetsu} {et~al.} 2009b). We then
select samples by applying the same CC/magnitude limits as for A1703,
A370 and RXJ1347-11.  The resulting redshift distributions of the
green (thin solid line), red (thick solid line), blue (dashed line)
and foreground (dash-dotted line), samples selected to match A370
samples using COSMOS are displayed in Fig.~\ref{redshift} (bottom).
The results for each sample using SDF or COSMOS are summarized in
table~\ref{tab:samples}.

Although SDF is deeper, COSMOS redshifts are based on more bands over
a wider spectral range.  However, as stated in
\nocite{2009ApJ...690.1236I}{Ilbert} {et~al.} (2009), COSMOS photo-$z$'s are reliable only to a
magnitude of $i'<25$, which is below our magnitude cutoff for the
sample selection.  We therefore make sure to limit our redshift
estimation for the green sample to $z<3$, to avoid mistakenly
including high-$z$ dropouts that lie above the cluster in CC due to
slight photometric offsets or unreliable photo-$z$'s. Note we are only
interested in the mean $D_{\rm ls}/D_{\rm s}$ of \textbf{background}
objects, and therefore estimate the depth of background galaxies
present in the green sample in the range $z_{cluster}<z<3$.  Overall,
the redshift distributions of both field surveys look quite similar,
and there is good agreement between values derived with either SDF or
COSMOS, to $\sim10\%$ level in the depth, which can be due to the
differences mentioned above. These depth values will be used later to
correct the lensing signal (see \S~\ref{dilution}).

Importantly, this redshift analysis independently supports our
assessment in both sections \ref{sec:samples} \& \ref{sec:tracks} that
the various regions of the CC space correspond to differing
populations. In particular the foreground population isolated in the
centre of the CC diagram (shown in Fig.~\ref{sample_analysis}, magenta
points, whose redshift distribution is shown in Fig.~\ref{redshift},
magenta dotted-dashed curve) corresponds to predominantly low redshift
galaxies $\bar z\simeq0.4$, most of which are in the foreground of the
clusters examined here, or at such low redshift behind the cluster
that the lensing signal is small by virtue of the small separation,
$D_{ls}$, between the lens and the source. In contrast, the red and
blue populations are much more distant lying well behind our clusters,
supporting our conclusion that the continuously rising WL signal of
these populations is not significantly contaminated by cluster
members. This also indicates that the predicted level of unlensed
foreground galaxies is negligible for the red sample, as expected, and
with only a possible $\sim10\%$ dilution in the blue.

\section{Weak Lensing Analysis}\label{dist_Sub}
\begin{figure}
  \centering \includegraphics[scale=.50]{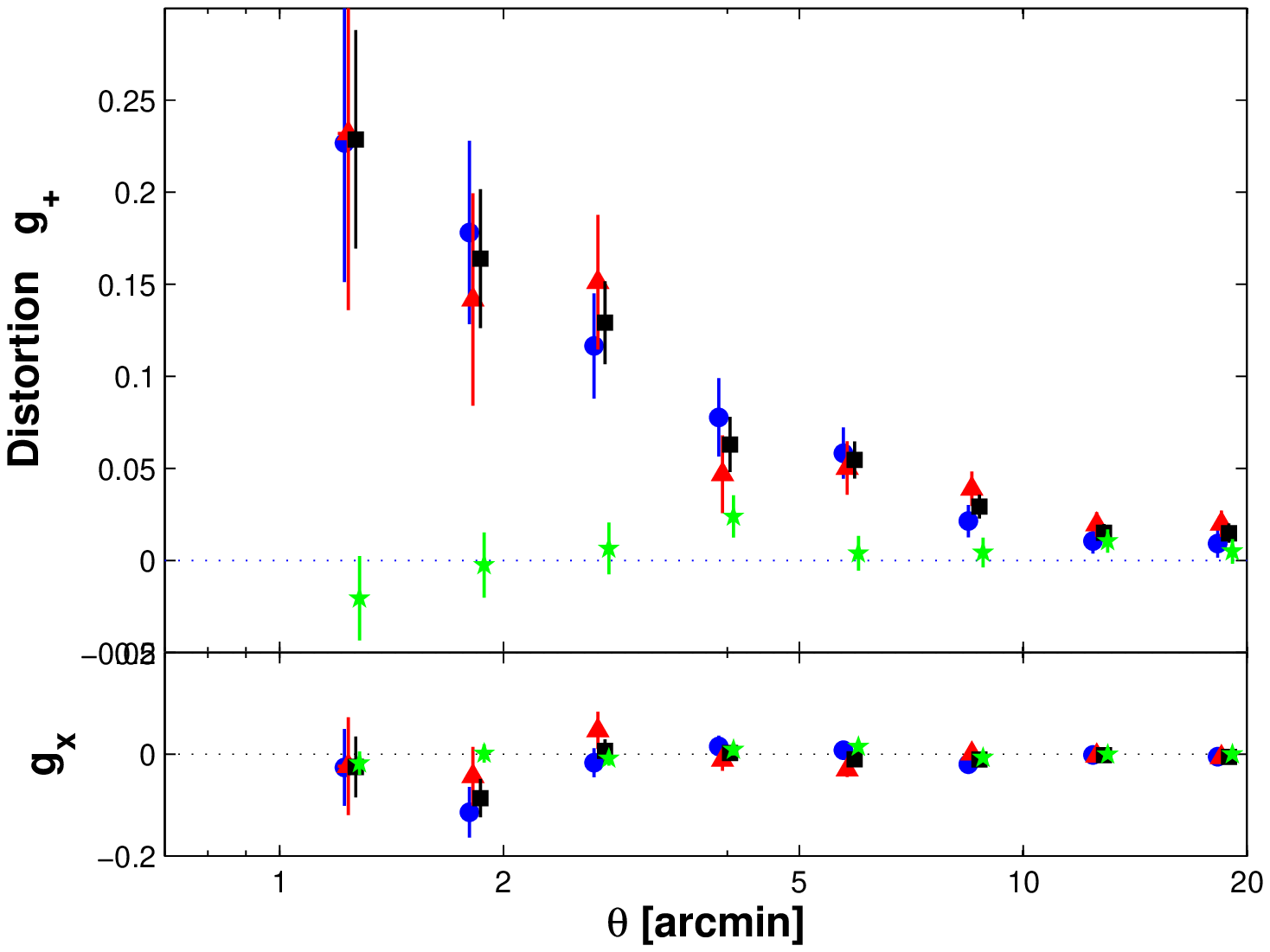}
  \includegraphics[scale=.50]{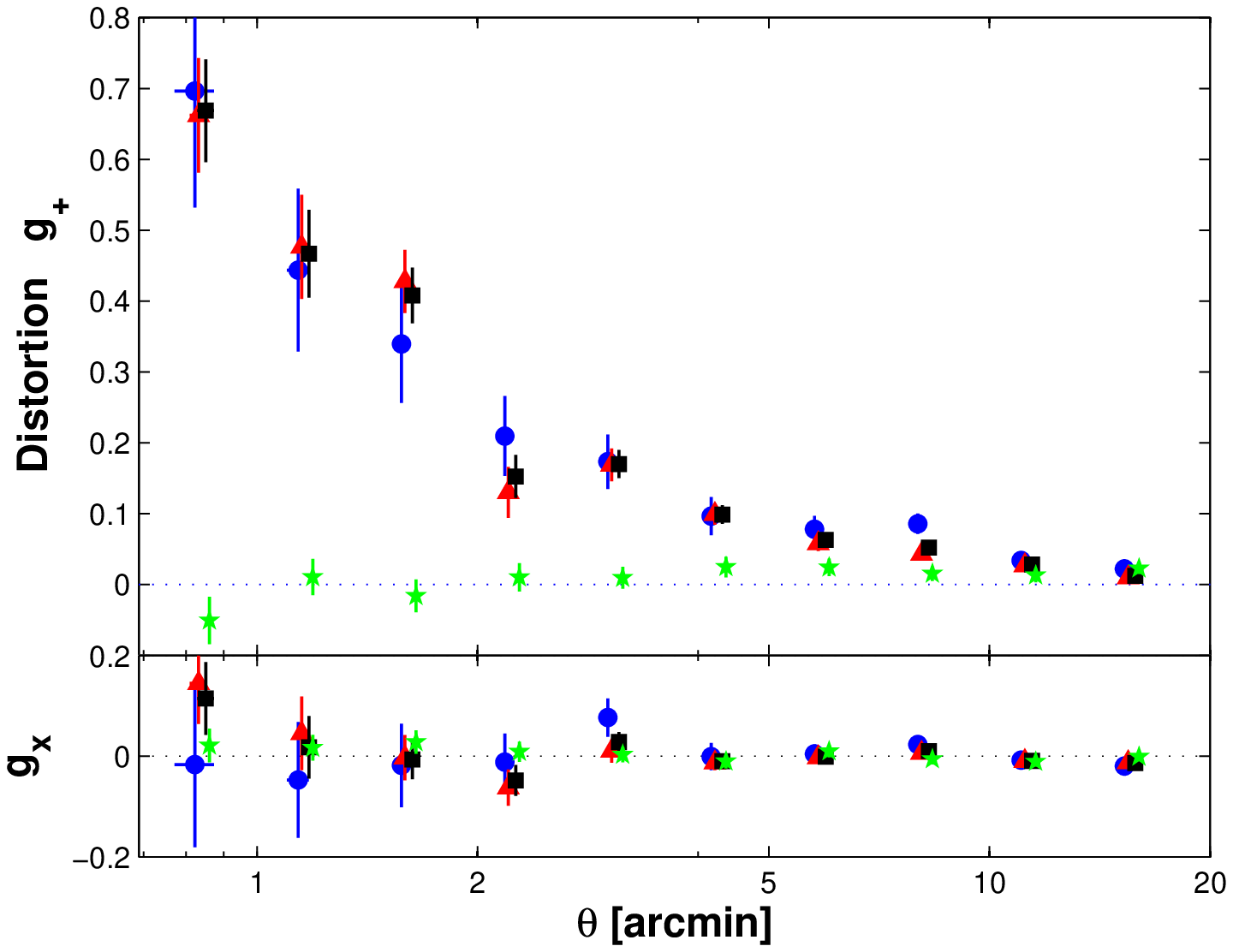}
  \includegraphics[scale=.50]{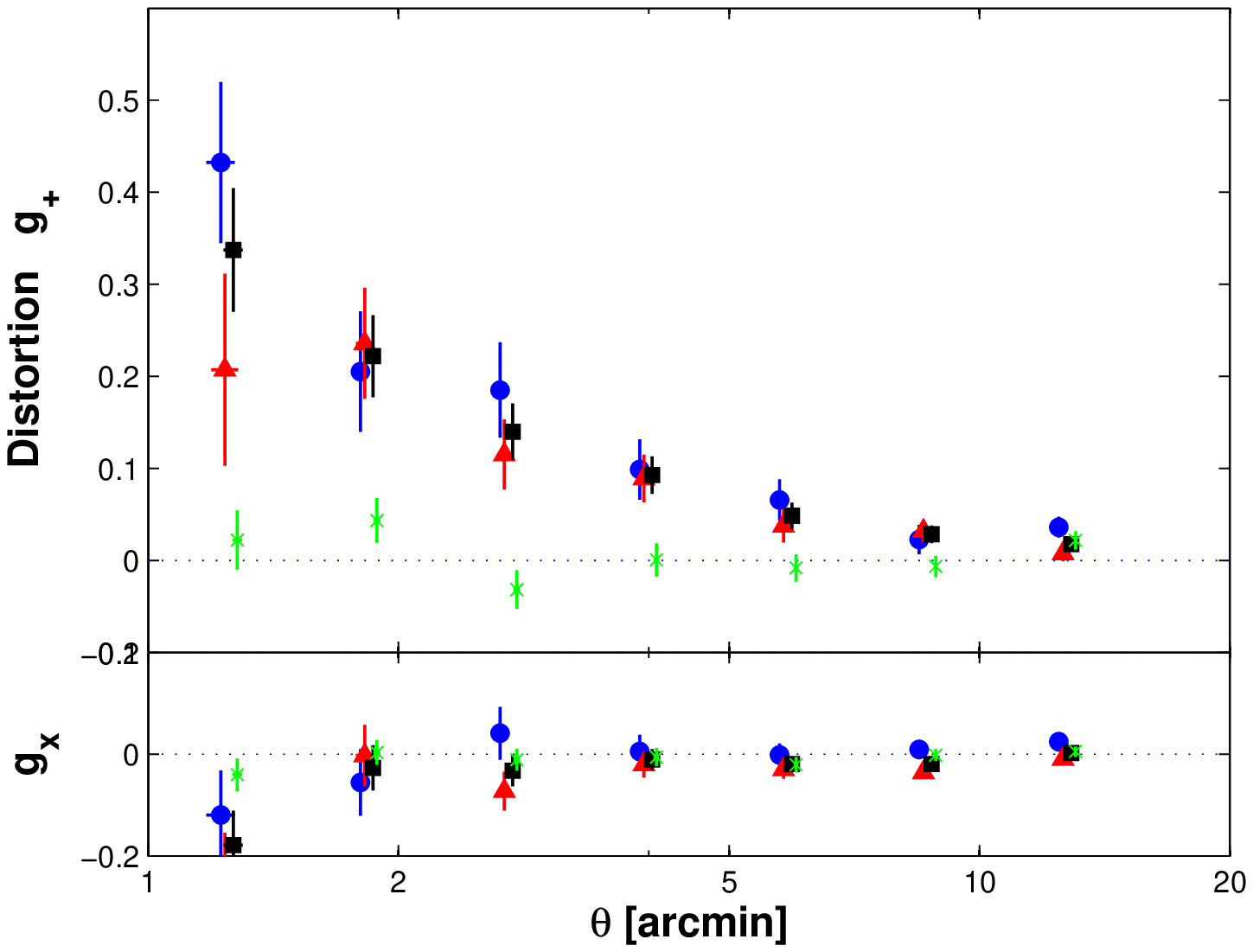}
  \caption{Tangential reduced shear vs. radius for A1703, A370 and
    RXJ1347-11 (top to bottom). The lower panels show the
    $45^{\circ}$-rotated ($g_{\rm X}$) component of the reduced shear.
    The green pentagrams represent the green sample, containing mostly
    cluster and some background galaxies whose colours are similar to
    the cluster. The black squares show the level of tangential
    distortion of the combined red+blue (triangles and circles,
    respectively) samples of the background.  The green pentagrams
    slightly rise to the background level at large radii, indicating
    the green sample still contains some level of background galaxies
    at the outskirts of the cluster, and is consistent with zero
    towards the cluster centre where cluster members dominate the
    green sample and dilute the lensing signal.}
  \label{gt_sub}
\end{figure}
\begin{figure}
  \centering \includegraphics[scale=.50]{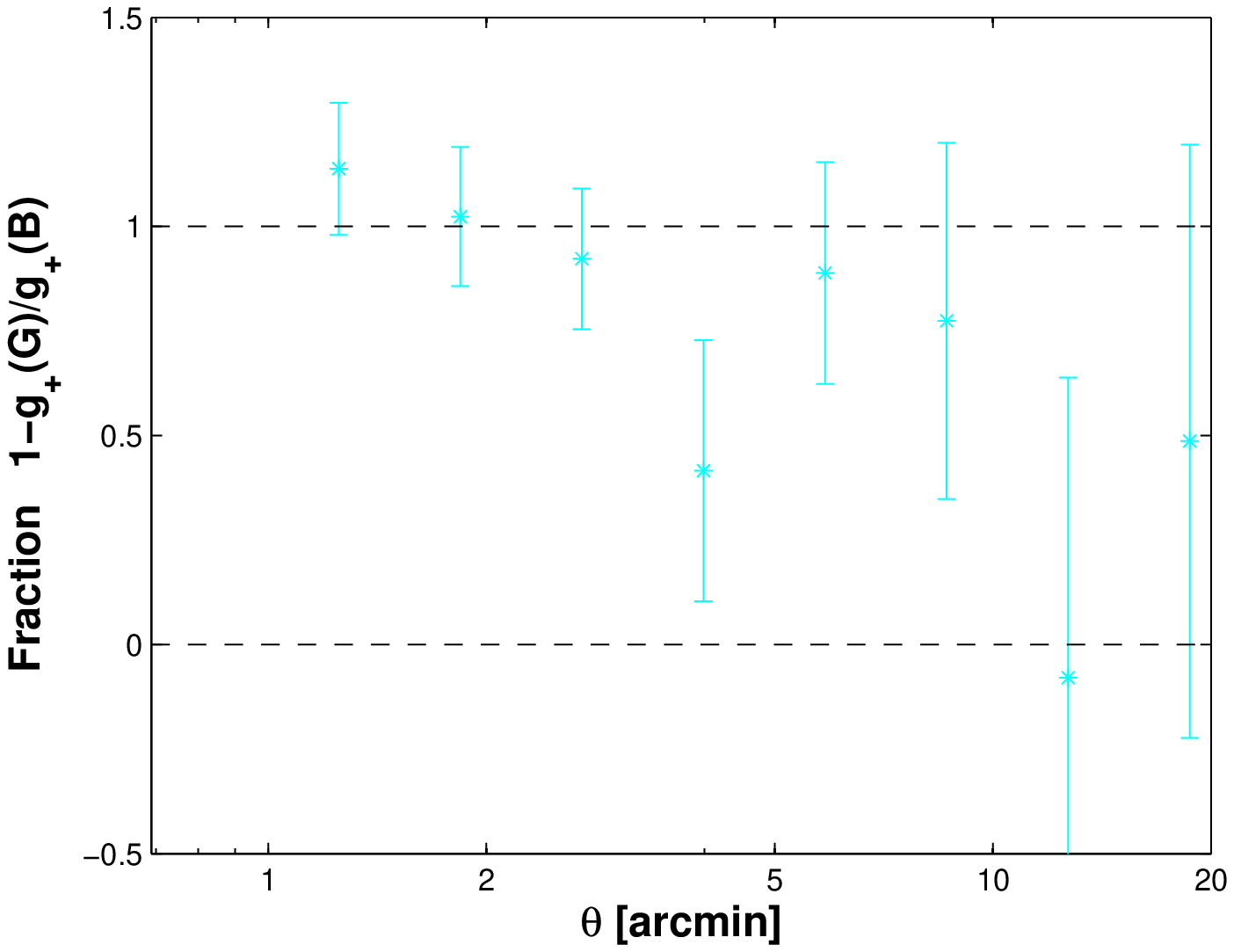}
  \includegraphics[scale=.50]{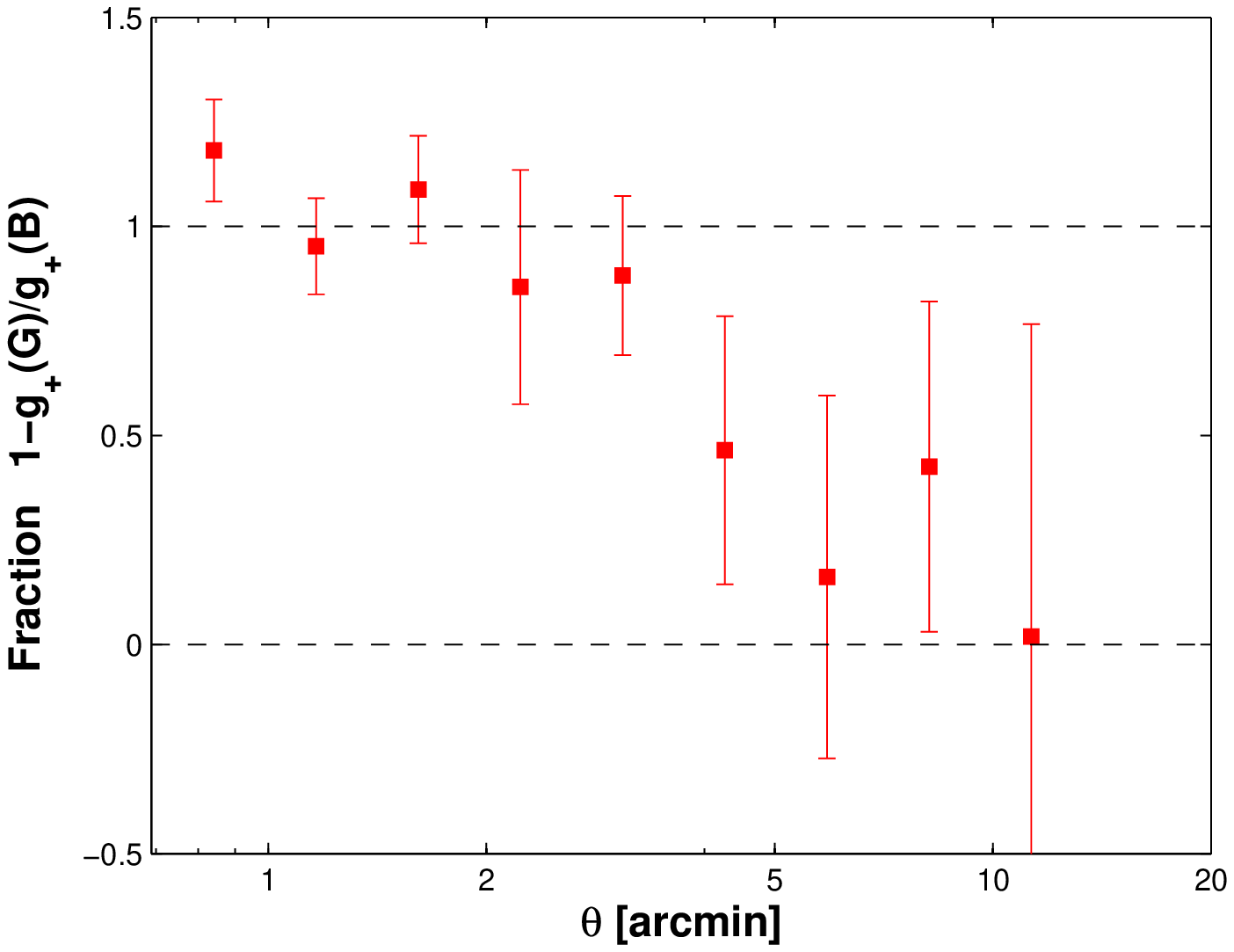}
  \includegraphics[scale=.50]{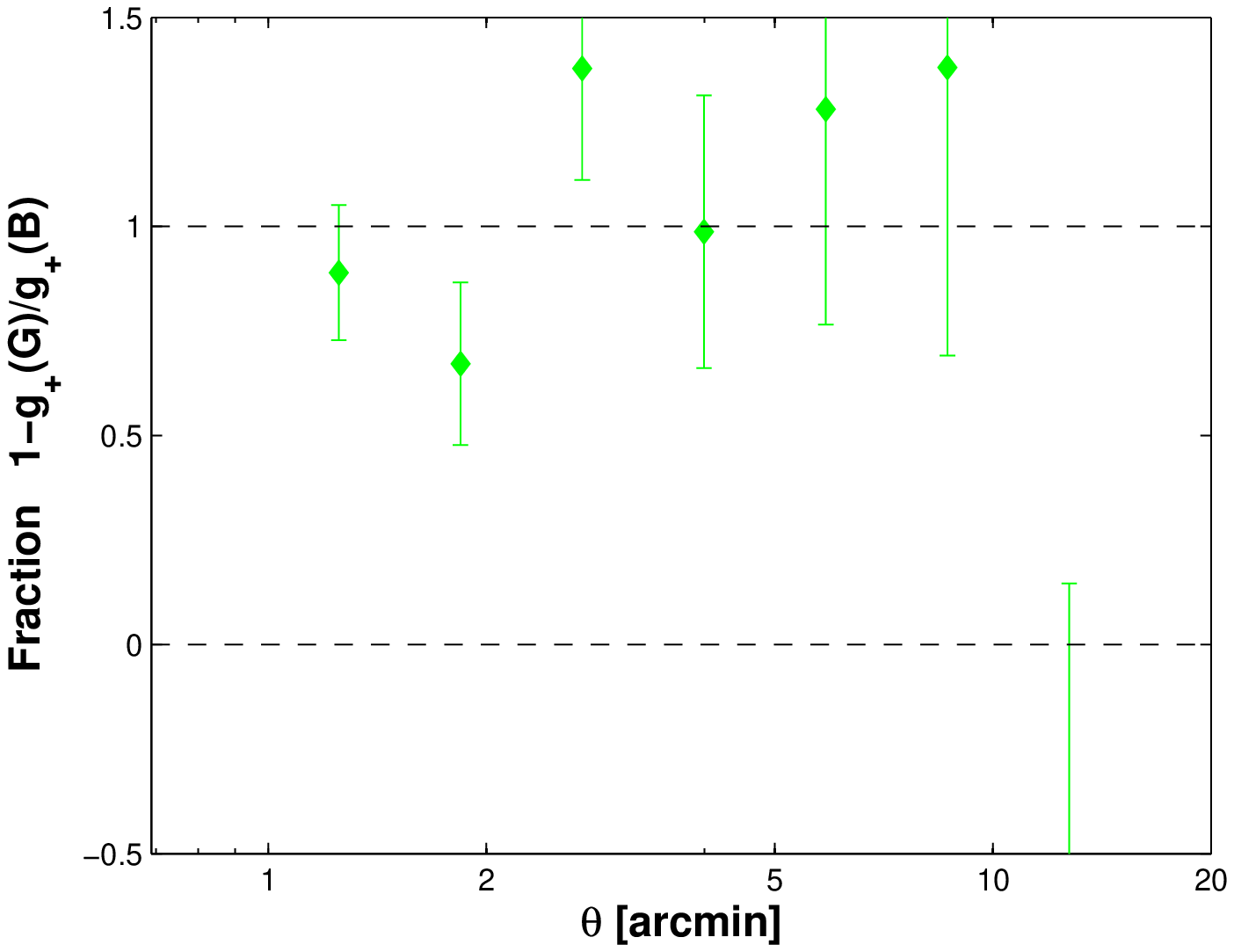}
  \caption{Fraction of cluster membership vs. radius for A1703, A370
    and RXJ1347-11 (top to bottom). Cluster membership is proportional
    to the dilution of the distortion signal of the green sample,
    relative to the expected distortion of the background galaxies set
    by the combined red and blue samples. The fraction departs from
    unity only at large radius.}
  \label{frac}
%\smallskip
\end{figure}

We use the IMCAT package developed by N. Kaiser
\footnote{http://www.ifa.hawaii.edu/~kaiser/imcat} to perform object
detection and shape measurements, following the formalism outlined in
\nocite{1995ApJ...449..460K}{Kaiser}, {Squires} \&  {Broadhurst} (1995, hereafter KSB). Our analysis pipeline is
described in \nocite{2009arXiv0908.0069U}{Umetsu} {et~al.} (2009b). We have tested our shape
measurement and object selection pipeline using STEP
\nocite{2006MNRAS.368.1323H}({Heymans} {et~al.} 2006) data of mock ground-based observations
(see \nocite{2009arXiv0908.0069U}{Umetsu} {et~al.} 2009b, \S~3.2). Full details of the
methods are presented in \nocite{2008ApJ...684..177U}{Umetsu} \& {Broadhurst} (2008),
\nocite{2009ApJ...694.1643U}{Umetsu} {et~al.} (2009a) and \nocite{2009ApJ...694.1643U}{Umetsu} {et~al.} (2009a). In short,
we find that we can recover the weak lensing signal with good
precision, typically, $m\sim5\%$ of the shear calibration bias, and
$c\sim10^{-3}$ of the residual shear offset which is about one-order
of magnitude smaller than the weak-lensing signal in the cluster
outskirts ($|g|\sim10^{-2}$).  We emphasize that this level of
calibration bias is sub-dominant compared to the statistical
uncertainty ($\sim 15\%$) in the mass due to the intrinsic scatter in
galaxy shapes.  Rather, the most critical source of systematic
uncertainty in cluster weak lensing is dilution of the distortion
signal due to the inclusion of unlensed foreground and cluster member
galaxies, which can lead to an underestimation of the true signal for
$R < 400$~kpc/h by a factor of $2-5$, as demonstrated in
\nocite{2005ApJ...619L.143B}{Broadhurst} {et~al.} (2005), \nocite{2007ApJ...663..717M}{Medezinski} {et~al.} (2007) and here.

The shape distortion of an object is described by the complex
reduced-shear, $g=g_1+ig_2$, where the reduced-shear is defined as:
\begin{equation}
g_{\alpha}\equiv\gamma_{\alpha}/(1-\kappa). 
\end{equation}
The tangential component of the
reduced-shear, $g_{T}$, is used to obtain the azimuthally averaged
distortion due to lensing, and computed from the distortion
coefficients $g_{1},g_{2}$:
\begin{equation}
g_{T}=-( g_{1}\cos2\theta +  g_{2}\sin2\theta),
\end{equation}
where $\theta$ is the position angle of an object with respect to the
cluster centre, and the uncertainty in the $g_{T}$ measurement is
$\sigma_{T} = \sigma_{g}/\sqrt{2}\equiv \sigma$ in terms of the RMS
error $\sigma_{g}$ for the complex reduced-shear measurement.  To
improve the statistical significance of the distortion measurement, we
calculate the weighted average of $g_{T}$ and its weighted error, as
\begin{eqnarray}
\label{eq:mean_gt}
\langle g_{T}(\theta_n)\rangle &=& \frac{\sum_i u_{g,i}\, g_{+,i}}
{\sum_i u_{g,i}},\\
\sigma_{T}(\theta_n) &=& \sqrt{
\frac{\sum_i u^2_{g,i}\sigma^2_i}{\left(\sum_i u_{g,i}\right)^2},
}
\end{eqnarray}
where the index $i$ runs over all of the objects located within the
$n$th annulus with a median radius of $\theta_n$, and $u_{g,i}$ is the
inverse variance weight for $i$th object,
$u_{g,i}=1/(\sigma_{g,i}^2+\alpha^2)$, softened with $\alpha$. We
choose $\alpha=0.4$, which is a typical value of the mean RMS
$\bar{\sigma}_g$ over the background sample.  The case with $\alpha=0$
corresponds to an inverse-variance weighting.  On the other hand, the
limit $\alpha\gg \sigma_{g,i}$ yields a uniform weighting.  We have
confirmed that our results are insensitive to the choice of $\alpha$
(i.e., inverse-variance or uniform weighting) with the adopted
smoothing parameters.

In Fig.~\ref{gt_sub} we plot the radial profile of $g_T$ of the green
(pentagrams), red (triangles) and blue (circles) samples defined
above. The black squares represent the red+blue combined sample,
showing the best estimate of the lensing signal.  The red and blue
samples profiles rise continuously toward the centre for each of the
clusters, and are in good agreement with each other, demonstrating
that both are dominated by background galaxies and are not
contaminated by the cluster at all radii.  The $g_T$ profile of the
green sample lies close to zero, especially at small radii, but shows
a small positive signal at large radius, close in amplitude to the
background signal measured above, meaning at beyond $\sim 5\arcmin$
there are few cluster members in comparison to ``green'' background
galaxies. Indeed, the virial radius derived from our best fitting NFW
model lies at about $10\arcmin-15\arcmin$ and so physically it is
reasonable that the cluster population is small here.  The measured
zero level of tangential distortion interior to this radius for the
green sample reinforces our CC selection.

\section{Weak Lensing Dilution}\label{dilution}

We can now estimate the fraction of cluster galaxies by taking the
ratio of the WL signal between the green sample and the background
sample, with the background including both red and blue galaxies as
explained in \S~\ref{sec:samples}.

For a given radial bin ($r_n$) containing objects in the green sample
(\S~\ref{sec:data}), the mean value of $g_T^{(G)}$
(eq.~\ref{eq:mean_gt}) is an average of the signal over cluster
members and some background galaxies.  Thus, its mean value $\langle
g_T^{(G)}\rangle$ will be lower than the true background level denoted
by $\langle g_T^{(B)}\rangle$ (Fig.~\ref{gt_sub}) in proportion to the
fraction of unlensed galaxies in the bin that lie in the cluster
(rather than in the background), since the cluster members on average
will add no net tangential signal. Therefore,
\begin{equation}\label{eq:frac}
  f_{cl}(r_n)\equiv\frac{N_{cl}}{N_{Green}}
  =1-\frac{\langle{}g_T(r_n)^{(G)}\rangle{}}{\langle{}g_T(r_n)^{(B)}\rangle{}}
  \frac{\langle D^{(B)}\rangle}{\langle D^{(G)}\rangle}
\end{equation}
is the cluster membership fraction of the green sample, where
${\langle D^{(B)}\rangle}$ and ${\langle D^{(G)}\rangle}$ are the mean
lensing depths, $D_{\rm ls}/D_{\rm s}$, of the background and green
galaxy samples, respectively, derived in \S~\ref{sec:photo-z}
\nocite{2007ApJ...663..717M}(see full derivation in the appendix of
 {Medezinski} {et~al.} 2007).
 
In Fig.~\ref{frac} we show this ratio for each of our new clusters.
In our previous analysis of A1689, where only colour-magnitude
selection was done, many background galaxies were present in the green
sample, so that only in the first radial bins the fraction of cluster
galaxies was high, and gradually declined with radius (due to
increasing fraction of background galaxies) at larger radii. Here,
since we select a tight region around the cluster in CC and are able
to better resolve it, our results show that the fraction is close to
unity throughout the radial range of the cluster.  Only in the outer
bins, usually at $r>5\arcmin-10\arcmin$ does the fraction decline, and
we see a smaller percentage of cluster members. Large errorbars are
introduced at outer radii since there are few cluster galaxies
present, making it harder to use this effect in order to determine
cluster membership.
% rephrase
% Another effect stemming from this is having big error-bars on our
% estimates at large radii. This stands to reason, since if there are no
% cluster galaxies and little background in the green sample at outer
% radii, small number statistics are causing large errors, and there is
% little background to compare with the red+blue background signal.  The
% significance of this is that up to a radius of $r\sim10\arcmin$ we can
% be confident of resolving the cluster, and beyond that we are likely
% unable to determine the cluster level to high accuracy since it is not
% present at high numbers.

\section{Cluster Light Profiles}\label{lum_sec}
\begin{figure}
  \centering
  \includegraphics[scale=.50]{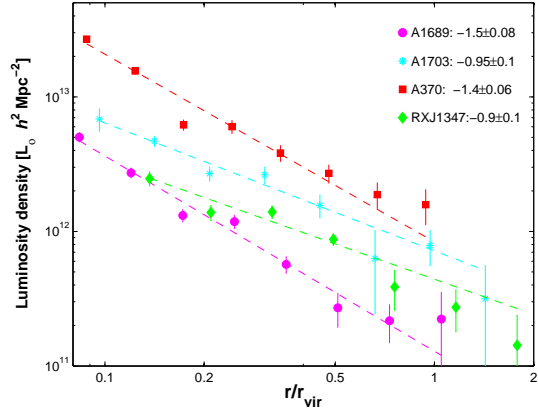}
  \caption{``g-Weighted'' luminosity density vs. $r/r_{vir}$ for A1689
    (magenta circles), A1703 (cyan asterisks), A370 (red squares) and
    RXJ1347-11 (green diamonds).  The flux $F_i$ (green sample) of
    each galaxy is weighted by its tangential distortion $g_{T,i}$
    with respect to the background distortion signal.  The dashed
    lines are the best fitting lines whose slopes are given in the
    legend. For clarity, the luminosity profiles have been shifted by
    a factor: A1703 shifted up by a factor of 2, A370 by a factor of
    3, and RXJ1347-11 by a factor of $1.2$. }
  \label{LPall}
\end{figure}
\begin{figure}
  \centering
  \includegraphics[scale=.50]{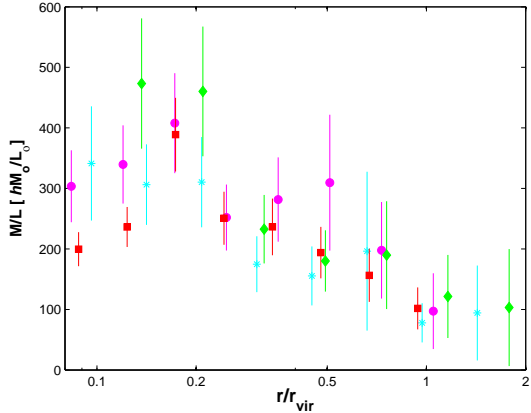}
  \caption{Mass-to-light ratio vs. $r/r_{vir}$ for A1689 (magenta
    circles), A1703 (cyan asterisks), A370 (red squares) and
    RXJ1347-11 (green diamonds). The profiles are similar to each
    other, peaking around $0.2r_vir$, and declining continuously to
    the virial radius, which seems to be a general property of all
    well-studied clusters.}
  \label{MLall}
\end{figure}

We now turn to translate the deduced profile of membership fraction to
a cluster luminosity profile.  To account for the background
contribution in the green sample and deduce only the cluster
luminosity, we first measure the flux of all the galaxies in the green
sample, $\displaystyle\sum_{i}{F_i^{(G)}}$, and then subtract the flux
of the $i$-th galaxy weighted by it's tangential distortion, $g_i$,
relative to the background sample distortion,
${\langle{}g_T(r_n)^{(B)}\rangle{}}$. When averaged over the entire
green sample this will have zero contribution from the unlensed
cluster members, but will remove the contribution of background
galaxies present in the green sample.
% We weight each ``green'' galaxy flux,
% $F_i$, by it's tangential distortion, $g_i$, and subtract this
% ``$g_T$-weighted'' luminosity contribution of each galaxy, which when
% averaged over the distribution will have zero contribution from the
% unlensed cluster members. 
% This will account for any difference in the
% brightness distributions of the cluster members to that of the
% background galaxies, in particular the skewness of the cluster
% sequence to lower magnitudes.
The total flux of the cluster in the
$n^{th}$ radial bin is therefore,
\begin{equation}
  F_{cl}(r_n)=\displaystyle\sum_{i}{F_i^{(G)}}
  -\frac{\langle{}D^{(B)}\rangle{}/\langle{}D^{(G)}\rangle{}}
  {\langle{}g_T(r_n)^{(B)}\rangle{}}
  \displaystyle\sum_{i}{F_i^{(G)}g_{T,i}^{(G)}}
\end{equation}
and the flux is then translated to luminosity.  First we calculate the
absolute magnitude,
\begin{equation}
  M=m-5\log{d_L}-K(z)+5,
\end{equation}
where the $K$-correction is evaluated for each radial bin according to
its colour. The luminosity is simply,
\begin{equation}
  L=10^{0.4(M_{\odot}-M)}L_{\odot},
\end{equation}
where $M_{\odot}$ is the absolute magnitude of the Sun (AB system) in
the relevant band.

The results for all four clusters, including A1689, are shown in
Fig.~\ref{LPall}.  As explained in \S~\ref{sec:samples}, the green
sample comprises mostly of cluster galaxies, therefore the correction
is not big for $r<10\arcmin$, and only present at large radii,
increasing the error. The cluster luminosity profiles mostly have a
steady linear decline with radius, and are well described by a
power-law fit, but with different slopes of $d\log(L)/d\log(r)\sim
-0.95\pm0.1, -1.4\pm0.06, -0.9\pm0.1$ for A1703, A370 and RXJ1347-11,
respectively.

\section{$M/L$ Profiles}\label{ML_sec}
%%%%%%%%%%%%%%%%%%%%%%%%%%%%%%%%%%%%%%%%%%%%
We may now derive a radial profile of the differential mass-to-light
ratio (often referred to as $\delta M(r)/\delta L(r)$) by dividing the
best-fit NFW projected mass density profile by the newly derived
luminosity density profile obtained in \S~\ref{lum_sec}.  The four
$M/L$ profiles are shown together in Fig.~\ref{MLall} to the limit of
our data, $\sim3$~Mpc/h.  The mass profiles for these clusters and
their NFW fits were derived in our earlier work,
\nocite{2005ApJ...619L.143B}{Broadhurst} {et~al.} (2005) for A1689 and in \nocite{2008ApJ...685L...9B}{Broadhurst} {et~al.} (2008)
for A370, A1703 and RXJ1347 based on the sample selection scheme
described in this paper.

For A1689, we found previously that the $M/L$ ratio peaks at
intermediate radius around $r\sim100h^{-1}$~kpc and then falls off to
larger radius. Here we find a similar behavior for the other clusters,
especially evident for A370, where $M/L$ peaks at about
$r\sim350h^{-1}$~kpc with values of $M/L_R\sim300-400~h(M/L)_{\odot}$,
and declines to values of $M/L_R\sim100~h(M/L)_{\odot}$ at the
outskirts. A decline in $M/L$ is also seen towards the centre of the
cluster, reaching $M/L_R\sim200~h(M/L)_{\odot}$ below
$r\lesssim200h^{-1}$~kpc.  The $M/L$ profile of A1703 also shows a
similar trend, where it peaks at a radius of $r\sim200-300h^{-1}$~kpc
with values of $M/L_{r'}\sim300~h(M/L)_{\odot}$ and declines gradually
to values of $M/L_{r'}\sim100~h(M/L)_{\odot}$ at large radii.  The
profile of RXJ1347-11 declines from values of
$M/L_R\sim400-500~h(M/L)_{\odot}$ at $r\sim200-300h^{-1}$~kpc to
values of $M/L_R\sim100~h(M/L)_{\odot}$ at large radius.
\begin{figure}
  \centering
  \includegraphics[scale=.50]{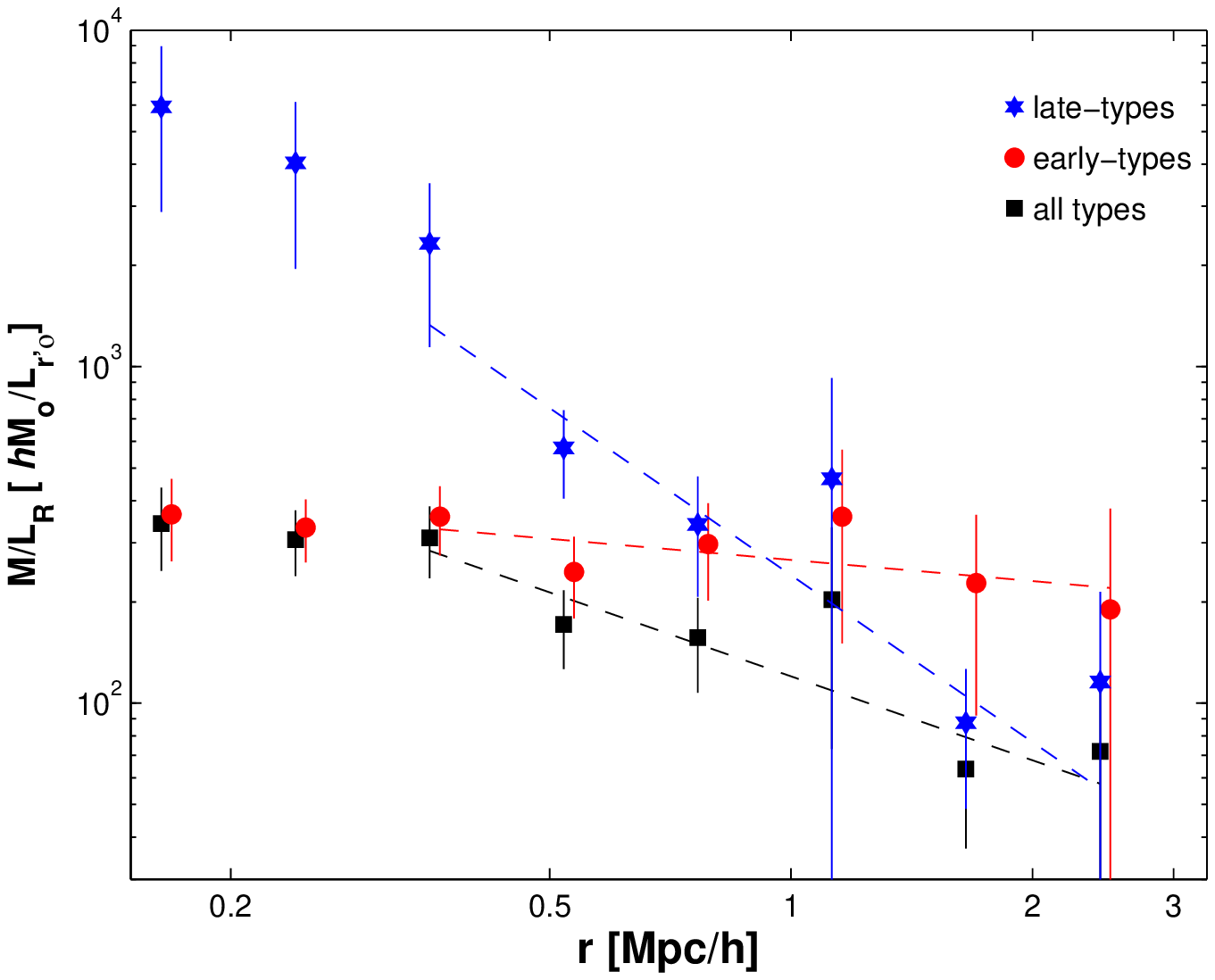}
  \includegraphics[scale=.50]{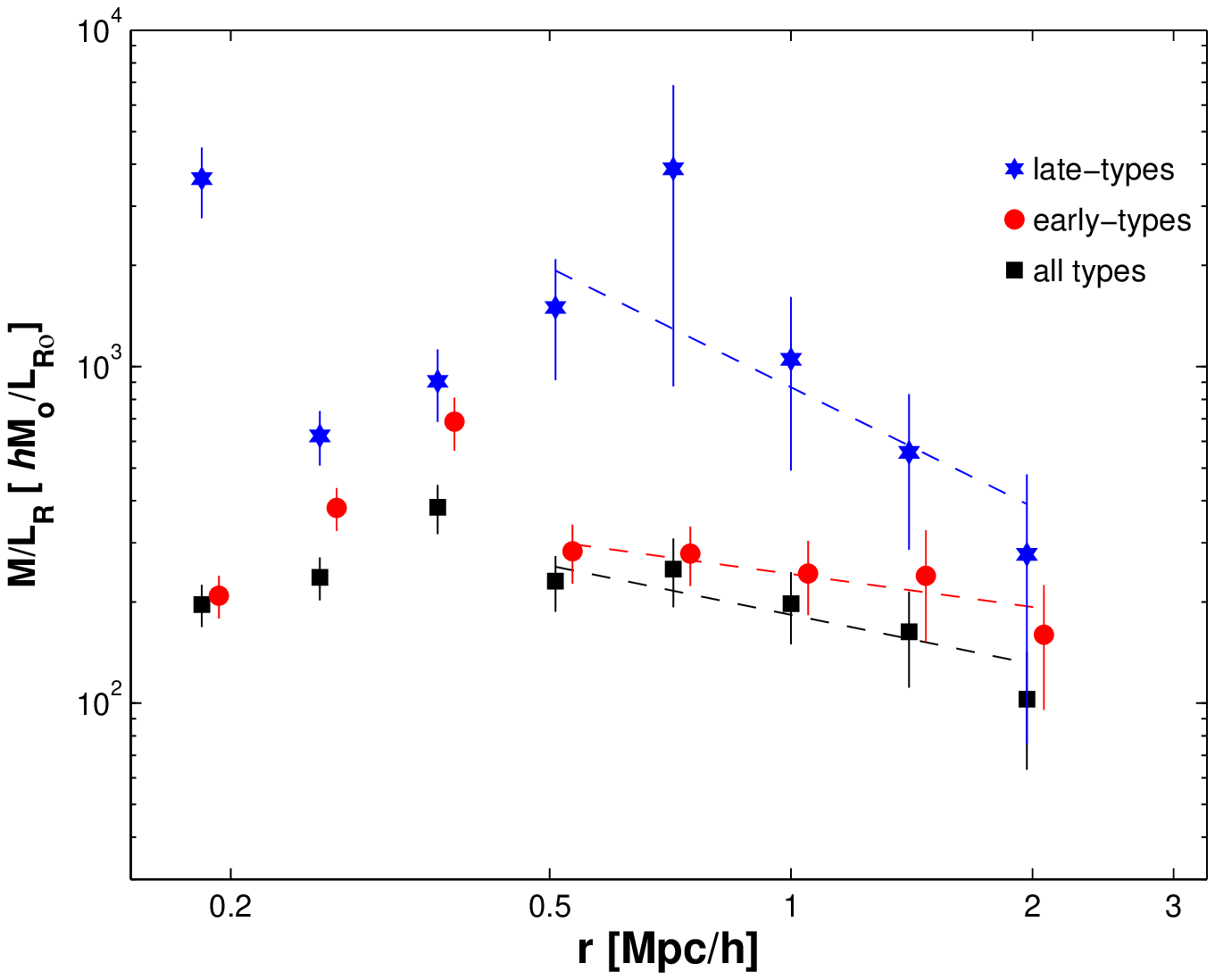}
  \includegraphics[scale=.50]{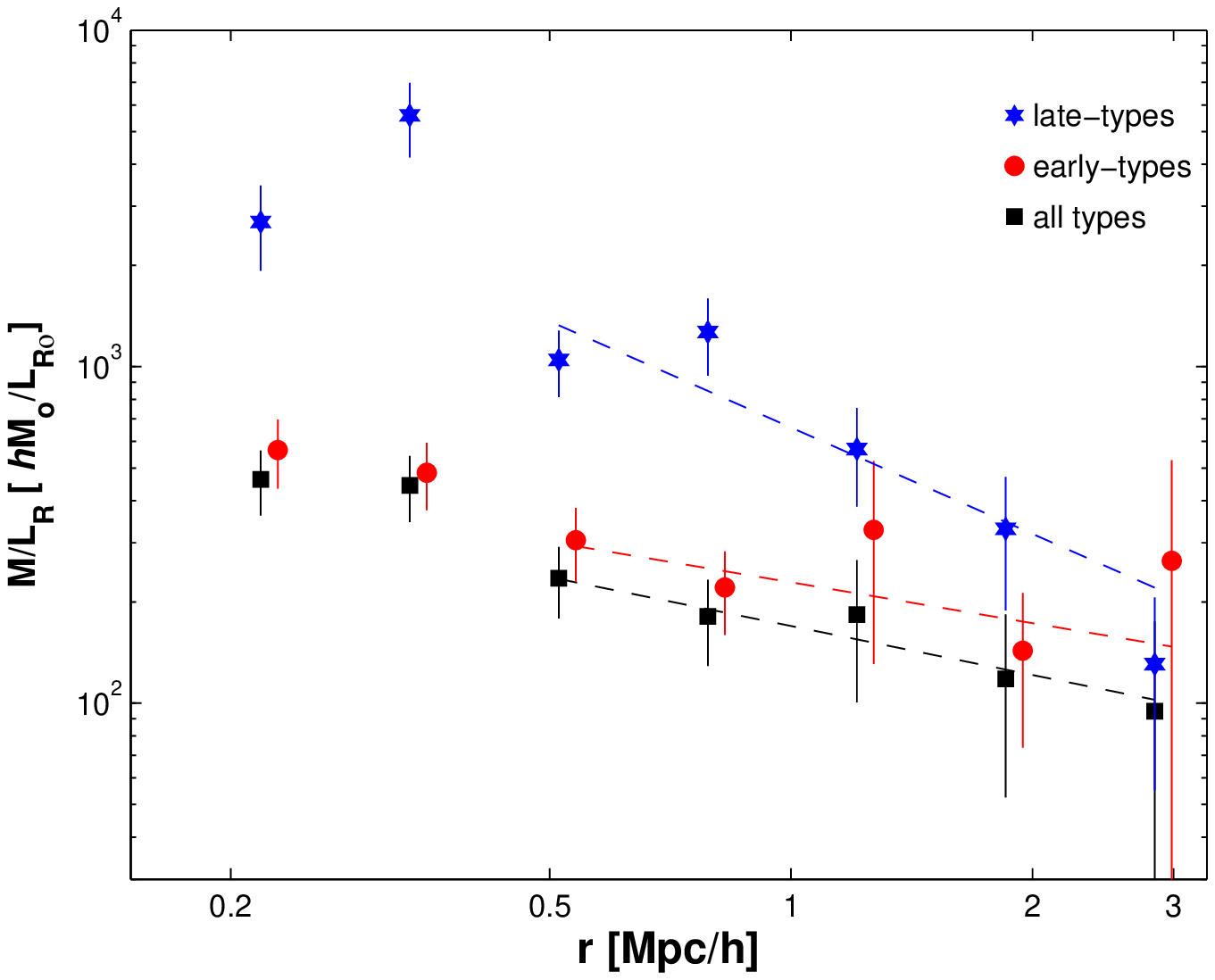}
  \caption{Mass-to-light ratio vs. radius for A1703, A370 and RXJ1347
    (top to bottom plots) divided into two different cluster
    sub-populations.  The black squares mark the entire green sample
    (same $M/L$ profiles shown above in Fig.~\ref{MLall}); the red
    circles mark a subsample from a tighter selection of cluster
    galaxies around the red sequence (white solid line marked in
    Fig.~\ref{colorcolor_ncell}), mostly early-type elliptical
    galaxies. The blue hexagrams correspond to later-type cluster
    members (the rest of the green population inside the white dashed
    line, but outside the solid white line marked in
    Fig.~\ref{colorcolor_ncell}). The slope of the declining outer
    profile is significantly steeper for the blue profiles, whereas
    for the red profiles the $M/L$ is much flatter.}
  \label{ML3pop}
\end{figure}
\begin{figure}
  \centering
  \includegraphics[scale=.50]{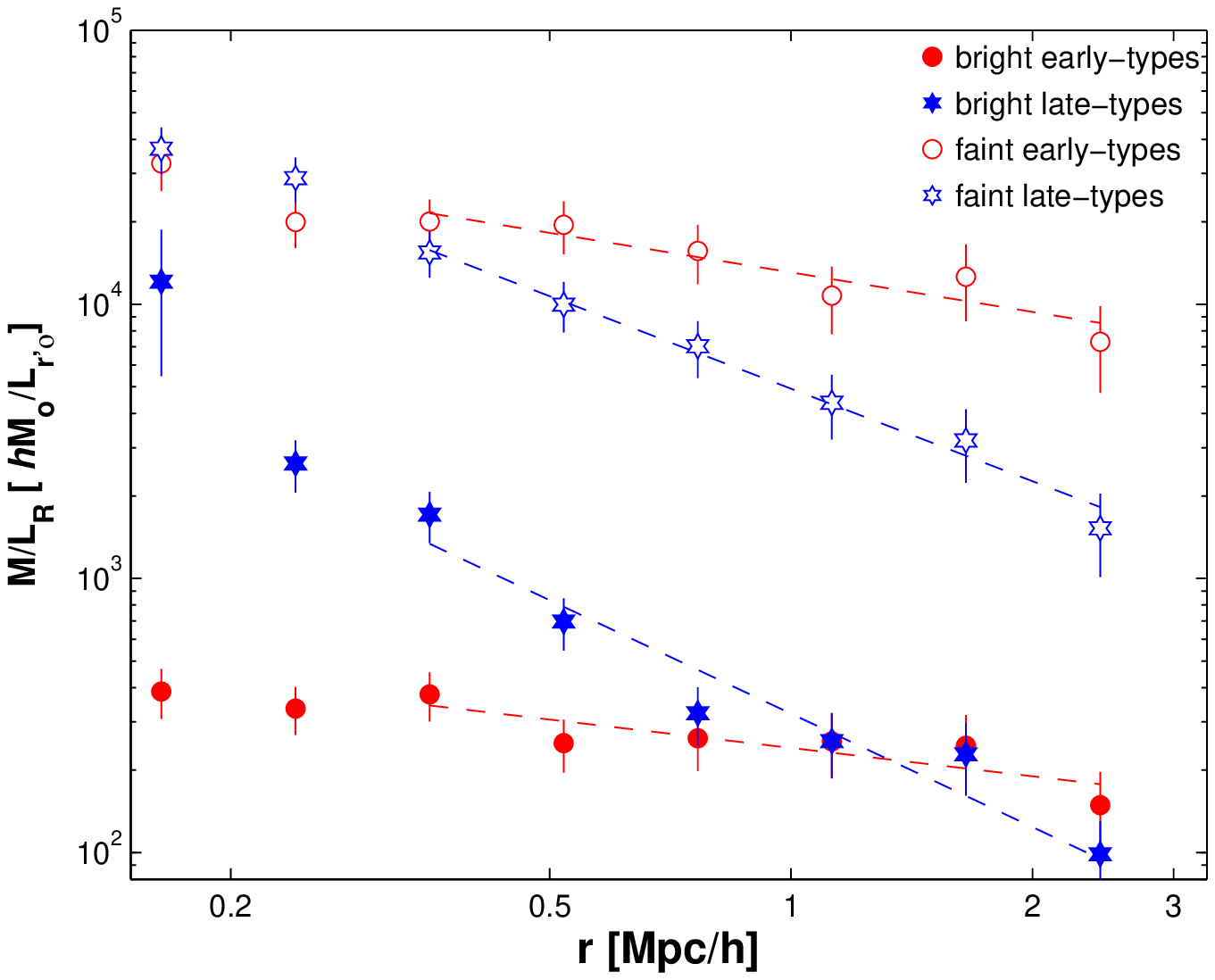}
  \includegraphics[scale=.50]{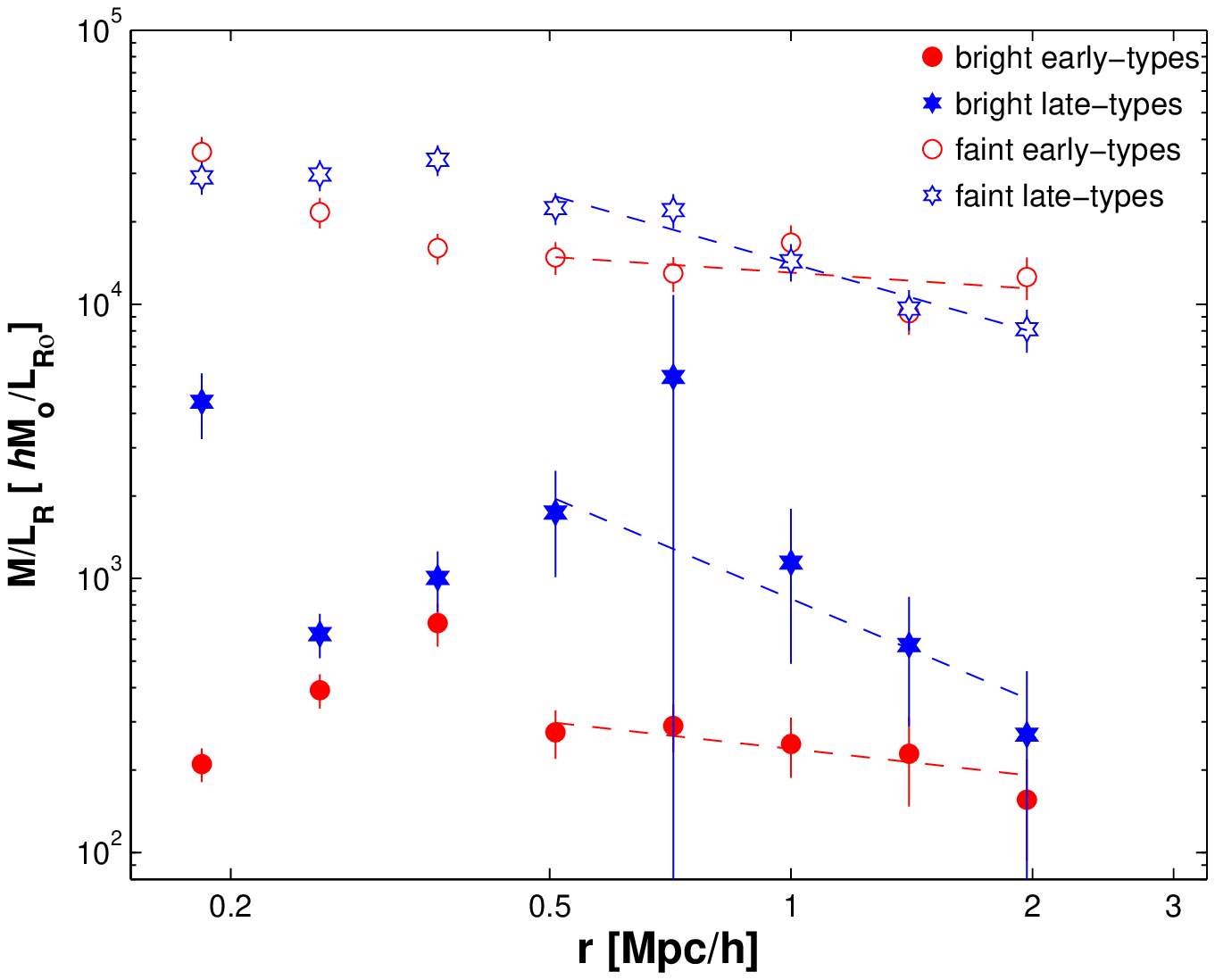}
  \includegraphics[scale=.50]{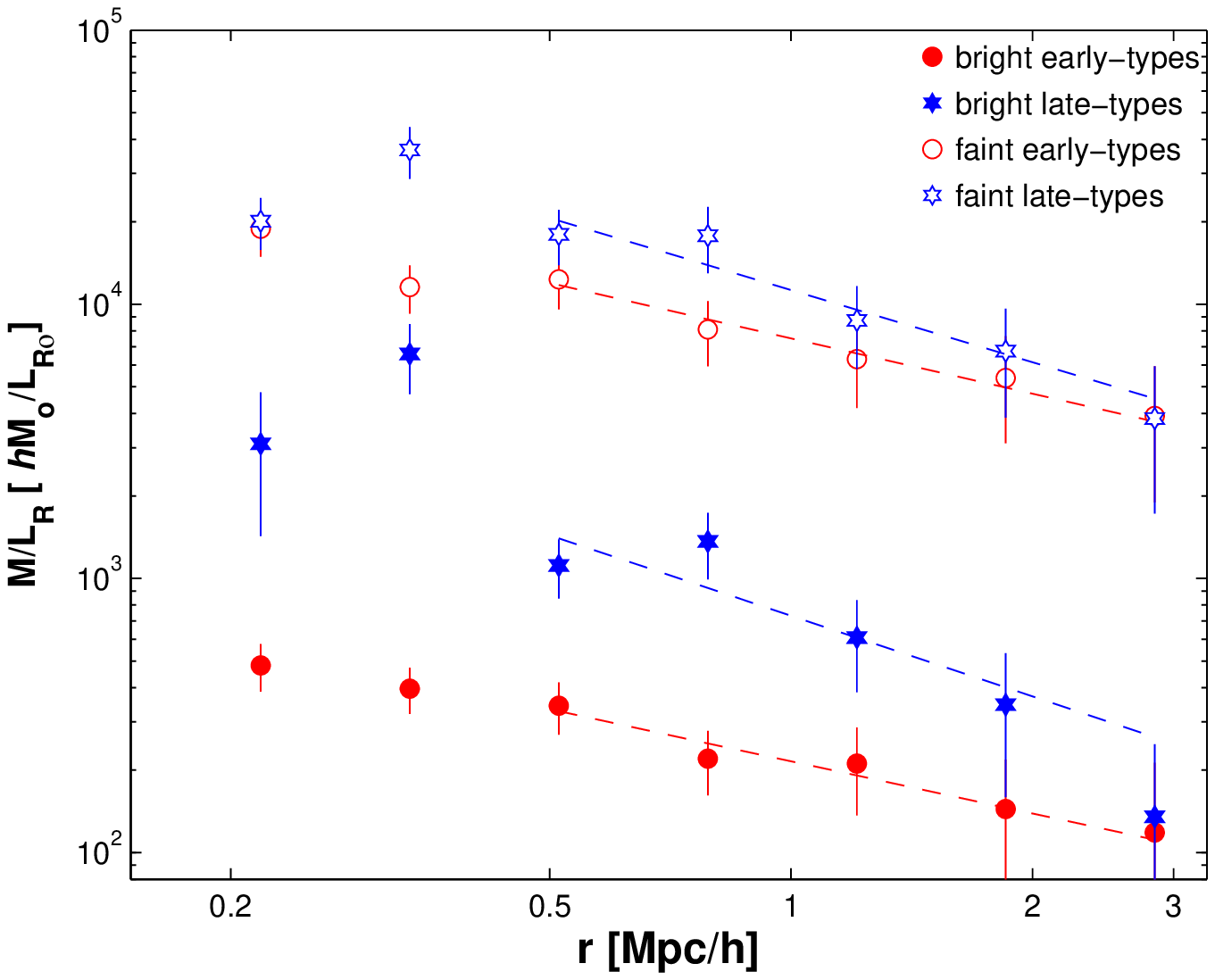}
  \caption{Mass-to-light ratio vs. radius for A1703, A370 and RXJ1347
    (top to bottom plots) for early (red circles) and late type (blue
    hexagrams) clusters members, as in Fig.~\ref{ML3pop}, but divided
    into high- (solid) and low-luminosity (empty) subsamples. Clearly
    the behavior of the early and later types is not a strong function
    of luminosity.}
  \label{ML2Lbins}
\end{figure}

The decrease in $M/L$ values toward the centre of the cluster is only
really evident for two clusters, A1689 and A370. For these clusters
the luminosity profiles are relatively steep, and it this is
continuation of a steep power-law towards the centre that causes $M/L$
to decline, as the mass profiles are not power-law but have a
continuously declining gradient towards the cluster centres
(NFW-like), so that the proportion of $M/L$ declines markedly interior
to the characteristic radius for these two clusters.

The decrease in $M/L$ toward large radius seen for all four clusters
is possibly explained by the increasing proportion of later type
galaxies at larger radii, a manifestation of the morphology-density
relation \nocite{1980ApJ...236..351D}({Dressler} 1980). Similar results have been shown
by \nocite{2000AJ....120.2338R}{Rines} {et~al.} (2000) and \nocite{2004ApJ...600..657K}{Katgert} {et~al.} (2004); in
\nocite{2000AJ....120.2338R}{Rines} {et~al.} (2000) where the differential $M/L$ profile is
similar to ours, with a decreasing $M/L$ with radius, and their
outermost value of $M/L_R ~100~h(M/L)_{\odot}$ is in good agreement
with the values we get for all our four clusters.
\nocite{2004ApJ...600..657K}{Katgert} {et~al.} (2004) show a decrease at small radius, and by
removing the brightest cluster members this decrease is shown to
disappear.

The declining behavior is not simply accounted for by the change of
stellar population with distance from the cluster centre. The galaxy
colours show a very moderate decline with radius. In the case of A370,
the radial colour gradient is $d(B-z')/d\log(r) = -0.13\pm0.04$,
equivalent to a total change of $\delta(B-z')=0.12$ over the virial
radius. This corresponds to a change in the mean stellar mass-to light
ratio, $M_*/L_R$, of only $\sim17\%$. Therefore, such a small
correction alone cannot explain the declining trend of $M/L$ with
radius.

\begin{figure}
  \centering
  \includegraphics[scale=.50]{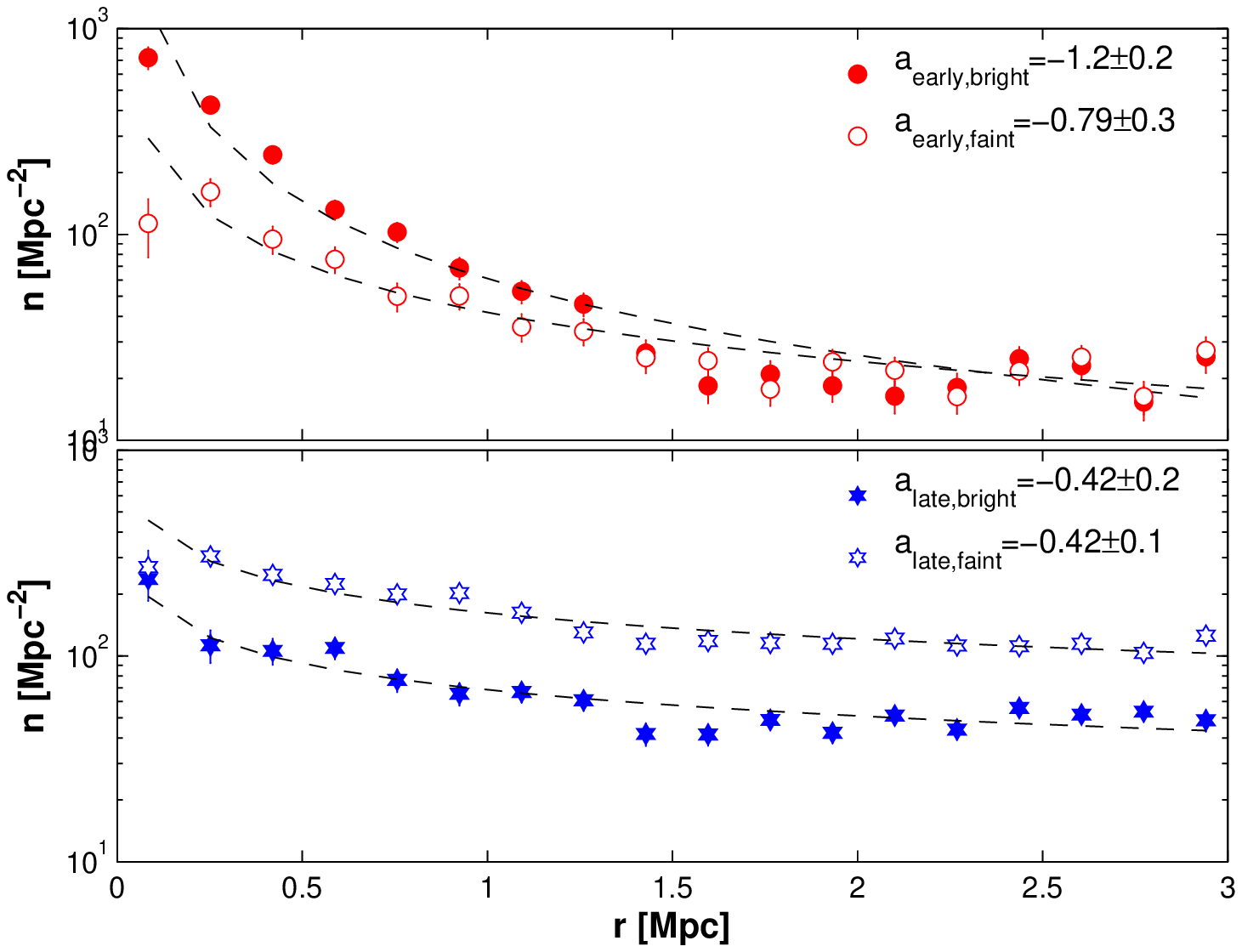}
  \includegraphics[scale=.50]{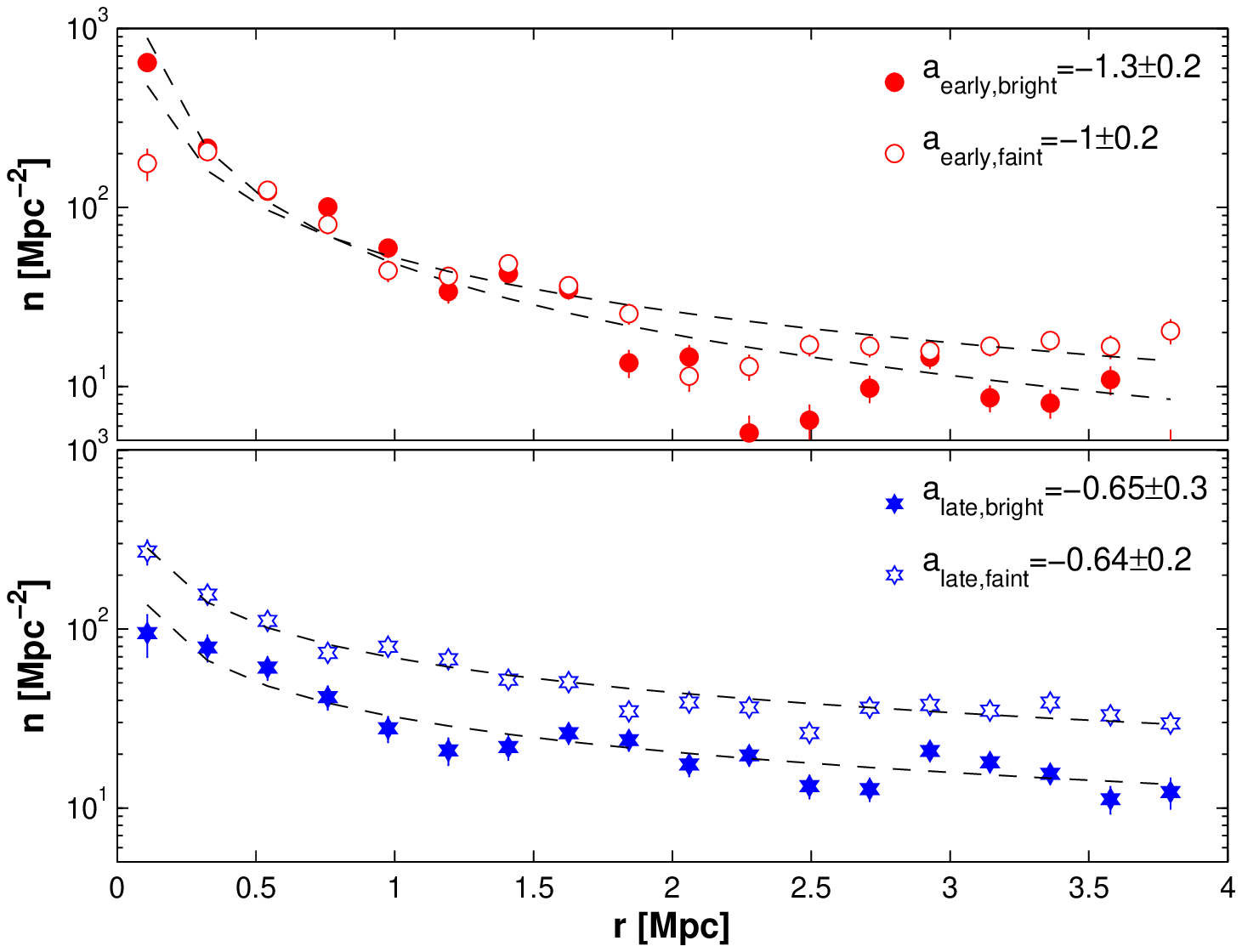}
  \includegraphics[scale=.50]{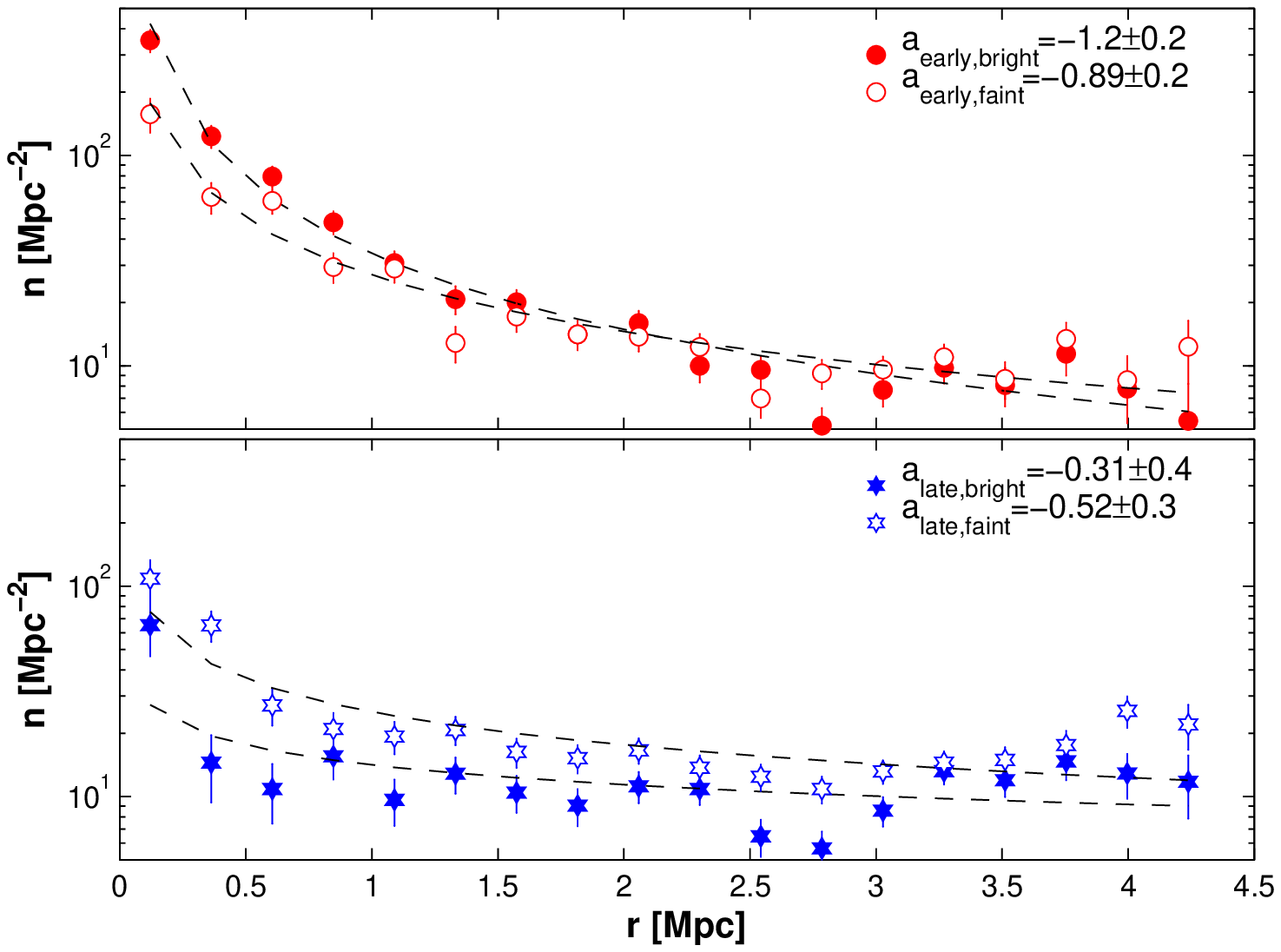}
  \caption{Number density profiles for A1703, A370 and RXJ1347-11 (top
    to bottom plots) for early (upper panels, red circles) and late
    (lower panels, blue hexagrams) type cluster members further
    divided into high (solid) and low-luminosity (empty) subsamples
    (as in Fig.~\ref{ML2Lbins}).  The redder profiles are more
    centrally concentrated than the bluer profiles, suggesting the
    proportion of early- to late-type members is a clear function of
    radius but not as much of luminosity, favouring early-types at
    smaller radius.}
  \label{n2Lbins}
\end{figure}

We now further examine the behaviour of $M/L$ by separating cluster
member galaxies by colour and by luminosity. There is a clear
distinction between the $M/L$ profiles based on colour selection as
shown in Fig.~\ref{ML3pop}. Tightly selecting just the sequence
galaxies (inside the white solid line marked in
Fig.~\ref{colorcolor_ncell}) the $M/L$ profiles are relatively flatter
for all the clusters (Fig.~\ref{ML3pop}, red circles), showing mild
decline with radius, whereas the remaining non-sequence objects (the
rest of the green population between the white dashed line and the
solid while line marked in Fig.~\ref{colorcolor_ncell}) have a steep
dependence (Fig.~\ref{ML3pop}, blue hexagrams) which is clearly
responsible for the majority of the trend of $M/L$ found above for the
cluster population as a whole (Fig.~\ref{ML3pop}, black squares). If
we now further divide these subsamples by luminosity into two broad
bins as shown in Fig.~\ref{ML2Lbins}, we see that this difference
between the sequence and non-sequence populations is not strongly
dependent on luminosity.  If we also look at the number-density
profile of each of these four subsamples (Fig.~\ref{n2Lbins}), we see
that the early-type galaxies (red circles) are more concentrated than
the later types (blue hexagrams), independently of luminosity.

It is evident from the above comparisons that later-type galaxies are
relatively rare in the cluster centre, an effect which most likely has
its origins in tidal and/or stripping of infalling later-type galaxies
and their transformation into redder and rounder galaxies
predominating in cluster centres.  We now explore tidal effects in
this context and comment on other related work.

%%%%%%%%%%%%%%%
\subsection{Tidal and Ram-Pressure Stripping}\label{sec:tidal}
In CDM-dominated cosmological models, the growth of structure is
hierarchical, with clusters evolving through many merger events during
which galaxies are dynamically affected to a varying degree based on
their formation times and spatial locations in their parent systems.
This process can be described by the excursion set formalism
\nocite{1993MNRAS.262..627L}({Lacey} \& {Cole} 1993) with more detailed insight gained from
hydrodynamical simulations. It has been clearly demonstrated that
during these merger events DM haloes suffer pronounced tidal stripping
\nocite{1996Natur.379..613M}(e.g., {Moore} {et~al.} 1996).  Detailed N-body simulations
indicate that up to 70\% of the mass may be stripped from halos
orbiting close to the centre of massive clusters when integrated over
the Hubble time, and with $\sim 20\%$ loss even near the virial radius
\nocite{2005ApJ...618..557N}({Nagai} \& {Kravtsov} 2005).  Statistical evidence for halo truncation
is claimed also from lensing based measurements
\nocite{2002ApJ...580L..17N,2009ApJ...693..970N}({Natarajan} {et~al.} 2002, 2009).

\begin{figure}
  \centering
  \includegraphics[scale=.50]{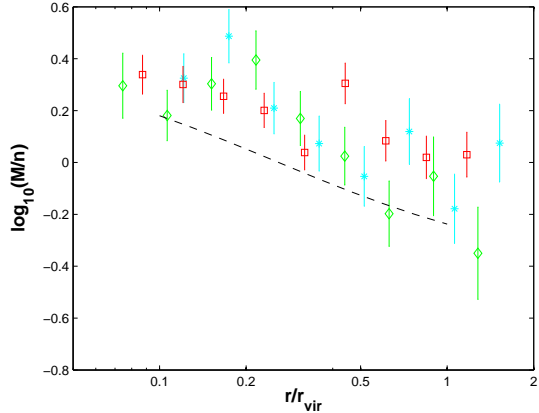}
  \caption{Radial profile of the ratio of the cluster mass density
    over the galaxy number density for A1703 (cyan stars), A370 (red
    squares) and RXJ1347-11 (green diamonds). Dashed line shows this
    ratio as calculated from the simulations of Nagai \& Kravtsov
    (2005). The observed trend is similar to the theoretical
    prediction for each of the clusters and can be attributed
    theoretically to tidal effects by the cluster acting on the galaxy
    DM halos. }
  \label{Mnall}
\end{figure}

As can be expected, simulations show that despite the substantial
effect of tidal forces in removing the outer DM from galaxy halos, the
stellar content of early-type galaxies is only weakly affected by
tidal interactions due to the centralised location of most stars
within the DM halos \nocite{2005ApJ...618..557N,2005MNRAS.359.1537R}({Nagai} \& {Kravtsov} 2005; {Reed} {et~al.} 2005).
Hence, the ratio of total mass to galaxy luminosity is predicted to be
approximately independent of radius in this context.  In detail, the
above simulations show a mild radial decline of the ratio of the total
mass density over the number density of DM halos, $M/n$, as shown in
Fig.~\ref{Mnall} (dashed curve). The relatively gentle declining trend
in this ratio is the result of tidal forces
\nocite{2005ApJ...618..557N,2005MNRAS.359.1537R}({Nagai} \& {Kravtsov} 2005; {Reed} {et~al.} 2005), which act over time
to reduce the number of sub halos in the inner region as compared to
the cluster outskirts.

This $M/n$ ratio is a very useful prediction of the simulations which
may be readily compared to our data.  For each cluster we indeed find
a gentle decline in the ratio of the total mass to the number of early
type galaxies, the mean slope of which is in very good agreement with
the prediction as shown in Fig.~\ref{Mnall}. The declining trend in
this ratio is seen in the region $0.1-1~r_{vir}$, in good agreement
with the prediction by \nocite{2005ApJ...618..557N}{Nagai} \& {Kravtsov} (2005).
\clearpage

As we showed in \S~\ref{ML_sec} the ratio of the total mass density to
bluer cluster members light density has a very steep decline with
radius -- much greater than the mild trend observed above for
early-type red sequence galaxies. The tendency towards an increasing
proportion of early-type galaxies near cluster centres (also evident
in Fig.~{~\ref{n2Lbins}}) is of course well known and is understood
empirically to result from the disruption of the gaseous disks of
infalling galaxies. Radio continuum observations reveal that
``wide-angle tails'' of gaseous material appearing to trail behind
disk galaxies \nocite{1999ApJS..120..147M}(e.g.,  {Marvel}, {Shukla} \&  {Rhee} 1999).  Recently
spectacular large trails of $H_\alpha$ emission have been detected
around disk galaxies in the Coma and Virgo clusters
\nocite{2000AJ....119..580B,2004AJ....127.3375V,2008ApJ...687L..69K}({Bravo-Alfaro} {et~al.} 2000; {Vollmer} {et~al.} 2004; {Kenney} {et~al.} 2008).
The most impressive object is a large spiral galaxy (NGC~4438) with
obvious disrupted arms and a long trail ($\sim120$~kpc) of
$H_{\alpha}$ emission revealing the orbit of this galaxy has been bent
around the massive central galaxy, M86, with a large relative velocity
of $\sim500$~km/s \nocite{2008ApJ...687L..69K}({Kenney} {et~al.} 2008).

Ram-pressure stripping has long been calculated to be important
\nocite{1972ApJ...176....1G}({Gunn} \& {Gott} 1972) for disk galaxies (in clusters), removing
preferentially the relatively low-density gas within a characteristic
stripping radius, the scale of which can be estimated from basic
considerations, and which was verified by careful simulations
\nocite{1999MNRAS.308..947A,2008A&A...481..337K}({Abadi}, {Moore} \&  {Bower} 1999; {Kronberger} {et~al.} 2008).  The loss of gas has
a transformative effect on both the morphology and the colours of such
objects, a process which is claimed as a natural explanation for the
cluster populations of S0 galaxies,
\nocite{2005ApJ...621..651H,2005ApJ...623..721P,2009ApJ...697L.137P,2009A&A...505...83H}({Homeier} {et~al.} 2005; {Postman} {et~al.} 2005; {Poggianti} {et~al.} 2009; {Huertas-Company} {et~al.} 2009)
and which is also related to the evolution of cluster gas metallicity
\nocite{2005A&A...435L..25S,2008ApJ...683L.111A}({Schindler} {et~al.} 2005; {Arieli}, {Rephaeli} \&  {Norman} 2008).

%\clearpage

\section{Cluster Luminosity Functions}\label{lum_func}

The data allow the LF to be usefully constructed in several
independent radial and magnitude bins, and hence we can examine the
form of the LF of cluster members as a function of projected distance
from the cluster centre.  By measuring the degree of dilution
separately in each magnitude bin we can construct LFs without
resorting to uncertain background subtraction.

\begin{figure*}
  \centering \includegraphics[width=15cm]{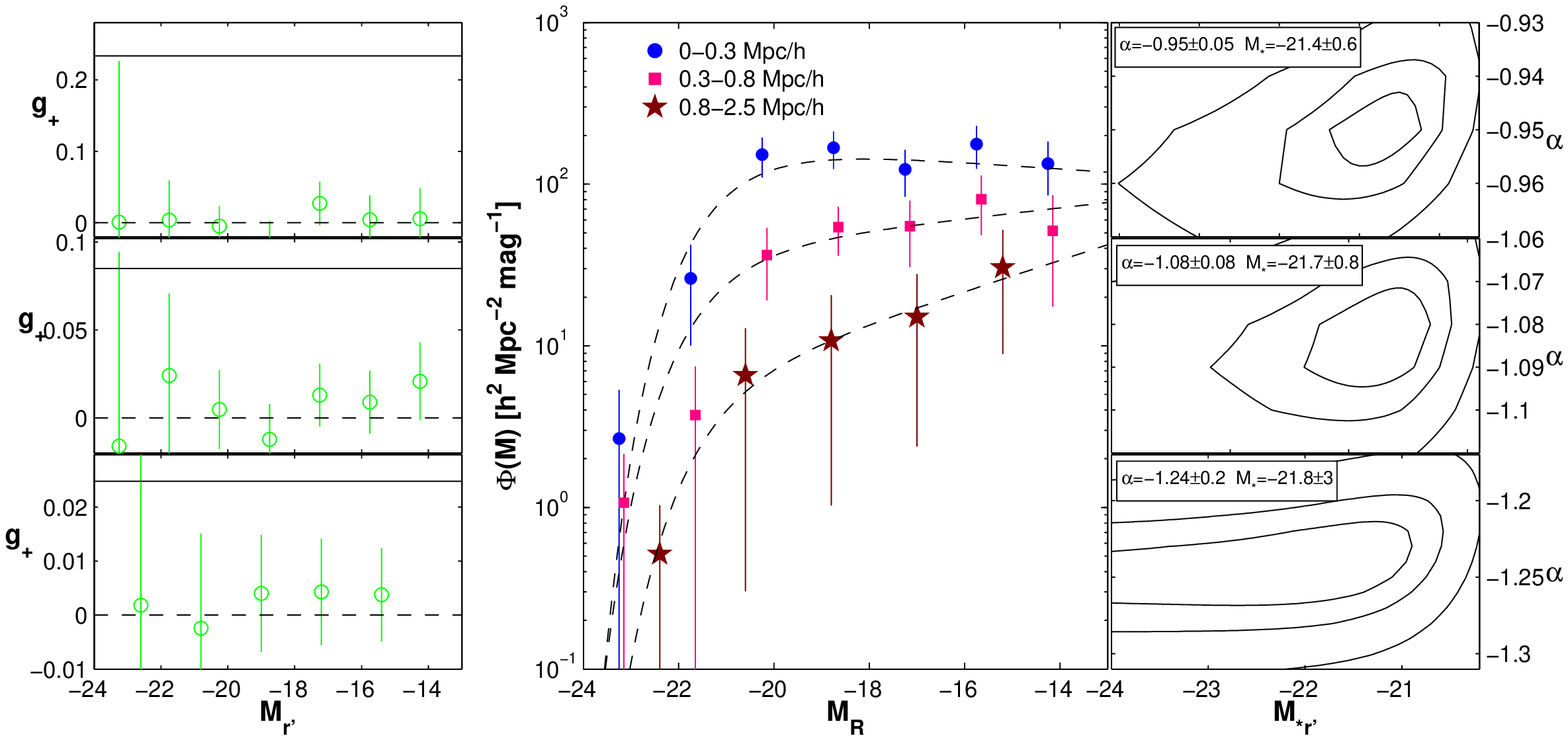}
  \centering \includegraphics[width=15cm]{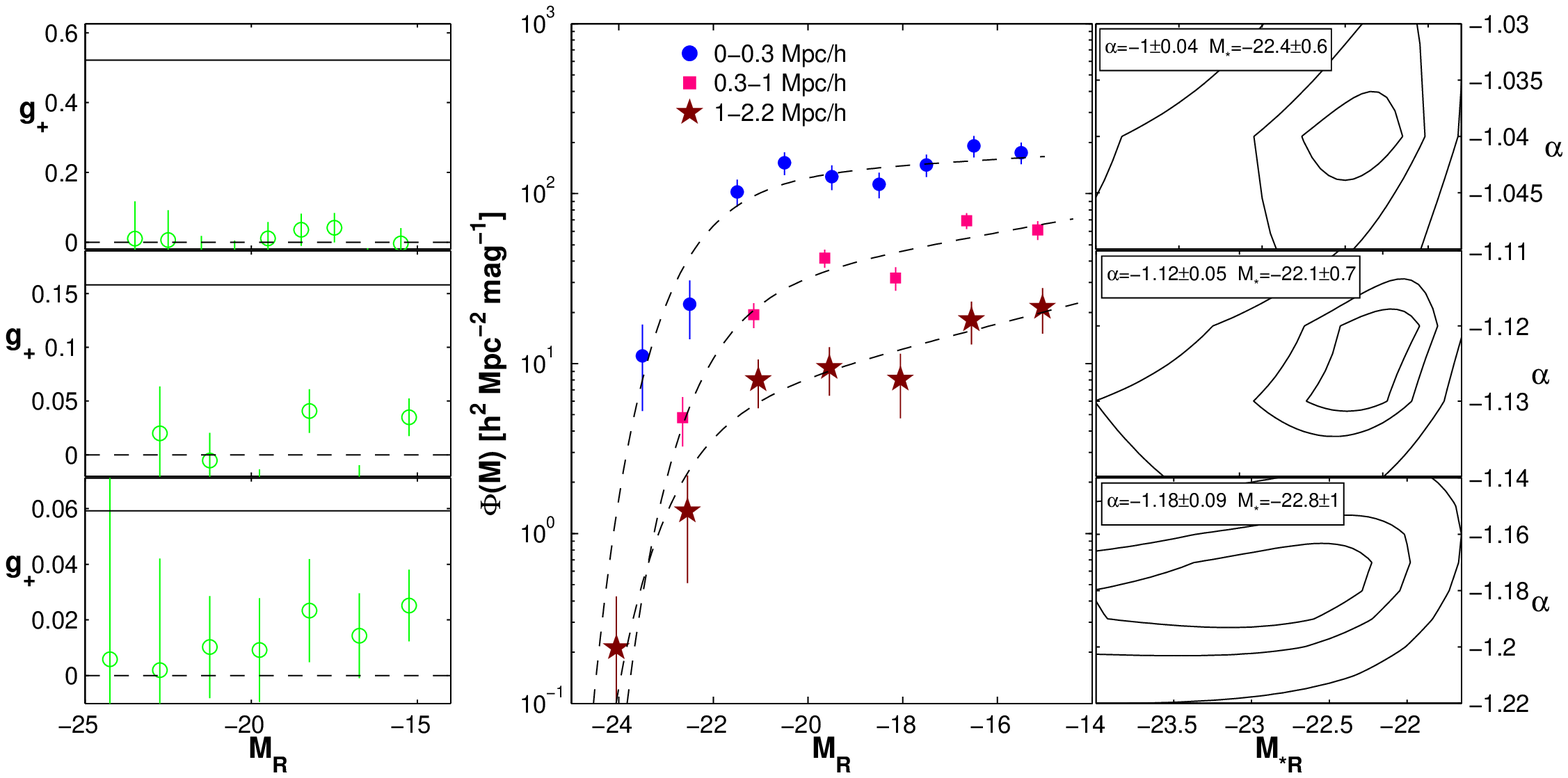}
  \centering \includegraphics[width=15cm]{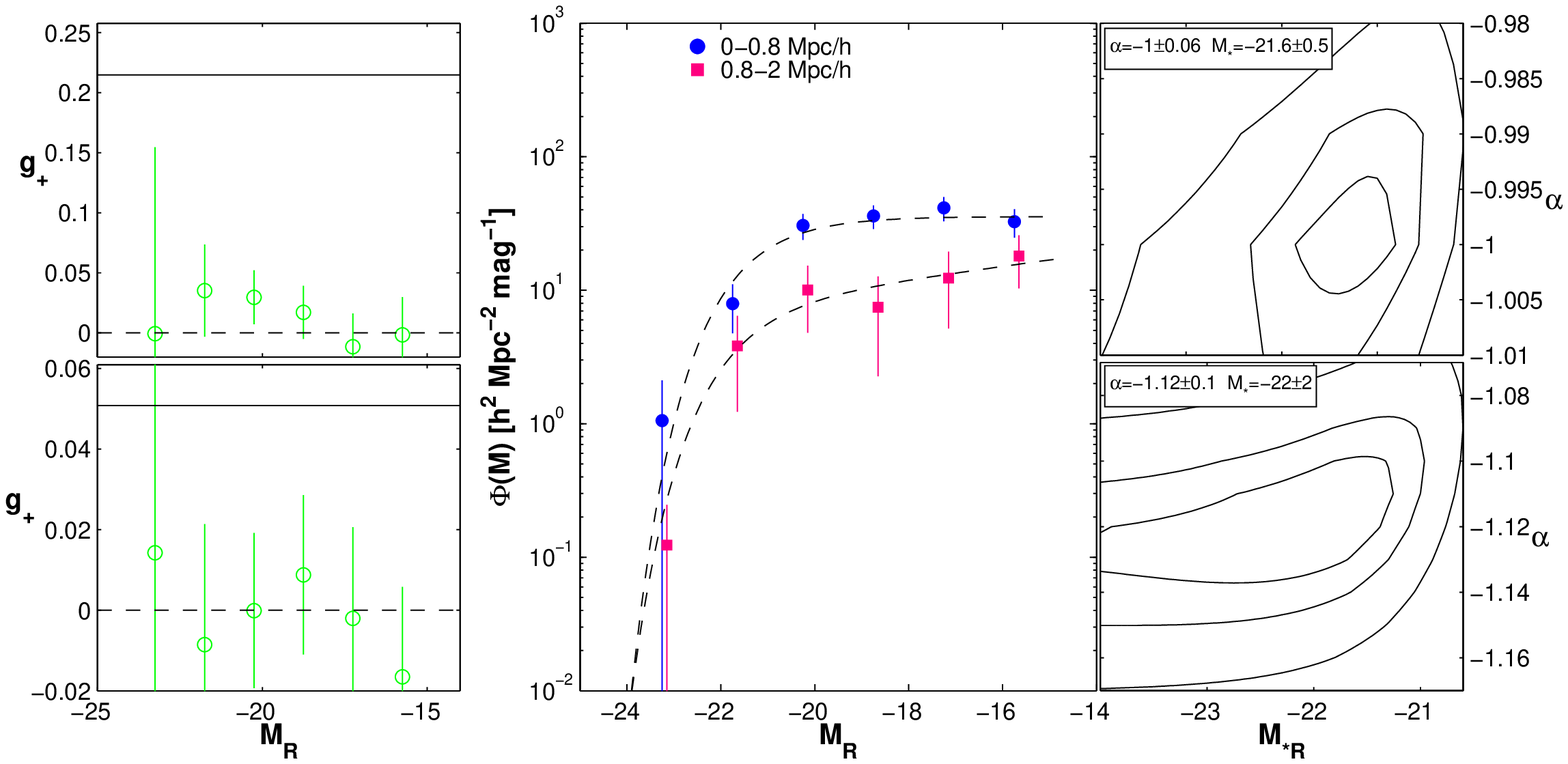}
  \caption{Luminosity functions of A1703 (top), A370 (middle) and
    RXJ1347-11 (bottom). {\it Centre part:} LFs and their Schechter
    fits are shown for several independent radial bins, indicating
    little trend of the shape of the LF with radius.  {\it Left part:}
    tangential distortion $g_T$ of the green sample as a function of
    magnitude used in the derivation of the corresponding LF.  {\it
      Right part:} $1\sigma,~2\sigma$, and $3\sigma$ contours for the
    Schechter function fits parameters $M_*$ and $\alpha$, for each of
    the corresponding LFs of each radial bin.}
  \label{LF}
\end{figure*}

Here, the dilution correction is evaluated as a function of magnitude
and radius. Left panels of Fig.~\ref{LF} show the tangential
distortion of the green sample versus absolute magnitude bin, and the
reference background distortion level is marked in the black solid
line; see \nocite{2007ApJ...663..717M}({Medezinski} {et~al.} 2007) for more details.  To correct
for this we apply equation~(\ref{eq:frac}) to each magnitude bin:
\begin{equation}
\Phi_{cl}(M_k)=\Phi(M_k)[1-\langle{}g_T^{(G)}(M_k)\rangle{}/\langle{}g_T^{(B)}(r)\rangle{}]
\end{equation}
(Note that the background signal is averaged over the whole range of
magnitudes at that radius).  The correction for background present in
the green sample is in fact not large for the brighter bins, but is
more essential at the fainter bins, with magnitude $M_R\gtrsim-18$.
This is similar to the observation we made in \S~\ref{dilution} for
the radial profile of the fraction correction, only there the
correction was a function of radius, and here it is a function of
magnitude.

We construct LFs in three broad radial annuli, except for RXJ1347-11
for which this is feasible -- due to its small angular size -- only in
two bins.  The results are shown in Fig.~\ref{LF} (centre panels).
Each LF is well fitted by a \nocite{1976ApJ...203..297S}{Schechter} (1976) function
(dashed lines), and the confidence contours for the Schechter function
parameters $M_*$ and $\alpha$ are shown on the right-hand panels of
Fig.~\ref{LF}. The Schechter fit parameters are also specified in
table~\ref{tab:LF}.  Clearly there is no evidence for a distinct
upturn, to the limits of the photometry of any of our clusters. The
slope of the Schechter function is well defined in each case, lying in
a tight range, from $\alpha=-0.95\pm0.05$ to $\alpha=-1.24\pm0.2$.
Also, there is no significant radial dependence in the shape of the
LFs with distance from the cluster centre. There may however be some
evidence that in the outer radial bin the LFs are somewhat steeper
than in the centre of the cluster for A1703, as can be seen in
Fig.~\ref{LF}, where the slope changes from $\alpha\sim-0.95$ to
$\alpha\sim-1.1$ to $\alpha\sim-1.25$ at larger radius. The knee of
the LFs, namely the bright-end break, is very distinct, and is around
$M_{*r'}\sim-21.5$ for A1703, $M_{*R}\sim-22.5$ for A370, and
$M_{*R}\sim-22$ for RXJ1347-11.

For A1689 we also saw no upturn at the faint magnitude limit, with
$\alpha=-1.05\pm0.07$ \nocite{2007ApJ...663..717M}({Medezinski} {et~al.} 2007).  There has been
some significant disagreement regarding the shapes and faint-end
slopes of LFs in clusters.  \nocite{2006A&A...445...29P}{Popesso} {et~al.} (2006) analyzing 69
RASS-SDSS clusters found a steep upturn at the faint-end by fitting a
composite of two Schechter functions, with the faint-end fit slope
being $\alpha\sim-2$.  \nocite{2008AJ....135.1837R}{Rines} \& {Geller} (2008) made a comprehensive
study of two local galaxy clusters, using spectroscopic data to
identify securely the cluster galaxies, for a more reliable
measurement of cluster LFs. They found no distinct upturn for either
of the clusters, with $\alpha=-1.13$ for A2199 and $\alpha=-1.28$ for
the Virgo cluster. When measuring the overall slope of the composite
cluster LF, i.e., up to the virial radius, we find slopes of
$\alpha\sim-1.1\pm0.1$ for all three clusters, in good agreement with
the results of \nocite{2008AJ....135.1837R}{Rines} \& {Geller} (2008).  We note that general
cluster-to-cluster variation in the total LF fits is not necessarily
unexpected, but probably depends on the specific characteristics of
individual clusters. Our clusters all show relatively flat LFs, with
only a weak radial dependence.
%\clearpage

\begin{table}
  \centering
  \caption{Parameters of Schechter fits to the Luminosity function}
  \begin{tabular}{cccc}
\hline
Cluster & radial range & $\alpha$ & $M_*$ \\
        &  (Mpc/h)   &           &  \\
\hline
A1703 & $0-0.3$ & $-0.95\pm0.05$ & $-21.4\pm0.6$\\
      & $0.3-0.8$  & $-1.08\pm0.08$ & $-21.7\pm0.8$\\
      & $0.8-2.5$ & $-1.24\pm0.2$ & $-21.8\pm3$\\
\hline
A370 & $0-0.3$ & $-1\pm0.04$ & $-22.4\pm0.6$\\
      & $0.3-1$  & $-1.12\pm0.05$ & $-22.1\pm0.7$\\
      & $1-2.2$ & $-1.18\pm0.09$ & $-22.8\pm1$\\
\hline
RXJ1347-11 & $0-0.8$ & $-1\pm0.06$ & $-21.6\pm0.5$\\
      & $0.8-2$ & $-1.12\pm0.1$ & $-22\pm2$\\
\hline
  \end{tabular}
  \label{tab:LF}
\end{table}

\section{Discussion and Conclusions}\label{discussion}

In this paper we have compared in detail the dark and luminous
properties of three high mass galaxy clusters, A1703, A370 and
RXJ1347-11, and examined the relation between mass and light all the
way out to the virial radius. We have further applied the WL dilution
approach previously developed for A1689 \nocite{2007ApJ...663..717M}({Medezinski} {et~al.} 2007) as
an alternative means of obtaining the LF and light profiles of galaxy
clusters. This based on the dilution of the lensing distortion signal
by unlensed cluster members, and has the advantage of being
independent of density fluctuations in the background.

We have examined clusters A1703, A370 and RXJ1347-11 for which data of
exceptional quality has been taken with Subaru/Suprime-Cam in multiple
passbands. We have shown how a careful separation of the cluster,
foreground, and background galaxies can be achieved using weak lensing
information galaxy colours. We also make use of the clustering of
cluster members, finding the cluster members are confined to a
distinct region of CC space with relatively small mean radial
distances. We have constructed a reliable background sample for WL
measurements, by selecting red background galaxies whose WL signal
rises all the way to the centre of the cluster, and adding to those
the faint cloud of very blue galaxies, whose WL signal we have shown
is in good agreement with the red galaxies. We have avoided dilution
of the WL signal by identifying and excluding the foreground
population in the CC diagram for which no clustering or WL signal is
detected.  Together, the red and blue samples WL measurements can be
used to reconstruct accurate mass maps of all four clusters
\nocite{2008ApJ...685L...9B}(see {Broadhurst} {et~al.} 2008).

By applying this sample selection and WL dilution correction, the LFs
are derived for each of the clusters, in several radial bins, without
the need of statistical background subtraction. The LFs extend to very
faint magnitude limits, $M_R\sim-14$, and each shows a very flat
faint-end slopes, $-0.95\leq\alpha\leq-1.24$, in agreement with our
previous study for A1689.  There are various claims in the literature
regarding the shape of the faint end of the galaxy LF
\nocite{2006A&A...445...29P,2008AJ....135.1837R,2008A&A...491..681A,
  2009ApJ...690.1158C}({Popesso} {et~al.} 2006; {Rines} \& {Geller} 2008; {Adami} {et~al.} 2008; {Crawford}, {Bershady} \&  {Hoessel} 2009), with relatively flat slopes measured and found
consistent with a simple Schechter function, and other work claiming a
distinctive upturn at faint magnitudes. Differences may conceivably
represent intrinsic differences between clusters, but a worry here is
the uncertain subtraction of the steep background counts of faint
galaxies when evaluating the background using control fields.
Uncertainties may result from inherent clustering of the background
galaxies and also variations in photometric incompleteness, or
differential extinction between the cluster and control fields.

% another LF discussion

Redshift measurements may remove concerns about the background for
complete surveys, but in practice spectroscopy can only reach
interestingly faint limiting luminosities for fairly local clusters,
requiring multiplexing capability over a wide-fields of view on large
telescopes.  \nocite{2008AJ....135.1837R}{Rines} \& {Geller} (2008) recently provided the tightest
constraints from spectroscopy out to a sizable fraction of the virial
radius, finding no significant upturn for the clusters studied.
Confidence is gaining in the use of photometric redshifts, from
careful deep multi-colour imaging, which in the work of
\nocite{2008A&A...491..681A}{Adami} {et~al.} (2008) a relatively steep LF is claimed for Coma,
but flatter slopes are measured for a sample of intermediate-redshift
clusters \nocite{2009ApJ...690.1158C}({Crawford} {et~al.} 2009).  Clearly, it is essential that
there is no significant confusion of background galaxies with cluster
members or foreground, as is achieved with our approach.

% summary of M/L behavior
The radial light profiles of these clusters are well characterized by
a power-law slope of approximately $d\log(L)/d\log(r)\simeq -1$ over
all radii. These are compared with the mass profiles from WL to
derive the radial profile of the $M/L$ ratio, a quantity which has
hitherto not been constrained in much detail, due to the difficulty of
measuring mass profiles with sufficient precision and light profiles
free of systematic effects.  Interestingly, we find that all our
clusters show a very similar radial trend of $M/L$, peaking in the
range $M/L_R\sim 500~h(M/L)_{\odot}$ at intermediate radius $\simeq
0.15R_{vir}$, and declining to approximately a limiting level of
$M/L_R\sim100~h(M/L)_{\odot}$ toward the virial radius of each
cluster.

% discussion of M/L decline 

The decline of $M/L$ towards large radius is very interesting and
seems to be a universal feature of clusters, or at least it is clearly
found in the clusters we have been able to study here in sufficient
detail and for A1689 \nocite{2007ApJ...663..717M}({Medezinski} {et~al.} 2007).  Additionally, in
the careful dynamical-based study of \nocite{2000AJ....120.2338R}{Rines} {et~al.} (2000) a
similar declining trend was found for A576. We have shown that such
behavior is not simply accounted for by the change in the stellar
content of galaxies with distance from the cluster centre.  The radial
trend of galaxy colours found is very mild and when corrected for
using standard stellar population synthesis falls well short of
explaining the marked trend of $M/L$ with radius.

% discussion of M/L decline (- late types?)
The variation of $M/L$ is however seen to be almost entirely stemming
from the much steeper density profile of blue cluster members compared
to the red sequence galaxies.  The higher relative abundance of blue
galaxies in the outer region of rich clusters is well known from local
morphological studies and deeper work with high resolution
\nocite{2000AJ....120.2338R,2004ApJ...600..657K,2005ApJ...623..721P}({Rines} {et~al.} 2000; {Katgert} {et~al.} 2004; {Postman} {et~al.} 2005)
and is part of the general morphology-density relation
\nocite{1980ApJ...236..351D}({Dressler} 1980). Such behaviour is commonly explained by
the continued conversion of disk galaxies into S0 galaxies, via tidal
and gas stripping processes \nocite{1999MNRAS.308..947A}({Abadi} {et~al.} 1999), and thought
to be consistent with the observed evolution of the numbers of S0
galaxies relative to other cluster populations measured in the
accessible redshift range $z<1$
\nocite{1980ApJ...236..351D,1997ApJ...483..582E,2005ApJ...623..721P,2009ApJ...697L.137P,2009A&A...508.1141S}({Dressler} 1980; {Ellis} {et~al.} 1997; {Postman} {et~al.} 2005; {Poggianti} {et~al.} 2009; {Simard} {et~al.} 2009).
Disk galaxies are expected to be relatively more affected than
early-type objects as their stellar content is relatively less bound
gravitationally and the gas is susceptible to ram pressure from the
ICM.

% more discussion on M/L (for early type?)
In contrast, for early-type galaxies the $M/L$ profile we obtained for
each cluster is fairly flat, with only a shallow radial decline for
$r>200h^{-1}$~kpc. Interestingly CDM based N-body simulations aimed at
evaluating the tidal influence of the cluster potential do claim to
predict a shallow radial decline in the ratio of cluster mass to the
number of sub-halos
\nocite{2005ApJ...618..557N,2005MNRAS.359.1537R,2009ApJ...693..970N}({Nagai} \& {Kravtsov} 2005; {Reed} {et~al.} 2005; {Natarajan} {et~al.} 2009),
matching well our observed trend. The simulations indicate that
although up to 70\% of a galaxy halo mass can be stripped over a
Hubble time for those objects with orbits bringing them close to the
cluster centre, the stellar content is not likely to be significantly
affected, at least for early-type galaxies, as the stars are more
tightly bound. Hence, a relatively flat relation is expected
reflecting the relatively collisionless behaviour of both galaxies and
DM.  Therefore, even though tidal effects are significant, the
luminous component that we measure when examining the $M/L$ profile of
early-type galaxies is expected to be only slightly affected, in good
agreement with the mild decline we observe for each cluster.

% future work
Further work on the tidal evolution of clusters is obviously needed.
This is particularly important in view of the interesting tension that
has now emerged between the relatively concentrated mass profiles
measured for many high mass clusters, compared to the much shallower
mass profiles predicted by standard $\Lambda$CDM
\nocite{2008MNRAS.390L..64D,2008ApJ...685L...9B,2009ApJ...699.1038O}({Duffy} {et~al.} 2008; {Broadhurst} {et~al.} 2008; {Oguri} {et~al.} 2009).
Steeper mass profiles lead to stronger tidal effects and compounded
with the greater ages implied by the relatively high concentrations
should lead to more significant tidal effects. Measuring directly the
tidal truncation of luminous cluster member galaxies should be
feasible via lensing in favorable cases where multiple images are
locally affected by individual halos. The statistical influence on the
general lensing deflection field has been claimed to be detected for
two clusters independently by \nocite{2009ApJ...693..970N}{Natarajan} {et~al.} (2009) for Cl0024+16
and by \nocite{2007ApJ...656..739H}{Halkola}, {Seitz} \&  {Pannella} (2007) for A1689.  These first results
encourage further deeper work with an emphasis on the degree of
correlation of the lensing signal with cluster member galaxies, a
formidable challenge but one well worthy of pursuit.

\section*{Acknowledgments}
We are grateful to N.~Kaiser for making the IMCAT package publicly
available. EM thanks Eran Ofek for his publicly available Matlab
scripts, and Assaf Horesh and Ido Finkelman for useful discussions. We
thank the anonymous referee for useful comments.  Work at Tel-Aviv
University was supported by Israel Science Foundation grant 214/02.
This work was also supported in part by National Science Council of
Taiwan under the grant NSC95-2112-M-001-074-MY2.MO and supported by
Department of Energy contract DE-AC02-76SF00515.

\bibliography{} 

%% --------------------------------------------------------------------
%% Sun Feb  7 19:02:58 2010
%%   This file was generated automagically from the files
%%   paper2_100204_color.bbl and paper2_100204_color.tex using
%%     nat2jour.pl
%%   This file should accompany paper2_100204_color_mnras.tex.
%% --------------------------------------------------------------------

\begin{thebibliography}{}

\bibitem[{Abadi}, {Moore}, \&  {Bower} 1999]{1999MNRAS.308..947A}
{Abadi} M.~G., {Moore} B., {Bower} R.~G., 1999, \mnras, 308, 947

\bibitem[{Adami}, {Ilbert}, {Pell{\'o}}, {Cuillandre},  {Durret}, {Mazure}, {Picat}, \& {Ulmer} 2008]{2008A&A...491..681A}
{Adami} C., {Ilbert} O., {Pell{\'o}} R., {Cuillandre} J.~C., {Durret} F.,  {Mazure} A., {Picat} J.~P., {Ulmer} M.~P., 2008, \aap, 491, 681

\bibitem[{Adelman-McCarthy}, {Ag{\"u}eros},  {Allam}, {Allende Prieto}, {Anderson}, {Anderson}, {Annis}, {Bahcall},  {Bailer-Jones}, {Baldry}, {Barentine}, {Bassett}, {Becker}, {Beers}, {Bell},  {Berlind}, {Bernardi}, {Blanton}, {Bochanski}, {Boroski}, {Brinchmann},  {Brinkmann}, {Brunner}, {Budav{\'a}ri}, {Carliles}, {Carr}, {Castander},  {Cinabro}, {Cool}, {Covey}, {Csabai}, {Cunha}, {Davenport}, {Dilday}, {Doi},  {Eisenstein}, {Evans}, {Fan}, {Finkbeiner}, {Friedman}, {Frieman},  {Fukugita}, {G{\"a}nsicke}, {Gates}, {Gillespie}, {Glazebrook}, {Gray},  {Grebel}, {Gunn}, {Gurbani}, {Hall}, {Harding}, {Harvanek}, {Hawley},  {Hayes}, {Heckman}, {Hendry}, {Hindsley}, {Hirata}, {Hogan}, {Hogg}, {Hyde},  {Ichikawa}, {Ivezi{\'c}}, {Jester}, {Johnson}, {Jorgensen}, {Juri{\'c}},  {Kent}, {Kessler}, {Kleinman}, {Knapp}, {Kron}, {Krzesinski}, {Kuropatkin},  {Lamb}, {Lampeitl}, {Lebedeva}, {Lee}, {Leger}, {L{\'e}pine}, {Lima}, {Lin},  {Long}, {Loomis}, {Loveday}, {Lupton}, {Malanushenko}, {Malanushenko},  {Mandelbaum}, {Margon}, {Marriner}, {Mart{\'{\i}}nez-Delgado}, {Matsubara},  {McGehee}, {McKay}, {Meiksin}, {Morrison}, {Munn}, {Nakajima}, {Neilsen},  {Newberg}, {Nichol}, {Nicinski}, {Nieto-Santisteban}, {Nitta}, {Okamura},  {Owen}, {Oyaizu}, {Padmanabhan}, {Pan}, {Park}, {Peoples}, {Pier}, {Pope},  {Purger}, {Raddick}, {Re Fiorentin}, {Richards}, {Richmond}, {Riess}, {Rix},  {Rockosi}, {Sako}, {Schlegel}, {Schneider}, {Schreiber}, {Schwope}, {Seljak},  {Sesar}, {Sheldon}, {Shimasaku}, {Sivarani}, {Smith}, {Snedden}, {Steinmetz},  {Strauss}, {SubbaRao}, {Suto}, {Szalay}, {Szapudi}, {Szkody}, {Tegmark},  {Thakar}, {Tremonti}, {Tucker}, {Uomoto}, {Vanden Berk}, {Vandenberg},  {Vidrih}, {Vogeley}, {Voges}, {Vogt}, {Wadadekar}, {Weinberg}, {West},  {White}, {Wilhite}, {Yanny}, {Yocum}, {York}, {Zehavi}, \&  {Zucker} 2008]{2008ApJS..175..297A}
{Adelman-McCarthy} J.~K., {Ag{\"u}eros} M.~A., {Allam} S.~S., {Allende Prieto}  C., {Anderson} K.~S.~J., {Anderson} S.~F., {Annis} J., {Bahcall} N.~A., {et al.}, 2008, \apjs,  175, 297

\bibitem[{Arieli}, {Rephaeli}, \&  {Norman} 2008]{2008ApJ...683L.111A}
{Arieli} Y., {Rephaeli} Y., {Norman} M.~L., 2008, \apjl, 683, L111

\bibitem[{Bardeau}, {Kneib}, {Czoske}, {Soucail},  {Smail}, {Ebeling}, \& {Smith} 2005]{2005A&A...434..433B}
{Bardeau} S., {Kneib} J.-P., {Czoske} O., {Soucail} G., {Smail} I., {Ebeling}  H., {Smith} G.~P., 2005, \aap, 434, 433

\bibitem[{Ben{\'{\i}}tez} 2000]{2000ApJ...536..571B}
{Ben{\'{\i}}tez} N., 2000, \apj, 536, 571

\bibitem[{Ben{\'{\i}}tez}, {Ford}, {Bouwens},  {Menanteau}, {Blakeslee}, {Gronwall}, {Illingworth}, {Meurer}, {Broadhurst},  {Clampin}, {Franx}, {Hartig}, {Magee}, {Sirianni}, {Ardila}, {Bartko},  {Brown}, {Burrows}, {Cheng}, {Cross}, {Feldman}, {Golimowski}, {Infante},  {Kimble}, {Krist}, {Lesser}, {Levay}, {Martel}, {Miley}, {Postman}, {Rosati},  {Sparks}, {Tran}, {Tsvetanov}, {White}, \& {Zheng} 2004]{2004ApJS..150....1B}
{Ben{\'{\i}}tez} N., {Ford} H., {Bouwens} R., {Menanteau} F., {Blakeslee} J.,  {Gronwall} C., {Illingworth} G., {Meurer} G., {et al.}, 2004, \apjs, 150, 1

\bibitem[{Bertin} 2006]{2006ASPC..351..112B}
{Bertin} E., 2006, in Astronomical Society of the Pacific Conference Series,  Vol. 351, Astronomical Data Analysis Software and Systems XV, {Gabriel} C.,  {Arviset} C., {Ponz} D., {Enrique} S., eds., pp. 112--+

\bibitem[{Bertin} \& {Arnouts} 1996]{1996A&AS..117..393B}
{Bertin} E., {Arnouts} S., 1996, \aaps, 117, 393

\bibitem[{Blumenthal}, {Faber}, {Flores}, \&  {Primack} 1986]{1986ApJ...301...27B}
{Blumenthal} G.~R., {Faber} S.~M., {Flores} R., {Primack} J.~R., 1986, \apj,  301, 27

\bibitem[{Bolzonella}, {Miralles}, \&  {Pell{\'o}} 2000]{2000A&A...363..476B}
{Bolzonella} M., {Miralles} J.-M., {Pell{\'o}} R., 2000, \aap, 363, 476

\bibitem[{Bravo-Alfaro}, {Cayatte}, {van Gorkom},  \& {Balkowski} 2000]{2000AJ....119..580B}
{Bravo-Alfaro} H., {Cayatte} V., {van Gorkom} J.~H., {Balkowski} C., 2000, \aj,  119, 580

\bibitem[{Broadhurst}, {Takada}, {Umetsu}, {Kong},  {Arimoto}, {Chiba}, \& {Futamase} 2005]{2005ApJ...619L.143B}
{Broadhurst} T., {Takada} M., {Umetsu} K., {Kong} X., {Arimoto} N., {Chiba} M.,  {Futamase} T., 2005, \apjl, 619, L143

\bibitem[{Broadhurst}, {Umetsu}, {Medezinski},  {Oguri}, \& {Rephaeli} 2008]{2008ApJ...685L...9B}
{Broadhurst} T., {Umetsu} K., {Medezinski} E., {Oguri} M., {Rephaeli} Y., 2008,  \apjl, 685, L9

\bibitem[{Broadhurst} \& {Barkana} 2008]{2008MNRAS.390.1647B}
{Broadhurst} T.~J., {Barkana} R., 2008, \mnras, 390, 1647

\bibitem[{Calzetti}, {Kinney}, \&  {Storchi-Bergmann} 1994]{1994ApJ...429..582C}
{Calzetti} D., {Kinney} A.~L., {Storchi-Bergmann} T., 1994, \apj, 429, 582

\bibitem[{Capak}, {Aussel}, {Ajiki}, {McCracken},  {Mobasher}, {Scoville}, {Shopbell}, {Taniguchi}, {Thompson}, {Tribiano},  {Sasaki}, {Blain}, {Brusa}, {Carilli}, {Comastri}, {Carollo}, {Cassata},  {Colbert}, {Ellis}, {Elvis}, {Giavalisco}, {Green}, {Guzzo}, {Hasinger},  {Ilbert}, {Impey}, {Jahnke}, {Kartaltepe}, {Kneib}, {Koda}, {Koekemoer},  {Komiyama}, {Leauthaud}, {Lefevre}, {Lilly}, {Liu}, {Massey}, {Miyazaki},  {Murayama}, {Nagao}, {Peacock}, {Pickles}, {Porciani}, {Renzini}, {Rhodes},  {Rich}, {Salvato}, {Sanders}, {Scarlata}, {Schiminovich}, {Schinnerer},  {Scodeggio}, {Sheth}, {Shioya}, {Tasca}, {Taylor}, {Yan}, \&  {Zamorani} 2007]{2007ApJS..172...99C}
{Capak} P., {Aussel} H., {Ajiki} M., {McCracken} H.~J., {Mobasher} B.,  {Scoville} N., {Shopbell} P., {Taniguchi} Y., {et al.}, 2007, \apjs, 172, 99

\bibitem[{Clowe} \& {Schneider} 2001]{2001A&A...379..384C}
{Clowe} D., {Schneider} P., 2001, \aap, 379, 384

\bibitem[{Coe}, {Ben{\'{\i}}tez}, {S{\'a}nchez}, {Jee},  {Bouwens}, \& {Ford} 2006]{2006AJ....132..926C}
{Coe} D., {Ben{\'{\i}}tez} N., {S{\'a}nchez} S.~F., {Jee} M., {Bouwens} R.,  {Ford} H., 2006, \aj, 132, 926

\bibitem[{Col{\'{\i}}n}, {Klypin}, {Kravtsov}, \&  {Khokhlov} 1999]{1999ApJ...523...32C}
{Col{\'{\i}}n} P., {Klypin} A.~A., {Kravtsov} A.~V., {Khokhlov} A.~M., 1999,  \apj, 523, 32

\bibitem[{Crawford}, {Bershady}, \&  {Hoessel} 2009]{2009ApJ...690.1158C}
{Crawford} S.~M., {Bershady} M.~A., {Hoessel} J.~G., 2009, \apj, 690, 1158

\bibitem[{De Lucia}, {Kauffmann}, {Springel},  {White}, {Lanzoni}, {Stoehr}, {Tormen}, \& {Yoshida} 2004]{2004MNRAS.348..333D}
{De Lucia} G., {Kauffmann} G., {Springel} V., {White} S.~D.~M., {Lanzoni} B.,  {Stoehr} F., {Tormen} G., {Yoshida} N., 2004, \mnras, 348, 333

\bibitem[{Dressler} 1980]{1980ApJ...236..351D}
{Dressler} A., 1980, \apj, 236, 351

\bibitem[{Duffy}, {Schaye}, {Kay}, \& {Dalla  Vecchia} 2008]{2008MNRAS.390L..64D}
{Duffy} A.~R., {Schaye} J., {Kay} S.~T., {Dalla Vecchia} C., 2008, \mnras, 390,  L64

\bibitem[{El-Zant}, {Shlosman}, \&  {Hoffman} 2001]{2001ApJ...560..636E}
{El-Zant} A., {Shlosman} I., {Hoffman} Y., 2001, \apj, 560, 636

\bibitem[{Ellis}, {Smail}, {Dressler}, {Couch},  {Oemler}, {Butcher}, \& {Sharples} 1997]{1997ApJ...483..582E}
{Ellis} R.~S., {Smail} I., {Dressler} A., {Couch} W.~J., {Oemler} A.~J.,  {Butcher} H., {Sharples} R.~M., 1997, \apj, 483, 582

\bibitem[{Fioc} \& {Rocca-Volmerange} 1997]{1997A&A...326..950F}
{Fioc} M., {Rocca-Volmerange} B., 1997, \aap, 326, 950

\bibitem[{Gao}, {White}, {Jenkins}, {Stoehr}, \&  {Springel} 2004]{2004MNRAS.355..819G}
{Gao} L., {White} S.~D.~M., {Jenkins} A., {Stoehr} F., {Springel} V., 2004,  \mnras, 355, 819

\bibitem[{Gavazzi} 2005]{2005A&A...443..793G}
{Gavazzi} R., 2005, \aap, 443, 793

\bibitem[{Ghigna}, {Moore}, {Governato}, {Lake},  {Quinn}, \& {Stadel} 1998]{1998MNRAS.300..146G}
{Ghigna} S., {Moore} B., {Governato} F., {Lake} G., {Quinn} T., {Stadel} J.,  1998, \mnras, 300, 146

\bibitem[{Ghigna}, {Moore}, {Governato}, {Lake},  {Quinn}, \& {Stadel} 2000]{2000ApJ...544..616G}
---, 2000, \apj, 544, 616

\bibitem[{Gunn} \& {Gott} 1972]{1972ApJ...176....1G}
{Gunn} J.~E., {Gott} J.~R.~I., 1972, \apj, 176, 1

\bibitem[{Halkola}, {Seitz}, \&  {Pannella} 2007]{2007ApJ...656..739H}
{Halkola} A., {Seitz} S., {Pannella} M., 2007, \apj, 656, 739

\bibitem[{Heymans}, {Van Waerbeke}, {Bacon}, {Berge},  {Bernstein}, {Bertin}, {Bridle}, {Brown}, {Clowe}, {Dahle}, {Erben}, {Gray},  {Hetterscheidt}, {Hoekstra}, {Hudelot}, {Jarvis}, {Kuijken}, {Margoniner},  {Massey}, {Mellier}, {Nakajima}, {Refregier}, {Rhodes}, {Schrabback}, \&  {Wittman} 2006]{2006MNRAS.368.1323H}
{Heymans} C., {Van Waerbeke} L., {Bacon} D., {Berge} J., {Bernstein} G.,  {Bertin} E., {Bridle} S., {Brown} M.~L., {et al.}, 2006, \mnras,  368, 1323

\bibitem[{Homeier}, {Demarco}, {Rosati}, {Postman},  {Blakeslee}, {Bouwens}, {Bradley}, {Ford}, {Goto}, {Gronwall}, {Holden},  {Jee}, {Martel}, {Mei}, {Menanteau}, {Zirm}, {Clampin}, {Hartig},  {Illingworth}, {Ardila}, {Bartko}, {Ben{\'{\i}}tez}, {Broadhurst}, {Brown},  {Burrows}, {Cheng}, {Cross}, {Feldman}, {Franx}, {Golimowski}, {Infante},  {Kimble}, {Krist}, {Lesser}, {Meurer}, {Miley}, {Motta}, {Sirianni},  {Sparks}, {Tran}, {Tsvetanov}, {White}, \& {Zheng} 2005]{2005ApJ...621..651H}
{Homeier} N.~L., {Demarco} R., {Rosati} P., {Postman} M., {Blakeslee} J.~P.,  {Bouwens} R.~J., {Bradley} L.~D., {Ford} H.~C., {et al.}, 2005,  \apj, 621, 651

\bibitem[{Huertas-Company}, {Foex}, {Soucail},  \& {Pell{\'o}} 2009]{2009A&A...505...83H}
{Huertas-Company} M., {Foex} G., {Soucail} G., {Pell{\'o}} R., 2009, \aap, 505,  83

\bibitem[{Ilbert}, {Capak}, {Salvato}, {Aussel},  {McCracken}, {Sanders}, {Scoville}, {Kartaltepe}, {Arnouts}, {Floc'h},  {Mobasher}, {Taniguchi}, {Lamareille}, {Leauthaud}, {Sasaki}, {Thompson},  {Zamojski}, {Zamorani}, {Bardelli}, {Bolzonella}, {Bongiorno}, {Brusa},  {Caputi}, {Carollo}, {Contini}, {Cook}, {Coppa}, {Cucciati}, {de la Torre},  {de Ravel}, {Franzetti}, {Garilli}, {Hasinger}, {Iovino}, {Kampczyk},  {Kneib}, {Knobel}, {Kovac}, {LeBorgne}, {LeBrun}, {F{\`e}vre}, {Lilly},  {Looper}, {Maier}, {Mainieri}, {Mellier}, {Mignoli}, {Murayama}, {Pell{\`o}},  {Peng}, {P{\'e}rez-Montero}, {Renzini}, {Ricciardelli}, {Schiminovich},  {Scodeggio}, {Shioya}, {Silverman}, {Surace}, {Tanaka}, {Tasca}, {Tresse},  {Vergani}, \& {Zucca} 2009]{2009ApJ...690.1236I}
{Ilbert} O., {Capak} P., {Salvato} M., {Aussel} H., {McCracken} H.~J.,  {Sanders} D.~B., {Scoville} N., {Kartaltepe} J., {et al.}, 2009, \apj,  690, 1236

\bibitem[{Kaiser}, {Squires}, \&  {Broadhurst} 1995]{1995ApJ...449..460K}
{Kaiser} N., {Squires} G., {Broadhurst} T., 1995, \apj, 449, 460

\bibitem[{Kashikawa}, {Shimasaku}, {Yasuda},  {Ajiki}, {Akiyama}, {Ando}, {Aoki}, {Doi}, {Fujita}, {Furusawa}, {Hayashino},  {Iwamuro}, {Iye}, {Karoji}, {Kobayashi}, {Kodaira}, {Kodama}, {Komiyama},  {Matsuda}, {Miyazaki}, {Mizumoto}, {Morokuma}, {Motohara}, {Murayama},  {Nagao}, {Nariai}, {Ohta}, {Okamura}, {Ouchi}, {Sasaki}, {Sato}, {Sekiguchi},  {Shioya}, {Tamura}, {Taniguchi}, {Umemura}, {Yamada}, \&  {Yoshida} 2004]{2004PASJ...56.1011K}
{Kashikawa} N., {Shimasaku} K., {Yasuda} N., {Ajiki} M., {Akiyama} M., {Ando}  H., {Aoki} K., {Doi} M., {et al.}, 2004, \pasj, 56,  1011

\bibitem[{Katgert}, {Biviano}, \&  {Mazure} 2004]{2004ApJ...600..657K}
{Katgert} P., {Biviano} A., {Mazure} A., 2004, \apj, 600, 657

\bibitem[{Kenney}, {Tal}, {Crowl}, {Feldmeier}, \&  {Jacoby} 2008]{2008ApJ...687L..69K}
{Kenney} J.~D.~P., {Tal} T., {Crowl} H.~H., {Feldmeier} J., {Jacoby} G.~H.,  2008, \apjl, 687, L69

\bibitem[{Kinney}, {Calzetti}, {Bohlin}, {McQuade},  {Storchi-Bergmann}, \& {Schmitt} 1996]{1996ApJ...467...38K}
{Kinney} A.~L., {Calzetti} D., {Bohlin} R.~C., {McQuade} K., {Storchi-Bergmann}  T., {Schmitt} H.~R., 1996, \apj, 467, 38

\bibitem[{Kneib}, {Hudelot}, {Ellis}, {Treu}, {Smith},  {Marshall}, {Czoske}, {Smail}, \& {Natarajan} 2003]{2003ApJ...598..804K}
{Kneib} J.-P., {Hudelot} P., {Ellis} R.~S., {Treu} T., {Smith} G.~P.,  {Marshall} P., {Czoske} O., {Smail} I., {et al.}, 2003, \apj, 598, 804

\bibitem[{Kotulla}, {Fritze}, {Weilbacher}, \&  {Anders} 2009]{2009MNRAS.396..462K}
{Kotulla} R., {Fritze} U., {Weilbacher} P., {Anders} P., 2009, \mnras, 396, 462

\bibitem[{Kronberger}, {Kapferer}, {Ferrari},  {Unterguggenberger}, \& {Schindler} 2008]{2008A&A...481..337K}
{Kronberger} T., {Kapferer} W., {Ferrari} C., {Unterguggenberger} S.,  {Schindler} S., 2008, \aap, 481, 337

\bibitem[{Lacey} \& {Cole} 1993]{1993MNRAS.262..627L}
{Lacey} C., {Cole} S., 1993, \mnras, 262, 627

\bibitem[{Lapi} \& {Cavaliere} 2009]{2009ApJ...695L.125L}
{Lapi} A., {Cavaliere} A., 2009, \apjl, 695, L125

\bibitem[{Limousin}, {Richard}, {Jullo}, {Kneib},  {Fort}, {Soucail}, {El{\'{\i}}asd{\'o}ttir}, {Natarajan}, {Ellis}, {Smail},  {Czoske}, {Smith}, {Hudelot}, {Bardeau}, {Ebeling}, {Egami}, \&  {Knudsen} 2007]{2007ApJ...668..643L}
{Limousin} M., {Richard} J., {Jullo} E., {Kneib} J.-P., {Fort} B., {Soucail}  G., {El{\'{\i}}asd{\'o}ttir} {\'A}., {Natarajan} P., {et al.}, 2007, \apj, 668, 643

\bibitem[{Limousin}, {Sommer-Larsen}, {Natarajan}, \&  {Milvang-Jensen} 2009]{2009ApJ...696.1771L}
{Limousin} M., {Sommer-Larsen} J., {Natarajan} P., {Milvang-Jensen} B., 2009,  \apj, 696, 1771

\bibitem[{Lu}, {Gilbank}, {Balogh}, {Milkeraitis},  {Hoekstra}, {Van Waerbeke}, {Wake}, {Edge}, \& {Bower} 2009]{2009arXiv0912.2356L}
{Lu} T., {Gilbank} D.~G., {Balogh} M.~L., {Milkeraitis} M., {Hoekstra} H., {Van  Waerbeke} L., {Wake} D.~A., {Edge} A.~C., {et al.}, 2009, ArXiv e-prints

\bibitem[{Marvel}, {Shukla}, \&  {Rhee} 1999]{1999ApJS..120..147M}
{Marvel} K.~B., {Shukla} H., {Rhee} G., 1999, \apjs, 120, 147

\bibitem[{Medezinski}, {Broadhurst}, {Umetsu},  {Coe}, {Ben{\'{\i}}tez}, {Ford}, {Rephaeli}, {Arimoto}, \&  {Kong} 2007]{2007ApJ...663..717M}
{Medezinski} E., {Broadhurst} T., {Umetsu} K., {Coe} D., {Ben{\'{\i}}tez} N.,  {Ford} H., {Rephaeli} Y., {Arimoto} N., {et al.}, 2007, \apj, 663, 717

\bibitem[{Miyazaki}, {Komiyama}, {Sekiguchi},  {Okamura}, {Doi}, {Furusawa}, {Hamabe}, {Imi}, {Kimura}, {Nakata}, {Okada},  {Ouchi}, {Shimasaku}, {Yagi}, \& {Yasuda} 2002]{2002PASJ...54..833M}
{Miyazaki} S., {Komiyama} Y., {Sekiguchi} M., {Okamura} S., {Doi} M.,  {Furusawa} H., {Hamabe} M., {Imi} K., {et al.}, 2002, \pasj, 54, 833

\bibitem[{Moore}, {Katz}, {Lake}, {Dressler}, \&  {Oemler} 1996]{1996Natur.379..613M}
{Moore} B., {Katz} N., {Lake} G., {Dressler} A., {Oemler} A., 1996, \nat, 379,  613

\bibitem[{Nagai} \& {Kravtsov} 2005]{2005ApJ...618..557N}
{Nagai} D., {Kravtsov} A.~V., 2005, \apj, 618, 557

\bibitem[{Natarajan}, {Kneib}, {Smail}, {Treu},  {Ellis}, {Moran}, {Limousin}, \& {Czoske} 2009]{2009ApJ...693..970N}
{Natarajan} P., {Kneib} J.-P., {Smail} I., {Treu} T., {Ellis} R., {Moran} S.,  {Limousin} M., {Czoske} O., 2009, \apj, 693, 970

\bibitem[{Natarajan}, {Loeb}, {Kneib}, \&  {Smail} 2002]{2002ApJ...580L..17N}
{Natarajan} P., {Loeb} A., {Kneib} J.-P., {Smail} I., 2002, \apjl, 580, L17

\bibitem[{Navarro}, {Frenk}, \&  {White} 1997]{1997ApJ...490..493N}
{Navarro} J.~F., {Frenk} C.~S., {White} S.~D.~M., 1997, \apj, 490, 493

\bibitem[{Oguri} \& {Blandford} 2009]{2009MNRAS.392..930O}
{Oguri} M., {Blandford} R.~D., 2009, \mnras, 392, 930

\bibitem[{Oguri}, {Hennawi}, {Gladders}, {Dahle},  {Natarajan}, {Dalal}, {Koester}, {Sharon}, \&  {Bayliss} 2009]{2009ApJ...699.1038O}
{Oguri} M., {Hennawi} J.~F., {Gladders} M.~D., {Dahle} H., {Natarajan} P.,  {Dalal} N., {Koester} B.~P., {Sharon} K., {et al.}, 2009, \apj, 699, 1038

\bibitem[{Okabe}, {Takada}, {Umetsu}, {Futamase}, \&  {Smith} 2009]{2009arXiv0903.1103O}
{Okabe} N., {Takada} M., {Umetsu} K., {Futamase} T., {Smith} G.~P., 2009, ArXiv  e-prints

\bibitem[{Ouchi}, {Shimasaku}, {Okamura}, {Furusawa},  {Kashikawa}, {Ota}, {Doi}, {Hamabe}, {Kimura}, {Komiyama}, {Miyazaki},  {Miyazaki}, {Nakata}, {Sekiguchi}, {Yagi}, \& {Yasuda} 2004]{2004ApJ...611..660O}
{Ouchi} M., {Shimasaku} K., {Okamura} S., {Furusawa} H., {Kashikawa} N., {Ota}  K., {Doi} M., {Hamabe} M., {et al.}, 2004,  \apj, 611, 660

\bibitem[{Poggianti}, {Fasano}, {Bettoni}, {Cava},  {Dressler}, {Vanzella}, {Varela}, {Couch}, {D'Onofrio}, {Fritz},  {Kjaergaard}, {Moles}, \& {Valentinuzzi} 2009]{2009ApJ...697L.137P}
{Poggianti} B.~M., {Fasano} G., {Bettoni} D., {Cava} A., {Dressler} A.,  {Vanzella} E., {Varela} J., {Couch} W.~J., {et al.}, 2009, \apjl, 697, L137

\bibitem[{Popesso}, {Biviano}, {B{\"o}hringer}, \&  {Romaniello} 2006]{2006A&A...445...29P}
{Popesso} P., {Biviano} A., {B{\"o}hringer} H., {Romaniello} M., 2006, \aap,  445, 29

\bibitem[{Postman}, {Franx}, {Cross}, {Holden},  {Ford}, {Illingworth}, {Goto}, {Demarco}, {Rosati}, {Blakeslee}, {Tran},  {Ben{\'{\i}}tez}, {Clampin}, {Hartig}, {Homeier}, {Ardila}, {Bartko},  {Bouwens}, {Bradley}, {Broadhurst}, {Brown}, {Burrows}, {Cheng}, {Feldman},  {Golimowski}, {Gronwall}, {Infante}, {Kimble}, {Krist}, {Lesser}, {Martel},  {Mei}, {Menanteau}, {Meurer}, {Miley}, {Motta}, {Sirianni}, {Sparks}, {Tran},  {Tsvetanov}, {White}, \& {Zheng} 2005]{2005ApJ...623..721P}
{Postman} M., {Franx} M., {Cross} N.~J.~G., {Holden} B., {Ford} H.~C.,  {Illingworth} G.~D., {Goto} T., {Demarco} R., {et al.}, 2005, \apj, 623,  721

\bibitem[{Reed}, {Governato}, {Quinn}, {Gardner},  {Stadel}, \& {Lake} 2005]{2005MNRAS.359.1537R}
{Reed} D., {Governato} F., {Quinn} T., {Gardner} J., {Stadel} J., {Lake} G.,  2005, \mnras, 359, 1537

\bibitem[{Rines} \& {Geller} 2008]{2008AJ....135.1837R}
{Rines} K., {Geller} M.~J., 2008, \aj, 135, 1837

\bibitem[{Rines}, {Geller}, {Diaferio}, {Kurtz}, \&  {Jarrett} 2004]{2004AJ....128.1078R}
{Rines} K., {Geller} M.~J., {Diaferio} A., {Kurtz} M.~J., {Jarrett} T.~H.,  2004, \aj, 128, 1078

\bibitem[{Rines}, {Geller}, {Diaferio}, {Mohr}, \&  {Wegner} 2000]{2000AJ....120.2338R}
{Rines} K., {Geller} M.~J., {Diaferio} A., {Mohr} J.~J., {Wegner} G.~A., 2000,  \aj, 120, 2338

\bibitem[{Sadeh} \& {Rephaeli} 2008]{2008MNRAS.388.1759S}
{Sadeh} S., {Rephaeli} Y., 2008, \mnras, 388, 1759

\bibitem[{Schechter} 1976]{1976ApJ...203..297S}
{Schechter} P., 1976, \apj, 203, 297

\bibitem[{Schindler}, {Kapferer}, {Domainko},  {Mair}, {van Kampen}, {Kronberger}, {Kimeswenger}, {Ruffert}, {Mangete}, \&  {Breitschwerdt} 2005]{2005A&A...435L..25S}
{Schindler} S., {Kapferer} W., {Domainko} W., {Mair} M., {van Kampen} E.,  {Kronberger} T., {Kimeswenger} S., {Ruffert} M., {et al.}, 2005, \aap, 435, L25

\bibitem[{Simard}, {Clowe}, {Desai}, {Dalcanton}, {von  der Linden}, {Poggianti}, {White}, {Arag{\'o}n-Salamanca}, {De Lucia},  {Halliday}, {Jablonka}, {Milvang-Jensen}, {Saglia}, {Pell{\'o}}, {Rudnick},  \& {Zaritsky} 2009]{2009A&A...508.1141S}
{Simard} L., {Clowe} D., {Desai} V., {Dalcanton} J.~J., {von der Linden} A.,  {Poggianti} B.~M., {White} S.~D.~M., {Arag{\'o}n-Salamanca} A., {et al.}, 2009, \aap, 508, 1141

\bibitem[{Springel}, {White}, {Tormen}, \&  {Kauffmann} 2001]{2001MNRAS.328..726S}
{Springel} V., {White} S.~D.~M., {Tormen} G., {Kauffmann} G., 2001, \mnras,  328, 726

\bibitem[{Umetsu}, {Birkinshaw}, {Liu},  {Wu}, {Medezinski}, {Broadhurst}, {Lemze}, {Zitrin}, {Ho}, {Huang}, {Koch},  {Liao}, {Lin}, {Molnar}, {Nishioka}, {Wang}, {Altamirano}, {Chang}, {Chang},  {Chang}, {Chen}, {Han}, {Huang}, {Hwang}, {Jiang}, {Kesteven}, {Kubo}, {Li},  {Martin-Cocher}, {Oshiro}, {Raffin}, {Wei}, \&  {Wilson} 2009a]{2009ApJ...694.1643U}
{Umetsu} K., {Birkinshaw} M., {Liu} G.-C., {Wu} J.-H.~P., {Medezinski} E.,  {Broadhurst} T., {Lemze} D., {Zitrin} A., {et al.}, 2009a, \apj, 694, 1643

\bibitem[{Umetsu} \& {Broadhurst} 2008]{2008ApJ...684..177U}
{Umetsu} K., {Broadhurst} T., 2008, \apj, 684, 177

\bibitem[{Umetsu}, {Medezinski},  {Broadhurst}, {Zitrin}, {Okabe}, {Hsieh}, \& {Molnar} 2009b]{2009arXiv0908.0069U}
{Umetsu} K., {Medezinski} E., {Broadhurst} T., {Zitrin} A., {Okabe} N., {Hsieh}  B., {Molnar} S.~M., 2009b, ArXiv e-prints

\bibitem[{Vollmer}, {Beck}, {Kenney}, \& {van  Gorkom} 2004]{2004AJ....127.3375V}
{Vollmer} B., {Beck} R., {Kenney} J.~D.~P., {van Gorkom} J.~H., 2004, \aj, 127,  3375

\bibitem[{Wuyts}, {Franx}, {Cox}, {Hernquist},  {Hopkins}, {Robertson}, \& {van Dokkum} 2009]{2009ApJ...696..348W}
{Wuyts} S., {Franx} M., {Cox} T.~J., {Hernquist} L., {Hopkins} P.~F.,  {Robertson} B.~E., {van Dokkum} P.~G., 2009, \apj, 696, 348

\bibitem[{Yagi}, {Kashikawa}, {Sekiguchi}, {Doi},  {Yasuda}, {Shimasaku}, \& {Okamura} 2002]{2002AJ....123...66Y}
{Yagi} M., {Kashikawa} N., {Sekiguchi} M., {Doi} M., {Yasuda} N., {Shimasaku}  K., {Okamura} S., 2002, \aj, 123, 66

\bibitem[{Zacharias}, {Monet}, {Levine}, {Urban},  {Gaume}, \& {Wycoff} 2004]{2004AAS...205.4815Z}
{Zacharias} N., {Monet} D.~G., {Levine} S.~E., {Urban} S.~E., {Gaume} R.,  {Wycoff} G.~L., 2004, in Bulletin of the American Astronomical Society,  Vol.~36, Bulletin of the American Astronomical Society, pp. 1418--+

\bibitem[{Zitrin}, {Broadhurst}, {Umetsu}, {Coe},  {Ben{\'{\i}}tez}, {Ascaso}, {Bradley}, {Ford}, {Jee}, {Medezinski},  {Rephaeli}, \& {Zheng} 2009]{2009MNRAS.396.1985Z}
{Zitrin} A., {Broadhurst} T., {Umetsu} K., {Coe} D., {Ben{\'{\i}}tez} N.,  {Ascaso} B., {Bradley} L., {Ford} H., {et al.}, 2009, \mnras, 396, 1985

\end{thebibliography}

\end{document}